\documentclass[aps,prb, twocolumn, times,superscriptaddress,showpacs,a4paper]{revtex4-1}
\usepackage{colortbl,marvosym}
\usepackage{amsfonts,amsmath,amssymb}
\usepackage{graphicx,wasysym}
\usepackage{tabularx,hhline,makecell}
\usepackage[dvipsnames]{xcolor}
\usepackage{multirow}% http://ctan.org/pkg/multirow
\usepackage[colorlinks,bookmarks=false,citecolor=blue,linkcolor=red,urlcolor=blue]{hyperref}
\usepackage{color}
\usepackage{tabularx,enumitem}
\usepackage{mathtools}
\usepackage[caption=false]{subfig}

\DeclareMathAlphabet{\mathpzc}{OT1}{pzc}{m}{it}

\makeatletter
\newcommand*{\rom}[1]{\expandafter\@slowromancap\romannumeral #1@}
\makeatother

\begin{document}
%%%%%%%%%%%%%%%%%%%%%%%%%%%%%%%%%%%%%%%%%%%%
\graphicspath{{./Images/}}
%%%%%%%%%%%%%%%%%%%%%%%%%%%%%%%%%%%%%%%%%%%%

%%%%%%%%%%%%%%%%%%%%%%%%%%%%%%%%%%%%%%%%%%%%
\title{Curie-law crossover in spin liquids}
%%%%%%%%%%%%%%%%%%%%%%%%%%%%%%%%%%%%%%%%%%%%

%%%%%%%%%%%%%%%%%%%%%%%%%%%%%%%%%%%%%%%%%%%%
\author{Rico Pohle}
\affiliation{Theory of Quantum Matter Unit, Okinawa Institute of Science and Technology 
Graduate University, Onna-son, Okinawa 904-0412, Japan}
\affiliation{
Department of Applied Physics, University of Tokyo, Hongo, 
Bunkyo-ku, Tokyo 113-8656, Japan}

%%%%%%%%%%%%%%%%%%%%%%%%%%%%%%%%%%%%%%%%%%%%
\author{Ludovic D. C. Jaubert}
\affiliation{CNRS, University of Bordeaux, LOMA, 
UMR 5798, F-33400 Talence, France}

%%%%%%%%%%%%%%%%%%%%%%%%%%%%%%%%%%%%%%%%%%%%
\date{\today} 
%%%%%%%%%%%%%%%%%%%%%%%%%%%%%%%%%%%%%%%%%%%%

%%%%%%%%%%%%%%%%%%%%%%%%%%%%%%%%%%%%%%%%%%%%
%
% 							ABSTRACT
%
%%%%%%%%%%%%%%%%%%%%%%%%%%%%%%%%%%%%%%%%%%%%
\begin{abstract}
%%%%%%%%%%%%%%%%%%%%%%%%%%%%%%%%%%%%%%%%%%%%
The Curie-Weiss law is widely used to estimate the strength of frustration in frustrated magnets. However, the Curie-Weiss law was originally derived as an estimate of magnetic correlations close to a mean-field phase transition, which -- by definition -- is absent in spin liquids. 
Instead, the susceptibility of spin liquids is known to undergo a Curie-law crossover between two magnetically disordered regimes. Here, we study the generic aspect of the Curie-law crossover by comparing a variety of frustrated spin models in two and three dimensions, using both classical Monte Carlo simulations and analytical Husimi tree calculations. Husimi tree calculations fit remarkably well the simulations for all temperatures and almost all lattices. We also propose a Husimi Ansatz for the reduced susceptibility $\chi T$, to be used in complement to the traditional Curie-Weiss fit in order to estimate the Curie-Weiss temperature $\theta_{\rm cw}$. Applications to materials are discussed.
%%%%%%%%%%%%%%%%%%%%%%%%%%%%%%%%%%%%%%%%%%%%
\end{abstract}
%%%%%%%%%%%%%%%%%%%%%%%%%%%%%%%%%%%%%%%%%%%%
\maketitle
%%%%%%%%%%%%%%%%%%%%%%%%%%%%%%%%%%%%%%%%%%%%

%%%%%%%%%%%%%%%%%%%%%%%%%%%%%%%%%%%%%%%%%%%%%
%
%					INTRODUCTION
%
%%%%%%%%%%%%%%%%%%%%%%%%%%%%%%%%%%%%%%%%%%%%%
\section{Introduction}
\label{sec:Intro}
%%%%%%%%%%%%%%%%%%%%%%%%%%%%%%%%%%%%%%%%%%
%

%%%%%%    history Curie-Weiss
%
The Curie-Weiss law is a simple and useful tool to estimate the behavior of the susceptibility $\chi$ for conventional magnets at high temperatures 
\cite{Curie1895, Weiss1907, Kittel, AshkroftMermin,mugiraneza2022}
%
%%%%%%%%%%%%%%%%%%%%%%%%%%
\begin{equation}
	\chi 	= \frac{C}{T - \theta_{\text{cw}}} 	\ ,
\label{eq:CW}
\end{equation}
%%%%%%%%%%%%%%%%%%%%%%%%%%
%
with $C$ the Curie constant, and $\theta_{\text{cw}}$ the Curie-Weiss temperature. 
In a Landau mean-field treatment \cite{Landau1937},  $|\theta_{\text{cw}}|$ represents the transition temperature.
The sign of $ \theta_{\text{cw}} $ indicates dominant ferromagnetic ($ \theta_{\text{cw}} > 0$) or 
antiferromagnetic ($ \theta_{\text{cw}} < 0$) 
interactions, while the limit $ \theta_{\text{cw}}  \to 0$ represents the susceptibility 
of a paramagnet, given by the Curie law, $\chi = C/T$. For more details about the calculation of the Curie-Weiss law in susceptibility measurements, we refer the reader to the recent tutorial by Mugiraneza \& Hallas [\onlinecite{mugiraneza2022}].

%%%%%%   application to frustrated magnets
%
In frustrated magnets, the Curie-Weiss temperature is often used to measure the ``frustration index'' \cite{Ramirez1994}
%
%%%%%%%%%%%%%%%%%%%%%%%%%%%
\begin{equation}
	f = \frac{| \theta_{\text{cw}} |}{T^{*}}     \,,
\label{eq:frustindex}
\end{equation}
%%%%%%%%%%%%%%%%%%%%%%%%%%%
%
by comparing the transition, or freezing, temperature of a material, $T^{*}$, to its mean-field expectation, $| \theta_{\text{cw}} |$, for an unfrustrated system. Large values of $f$ account for strong frustration in the system. For a spin liquid where $T^{*}\rightarrow 0^{+}$ theoretically, the frustration index diverges. Being a priori readily accessible to experiments, this quantity $f$ has become a convenient tool to gauge how frustrated a system is.

But as many successful, broadly used indicators, a few shortcomings are inevitable. Deviations from the standard Curie-Weiss law have been studied in a variety of magnetic systems, such as spin glasses  \cite{Nagasawa67a,Morgownik81a}, the pyrochlore molybdate Y$_{2}$Mo$_{2}$O$_{7}$ \cite{Silverstein14a}, the valence bond glass Ba$_2$YMoO$_6$ \cite{deVries10a}, or Kitaev materials with strong spin-orbit coupling \cite{Li2021}, to cite but a few. For example in anisotropic lattices, the high-temperature Curie constant and low-temperature transition temperature may be set by different energy scales, giving rise to an artificially large parameter $f$ even when the system is barely frustrated \cite{Pohle2016,Schmidt17a}.

In spin liquids, this deviation has been rationalized as the onset of a Curie-law crossover \cite{Jaubert09c, Jaubert2013} between the universal high-temperature Curie law and a low-temperature, model-specific, spin-liquid Curie law \cite{Isakov04,Isoda08a,Macdonald11a,Jaubert2013}. The problem is that fitting the susceptibility of spin liquids with a Curie-Weiss law always gives an answer, but not necessarily the right one, as illustrated for the Ising kagome antiferromagnet in Fig.~\ref{fig:CurieLaw.Crossover}. Beyond the traditional difficulties to measure the Curie-Weiss temperature \cite{mugiraneza2022,Li2021}, frustration precisely prevents the phase transition in spin liquids that would justify the Curie-Weiss fit as a mean-field approximation of a scaling law with critical exponent $\gamma=1$. Eq.~(\ref{eq:CW}) is only valid at high temperature as a first order perturbation of the Curie law. And whether this high-temperature regime is accessible to experiments then becomes an important question \cite{mugiraneza2022,Li2021}. Internal energy scales such as a single-ion crystal field, a band gap, the structural distortion of the lattice or even the melting of the materials might prevent access to the necessary high temperatures. In that case, the values of the Curie constant and Curie-Weiss temperature strongly depend on the temperature range of the fitting procedure \cite{Jaubert2013,Nag2017}. The latter can even change sign when the exchange coupling is particularly small (see e.g. Refs.~[\onlinecite{Bramwell2000, Lummen2008}] for Dy$_{2}$Ti$_{2}$O$_{7}$). And as a high-temperature expansion of the susceptibility, the Curie-Weiss fit is not designed to capture the spin-liquid behavior at low temperatures. 

To summarise the issue, applying the Curie-Weiss fit to frustrated magnets means applying a method that has been derived around a mean-field critical point, to a class of systems where this critical point is absent by definition.

Our goal in this paper is to rationalise this conceptual divergence of viewpoints and to build a generic understanding of the Curie-law crossover in spin liquids. Is it possible to quantify how the magnetic susceptibility deviates from the Curie-Weiss law, not just for a specific model but for frustrated magnets in general ? In particular, can we identify generic features ? Practically, understanding the limits of the Curie-Weiss fit will help estimate the appropriate temperature window to measure the Curie-Weiss temperature, and what to do when this window is not experimentally available.

%%%%%%%%%%%%%%%%%%%%%%%%%%%%%%%%%%%%%%%%%%%%%
%
%			   		SUMMARY
%
%%%%%%%%%%%%%%%%%%%%%%%%%%%%%%%%%%%%%%%%%%%%%
\section{Summary of results}
 \label{sec:Summary}
\begin{figure*}[t]
\centering
\captionsetup[subfigure]{justification=justified, singlelinecheck=false, position=top}
	\subfloat[\label{fig:MC_Triangular} triangular] {\includegraphics[width=0.2\textwidth]{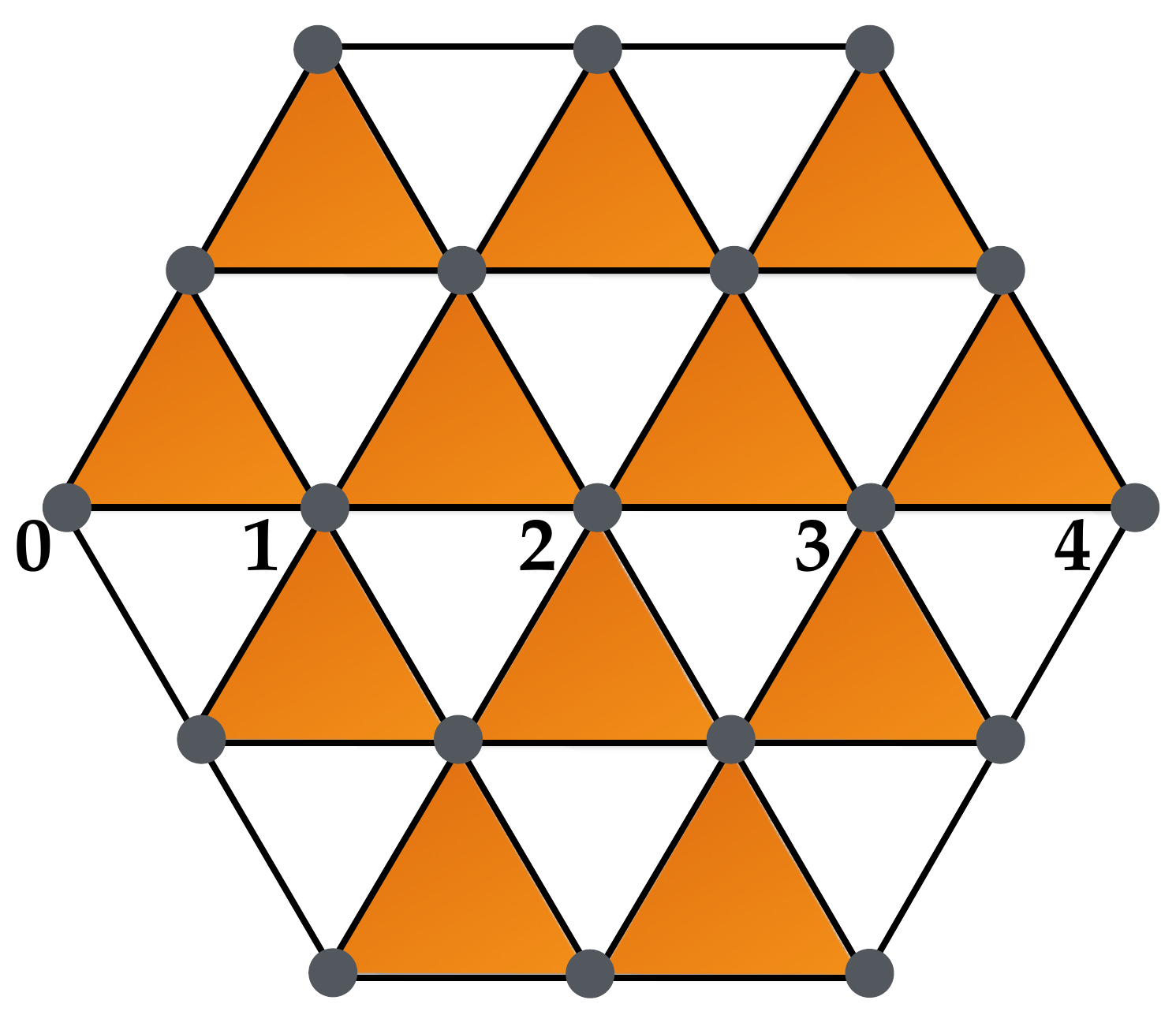}}
	\subfloat[\label{fig:MC_Kagome} kagome] {\includegraphics[width=0.2\textwidth]{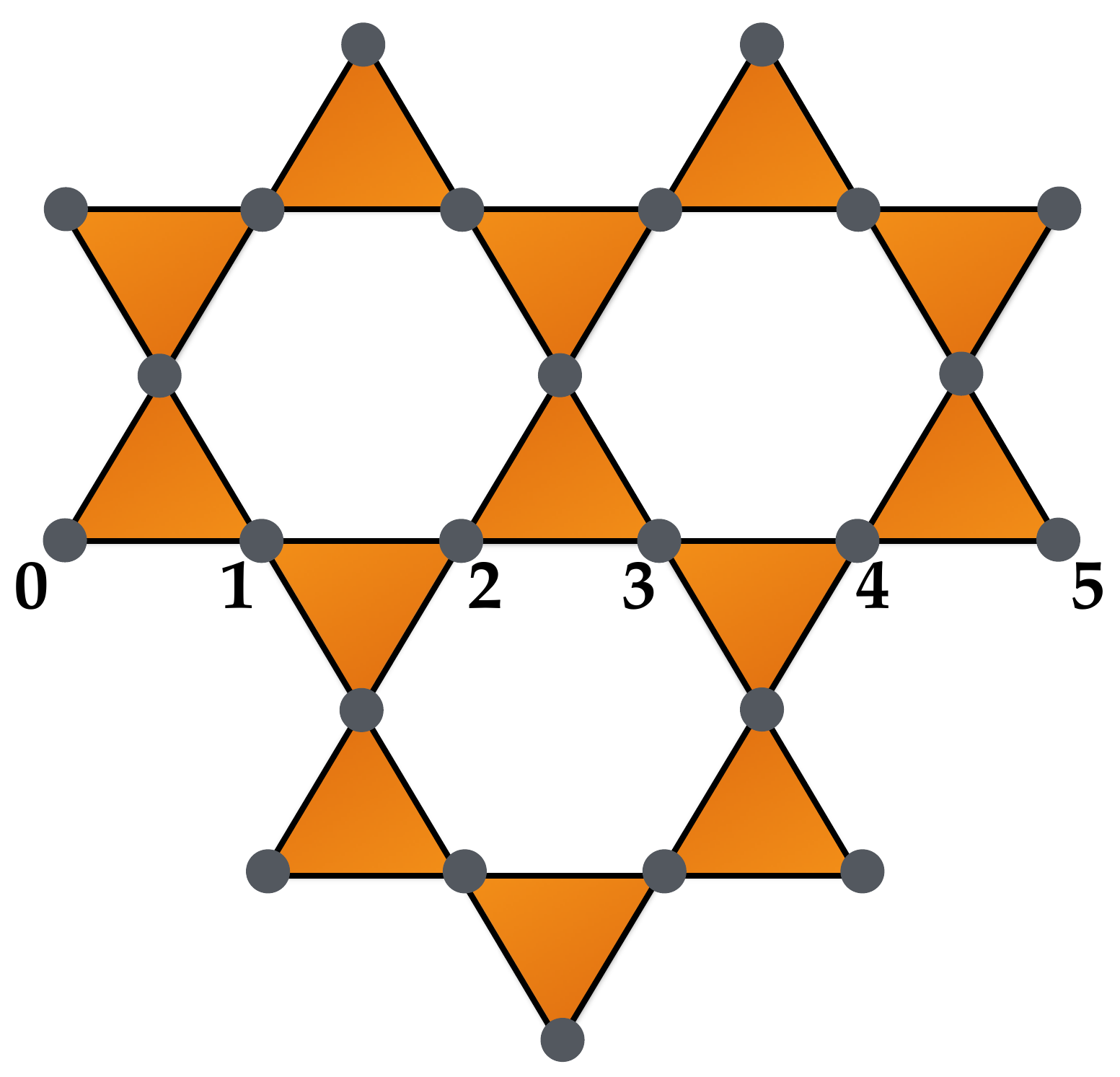}}		
	\subfloat[\label{fig:MC_Square-kagome} square-kagome] {\includegraphics[width=0.2\textwidth]{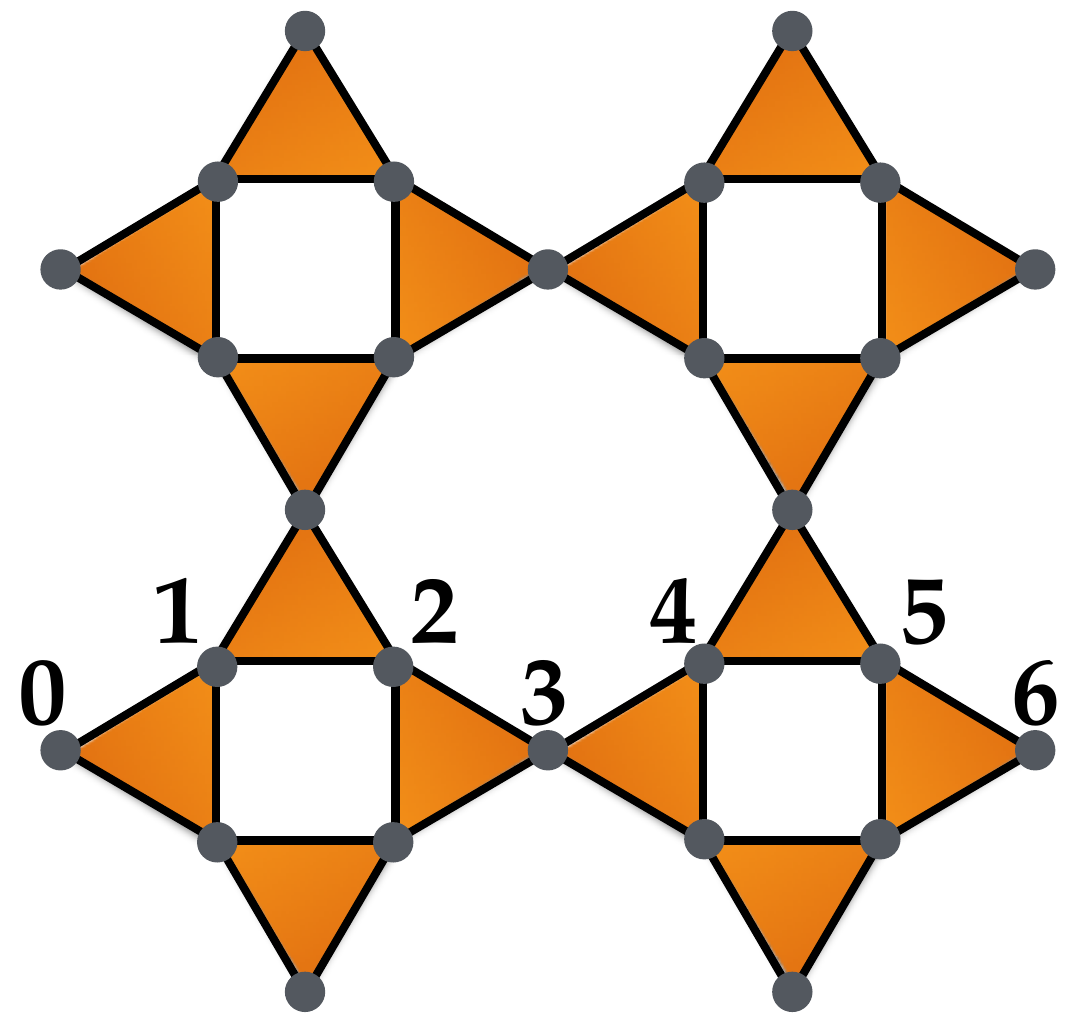}} 
	\subfloat[\label{fig:MC_Checker} checkerboard] {\includegraphics[width=0.2\textwidth]{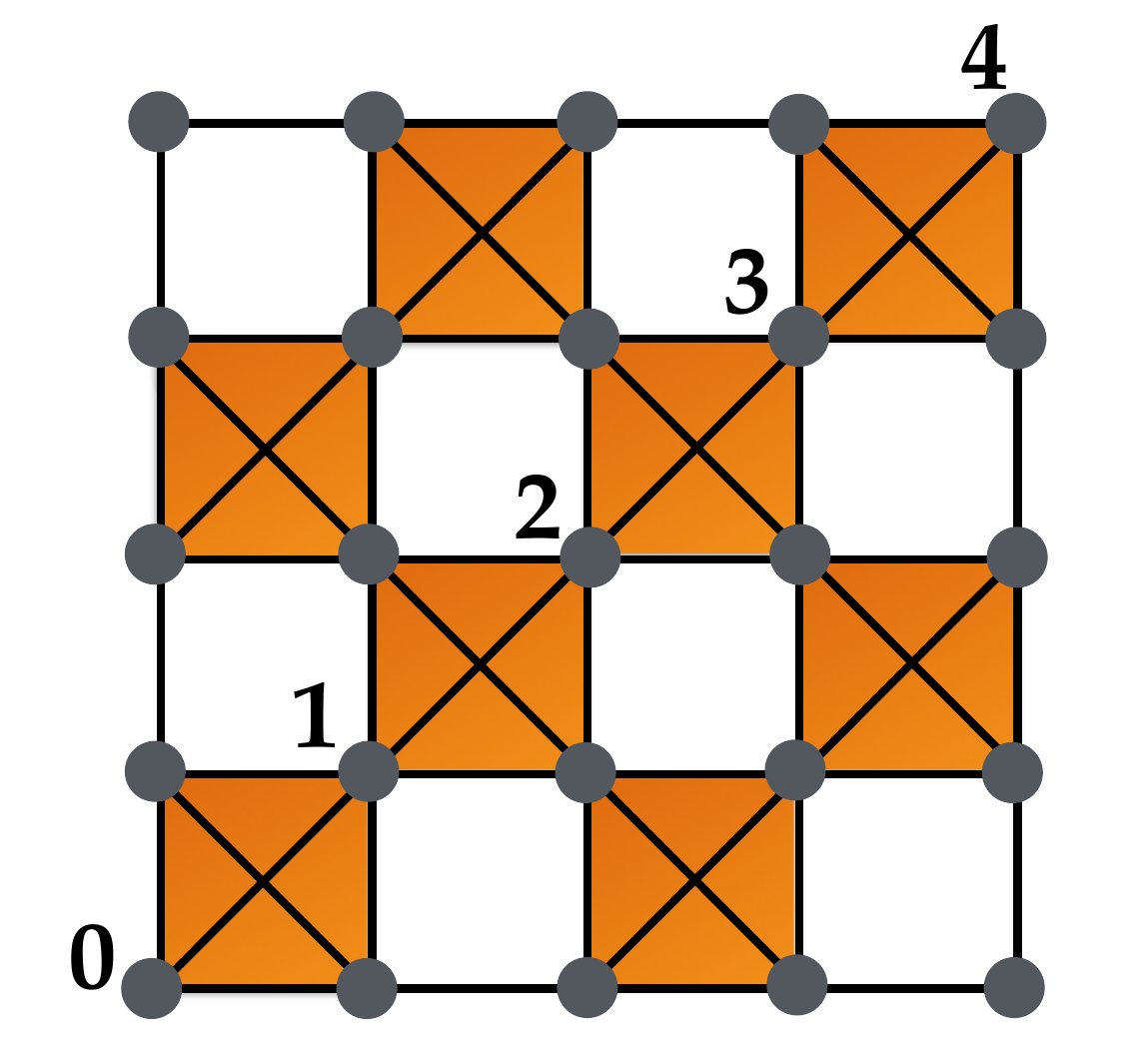}} 
	\subfloat[\label{fig:MC_Ruby} ruby] {\includegraphics[width=0.2\textwidth]{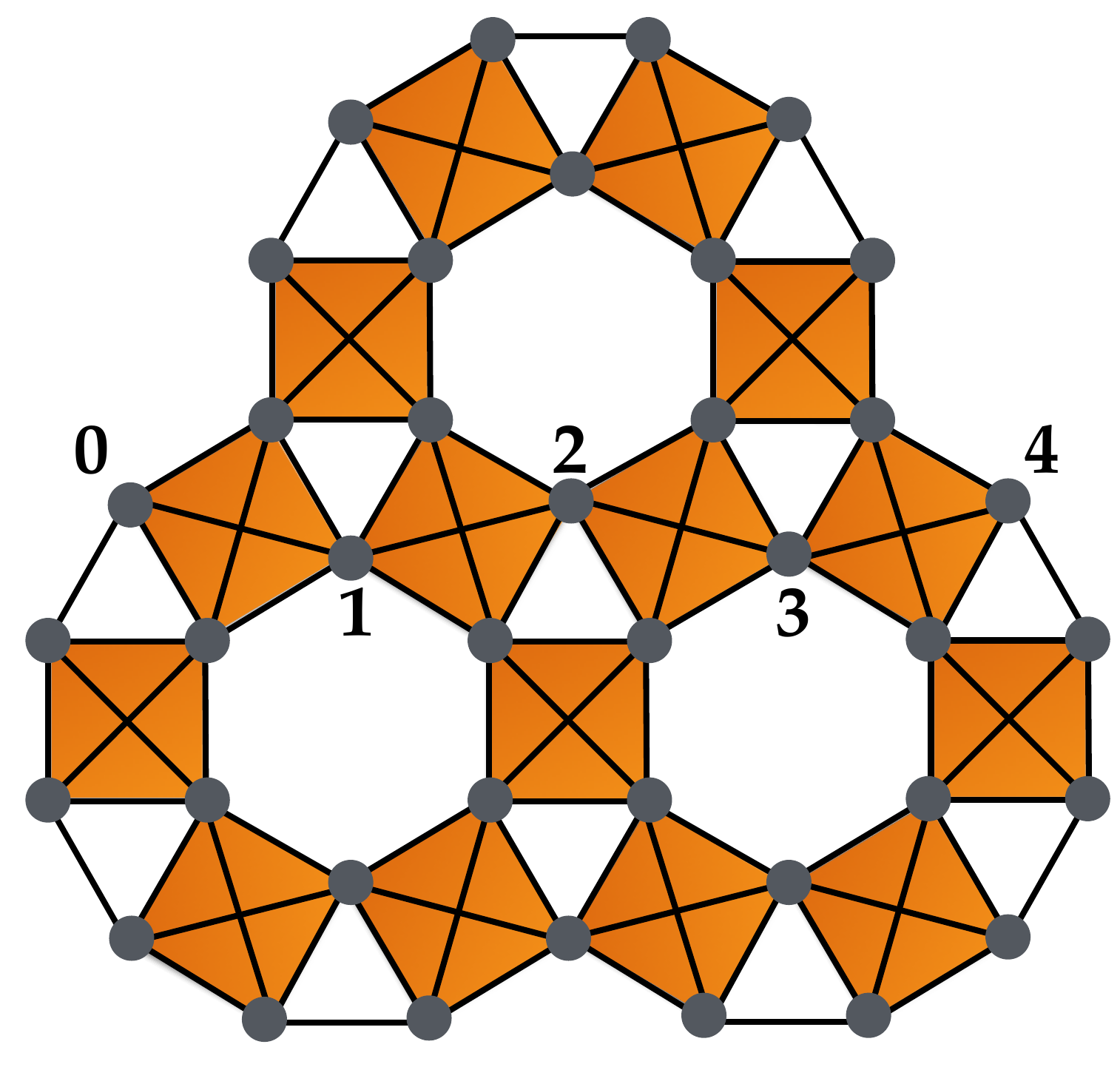}} \\
	\subfloat[\label{fig:MC_Trillium} trillium] {\includegraphics[width=0.25\textwidth]{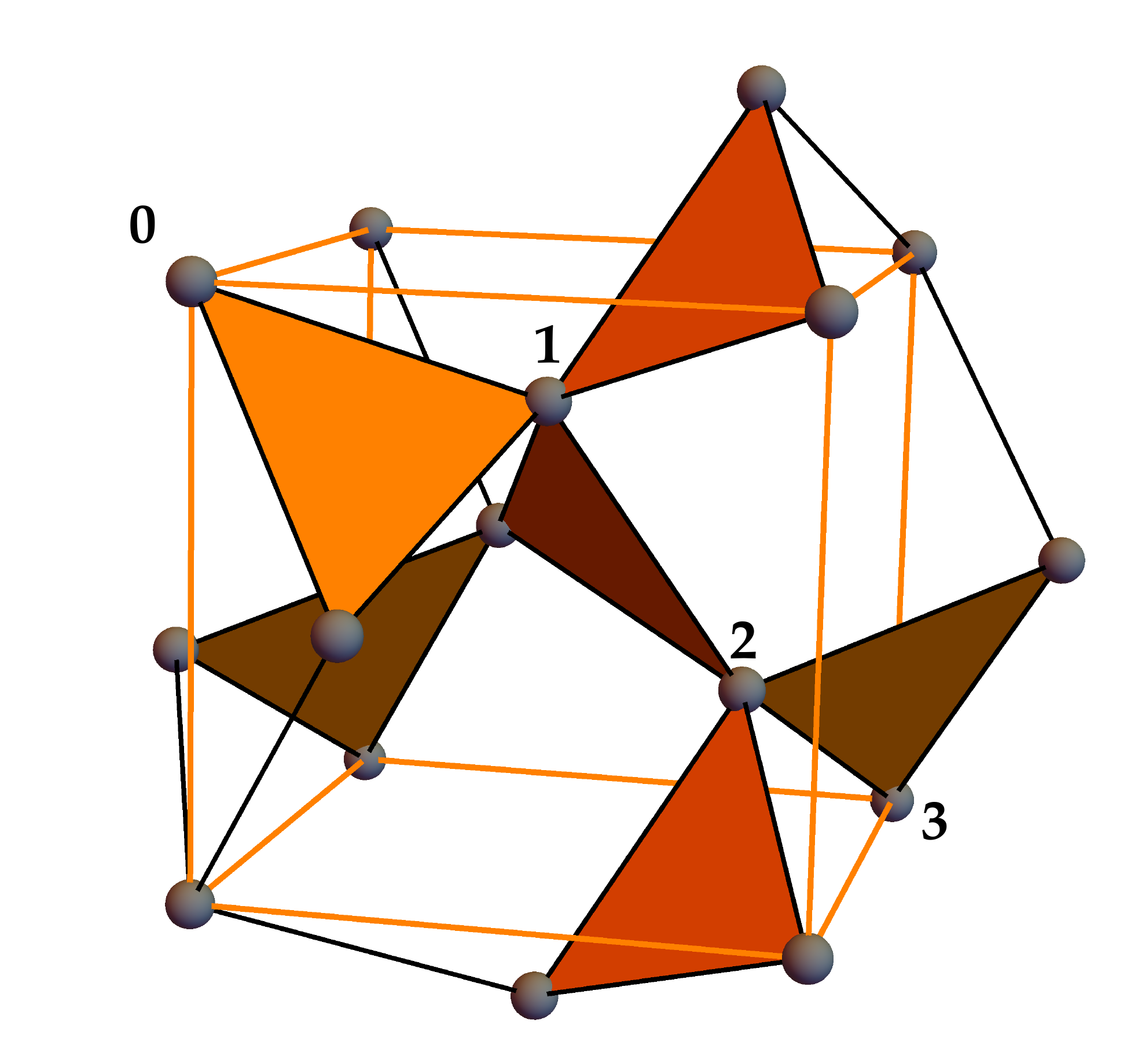}}
	\hspace{1cm}
	\subfloat[\label{fig:MC_HyperKag} hyperkagome] {\includegraphics[width=0.25\textwidth]{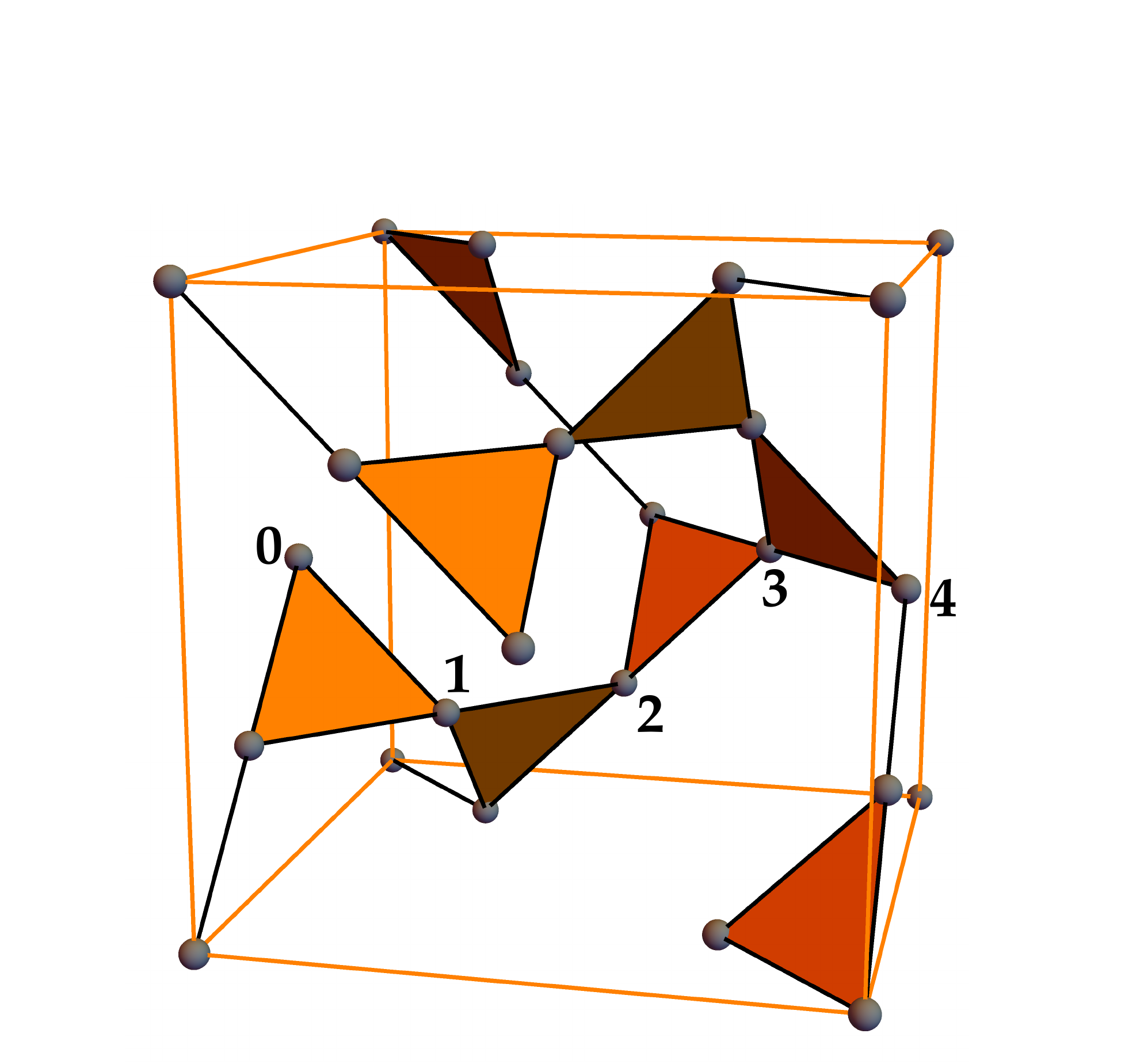}}
	\hspace{1cm}
	\subfloat[\label{fig:MC_Pyrochlore} pyrochlore] {\includegraphics[width=0.25\textwidth]{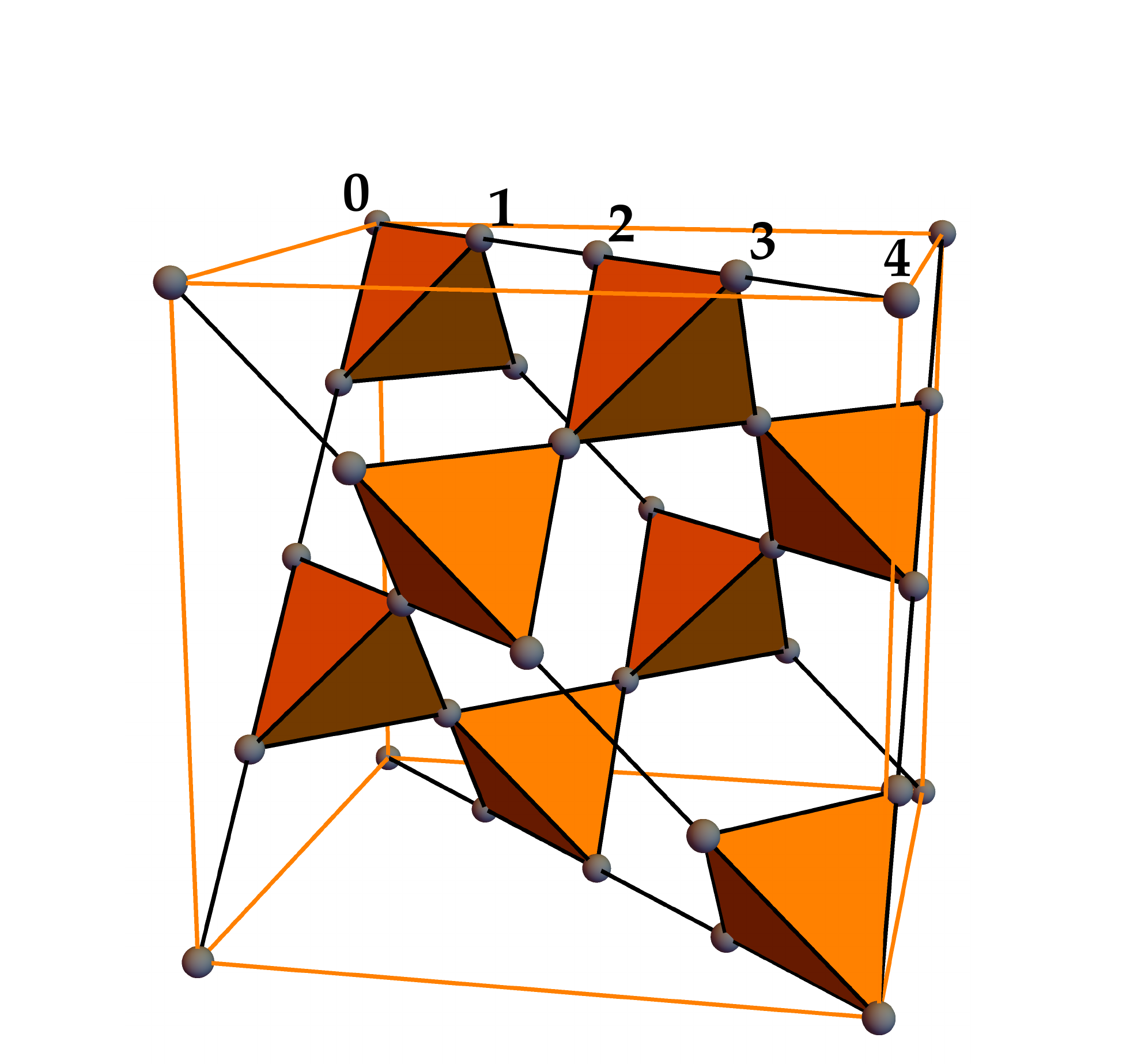}}
	\caption{
	Corner-sharing lattices in two and three dimensions involve different lengths of minimal loops $\mathcal{L}$
	between frustrated cells. 
	(a) triangular lattice ($\mathcal{L} = 3$), (b) kagome lattice ($\mathcal{L} = 6$), (c) square-kagome lattice ($\mathcal{L} = 4$), 
	(d) checkerboard lattice ($\mathcal{L} = 4$), (e) ruby lattice ($ \mathcal{L} = 3$), (f) trillium lattice ($\mathcal{L} = 5$), 
	(g) hyperkagome lattice ($ \mathcal{L} = 10$), and (f) pyrochlore lattice ($\mathcal{L} = 6$).  
	Thermodynamic observables for each lattice (see Fig.~\ref{fig:MC_HT_Thermodyn}) have been obtained numerically with 
	classical Monte Carlo simulations of Hamiltonian  $\mathpzc{H}$ [Eq.~(\ref{eq:Ham})], as described in Appendix \ref{sec:AppMC}. 
	While commonly referred to ``edge sharing'' in the literature, we describe the triangular lattice as corner sharing 
	to emphasise its analogy with the trillium lattice in three dimensions and the corresponding Husimi tree HT(3,3)
	in Fig.~\ref{fig:HT_Lattices}(c).
	Numbers on lattice sites indicate the Manhattan distance $\ell$, used in Fig.~\ref{fig:SpinCorr.realSpace}.
	}
\label{fig:MC_Lattices}
\end{figure*}

%%%%%%%%%%%%%%%%%%%%%%%%%%%%%%%%%%%%%%%%%%
Since our goal is to build a generic picture of the Curie-law crossover for spin liquids, we will study a variety of frustrated lattices in two and three dimensions [Fig.~\ref{fig:MC_Lattices}], first considering Ising spins, then extending the analysis to include continuous Heisenberg spins, anisotropic exchange, and finally analyzing experimental data of materials with quantum spins. Our motivation here is not to study each model individually. That has already been done extensively in the literature; see e.g. the following references for the two dimensional  triangular \cite{stephenson64a,rastelli77a,Isoda08a}, kagome \cite{Garanin99a,Macdonald11a}, square-kagome \cite{Pohle2016,jurcisinova18a,richter2022}, checkerboard \cite{Garanin02a,Henley2010} and ruby \cite{Rehn17a} lattices, and for the three dimensional trillium \cite{Redpath2010}, hyperkagome \cite{Hopkinson07b}, and pyrochlore \cite{Canals2001,Isakov04,Jaubert2013,Bovo2013,Bovo18a} lattices. Instead, we will compare these models together, understand why similarities appear between some of them, and build an overall intuition for the phenomenon of  the Curie-law crossover in spin liquids.

%%%%%%%%%%%%%%%%%%%%%%%%%%%%%%%%%%%%%%%%%%
% Fig.     CURIE LAW CROSSOVER 
%%%%%%%%%%%%%%%%%%%%%%%%%%%%%%%%%%%%%%%%%%
\begin{figure}[t]
\centering
\captionsetup[subfigure]{justification=justified, singlelinecheck=false, position=top}	
	\includegraphics[width=0.45\textwidth]{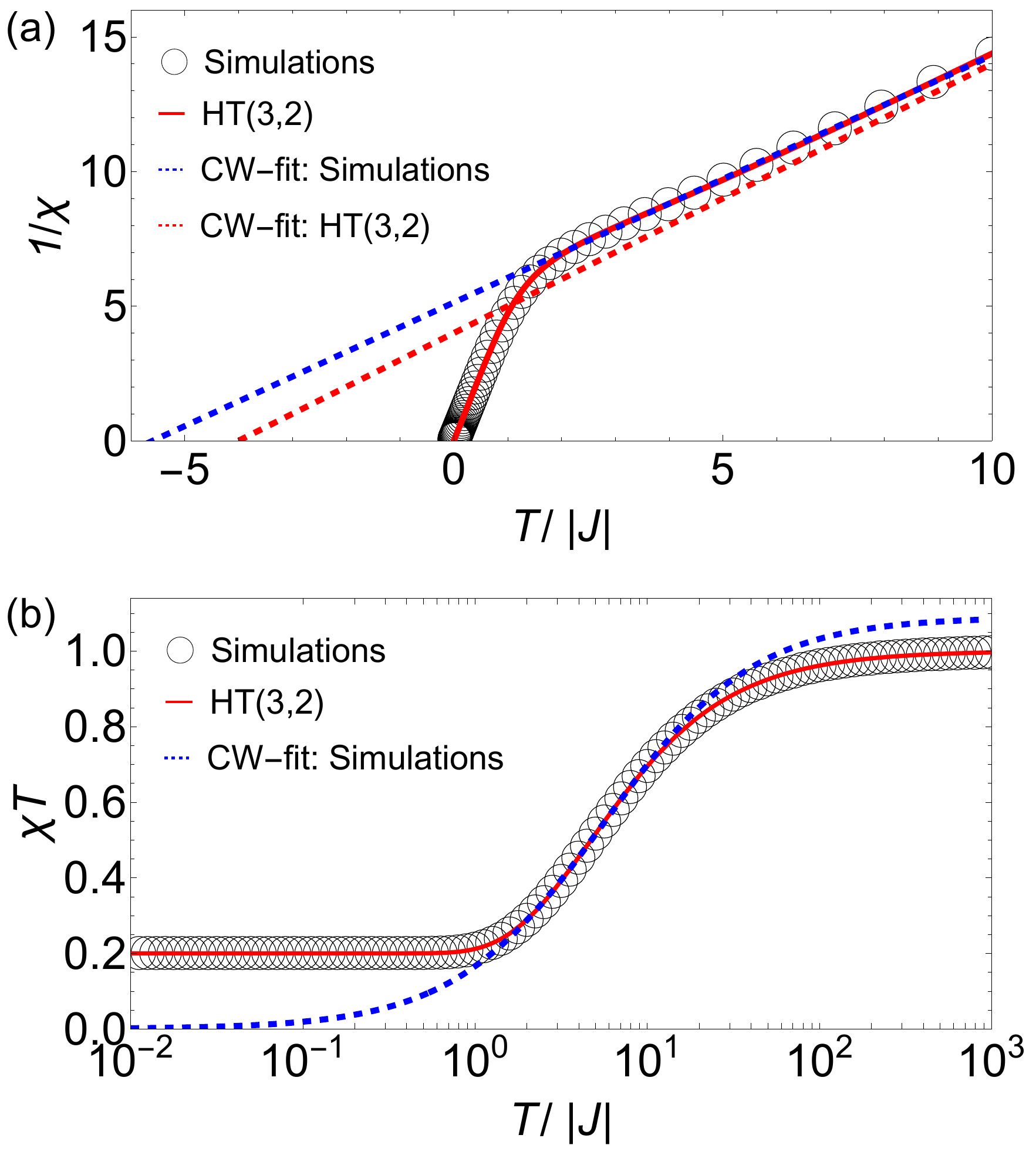}
	\caption{Curie-law crossover in spin liquids. 
	Both panels compare susceptibility results from Monte Carlo (MC) simulations of the Ising antiferromagnet on the kagome lattice [Eq.~(\ref{eq:Ham}) and Fig.~\ref{fig:MC_Lattices}(b)] (open black circles) to their corresponding results on the Husimi tree ``HT(3,2)'' [Eq.~(\ref{eq:HT.32.a}) and Fig.~\ref{fig:HT_Lattices}(a)] (solid red line).
	(a) Inverse susceptibility $1 / \chi $ on a linear temperature scale. The Curie Weiss fit has been obtained from fitting data for $2 <T/J < 10$ (blue dashed line), giving $\theta_{\text{cw}}^{\rm fit} \approx -5.6 J$ different from the known exact value of $-4 J$ (red dashed line, obtained from a Curie-Weiss fit of the HT(3,2) curve).
	(b) Same results are plotted for the reduced susceptibility $\chi T $ on a semi-logarithmic plot. The Husimi tree ``HT(3,2)'' result matches quantitatively with MC simulations, and shows the crossover between two different Curie constants at high-T ($C_{\infty}=1$ in paramagnetic phase) and low-T ($C_{0}=0.2$ in spin liquid phase), corresponding to two different Curie laws. If the fit is done in the intermediate crossover region ($2-10 |J|$), which is typically the region accessible to experiments (see Section \ref{sec:experiments}), the resulting Curie-Weiss law quickly deviates from simulations.
  	}
\label{fig:CurieLaw.Crossover}
\end{figure}
%%%%%%%%%%%%%%%%%%%%%%%%%%%%%%%%%%%%%%%%%%

On the theoretical front, comparing unbiased classical Monte Carlo simulations to the analytical Husimi-Tree approximation shows that thermodynamic quantities are, to a large extent, independent of the lattice dimension, and even of the structure of the lattice beyond the minimal frustrated unit cells [Fig.~\ref{fig:MC_HT_Thermodyn}]. What essentially matters is simply the type of frustrated unit cell (triangle, tetrahedron, ...) and the local connectivity between them. In addition, we also show how to compute correlations on Husimi trees with a non-trivial distribution of sublattices and local easy-axes on trillium and hyperkagome lattices. [Appendix ~\ref{sec:AppHT.reducedSus}].

%%%%%%%%%%%%%%%%%%%%%%%%%%%%%%%%%%%%%%%%%%
% Fig.     REAL LATTICES
%%%%%%%%%%%%%%%%%%%%%%%%%%%%%%%%%%%%%%%%%%
%
%%%%%%%%%%%%%%%%%%%%%%%%%%
%

On the experimental front, one of our take-home messages is that the reduced susceptibility $\chi T$ (that is frequently used by chemists) is a very useful complement to the inverse susceptibility $1/\chi$ for frustrated magnets. The Curie-law crossover is especially transparent in this quantity, between two horizontal asymptotic lines. $\chi T$ thus immediately tells us (i) how far we are from the high-temperature Curie law, and (ii) the presence or absence of a low-temperature spin-liquid Curie law. This is especially useful because some frustrated materials may ultimately order or freeze at a very low temperature $T^{*}$. But if the reduced susceptibility $\chi T$ reaches a low-temperature plateau at $T_p \gg T^{*}$, then it is a solid indication that a strongly correlated regime characteristic of a spin liquid exists over a finite temperature window $T^*<T<T_p$.

In order to describe the Curie-law crossover in its entirety, we introduce the following fitting Ansatz [Fig.~\ref{fig:CurieLaw.Crossover}(b)]:
%
%%%%%%%%%%%%%%%%%%%%%%%%%%%
\begin{equation}
\chi T|^{\sf fit} = \frac{1 + b_1 \ {\sf exp}[c_1 / T] }{a + b_2 \ {\sf exp}[c_2 / T] }   \ ,
\label{eq:HT.exp.fit_0}
\end{equation}
%%%%%%%%%%%%%%%%%%%%%%%%%%%
%
inspired by the above analogy between disparate models and Husimi-tree calculations. This empirical Ansatz provides a complementary estimate of the Curie constant and Curie-Weiss temperature,
%
%%%%%%%%%%%%%%%%%%%%%%%%%%%
\begin{equation}
C = \frac{1 + b_1}{a + b_2}\,
\quad\&\quad\label{eq:Exp.values.fit_0}
\theta_{\text{cw}} = \frac{b_1 c_1}{1 + b_1}  - \frac{b_2 c_2}{a + b_2}\, ,
\end{equation}
%%%%%%%%%%%%%%%%%%%%%%%%%%%
%
that is not based on a high-temperature expansion. Hence, if Eq.~(\ref{eq:Exp.values.fit_0}) agrees with values obtained from a Curie-Weiss fit, then it is reasonable to consider them as an accurate description of the material. On the other hand if there is a noticeable mismatch, then it is likely that experimental data have not reached the high-temperature regime where the Curie-Weiss law is valid.\\

\begin{figure*}[t]
\centering
\captionsetup[subfigure]{justification=justified, singlelinecheck=false, position=top}
	\subfloat[\label{fig:HT_Kagome} HT(3,2)]{\includegraphics[height=0.17\textheight]{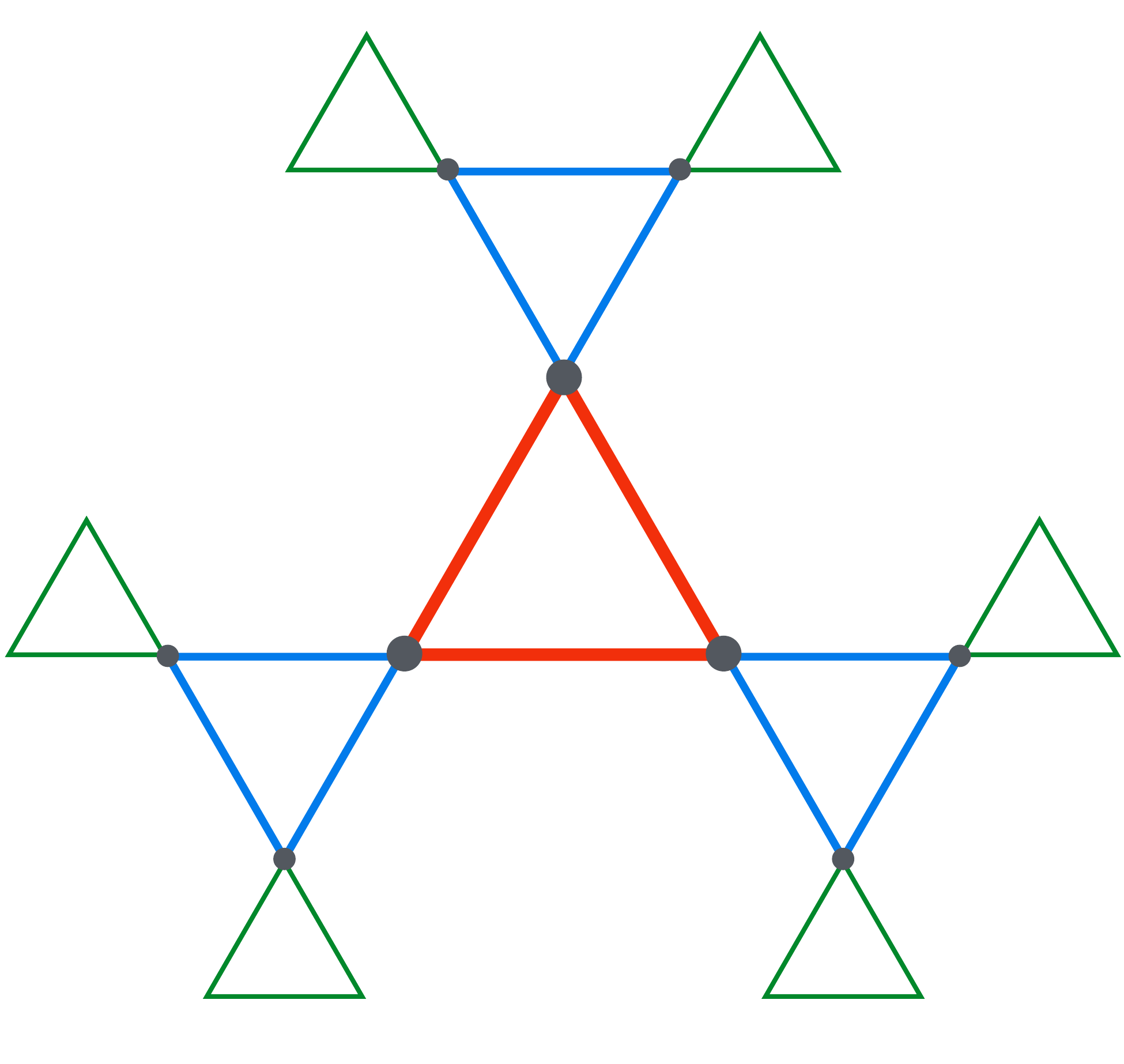}}
	\subfloat[\label{fig:HT_Shuriken} HTS] {\includegraphics[height=0.17\textheight]{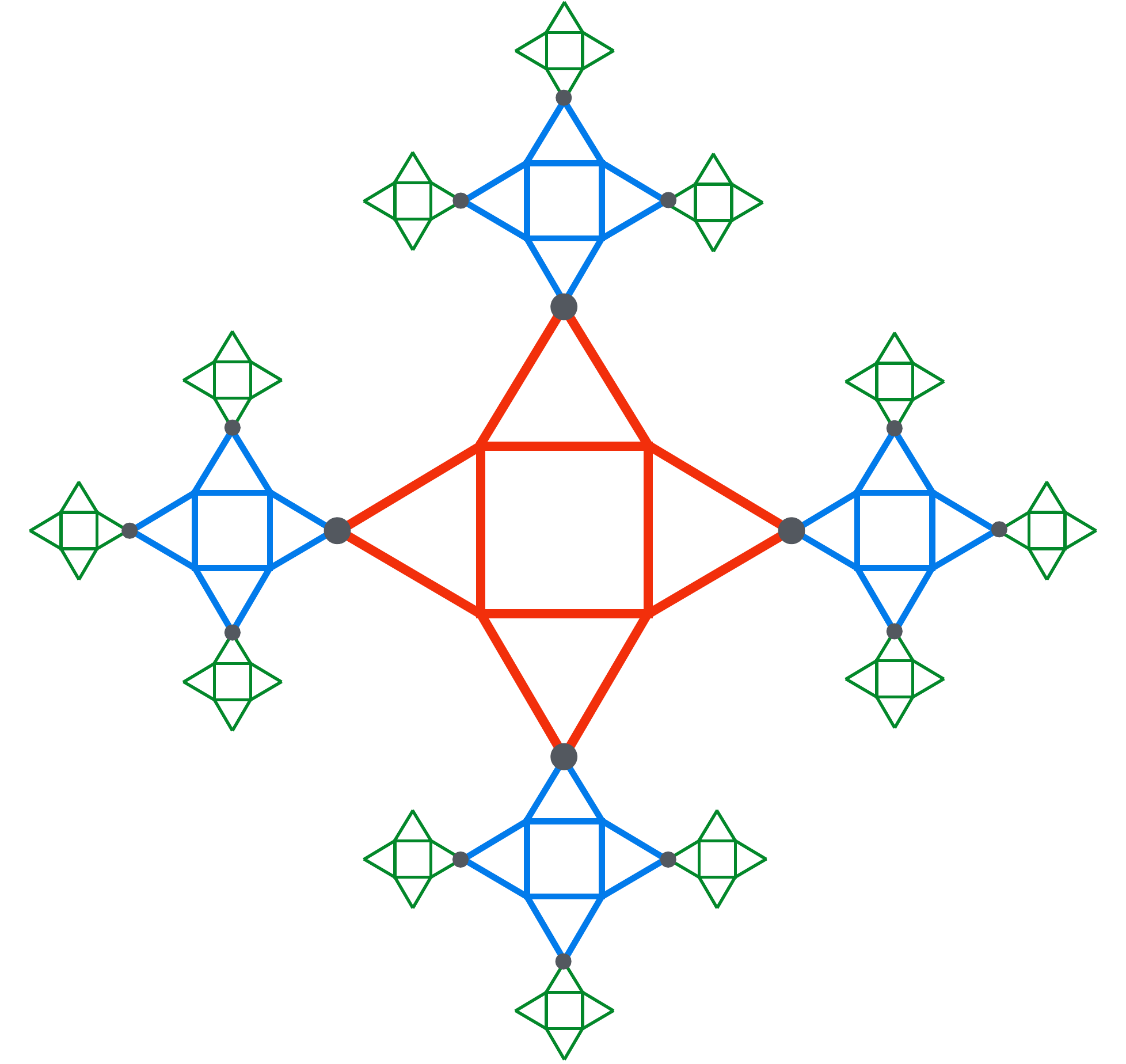}}	
	\subfloat[\label{fig:HT_Trillium} HT(3,3)] {\includegraphics[height=0.17\textheight]{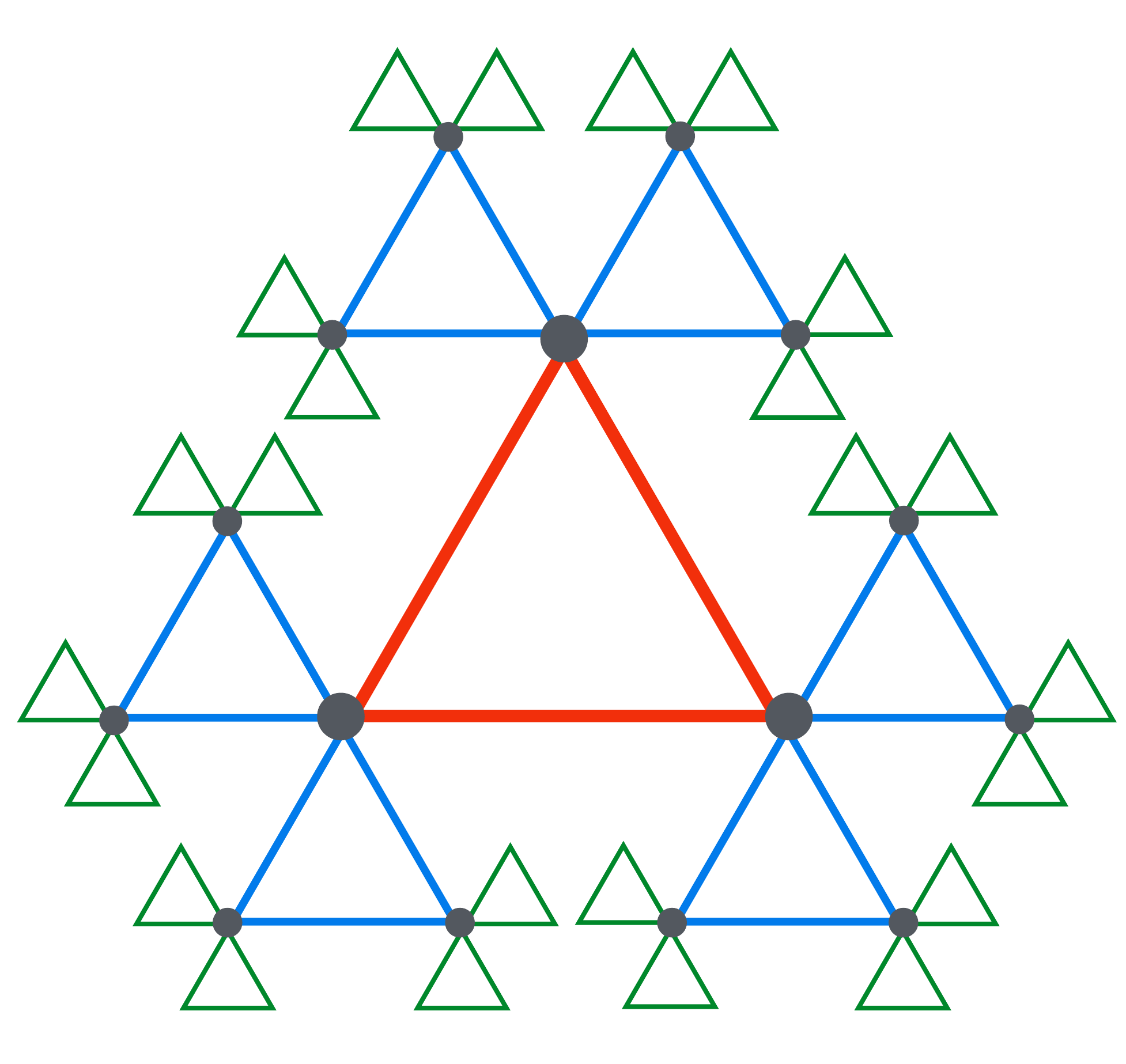}} 
	\quad
	\subfloat[\label{fig:HT_CheckerBoard} HT(4,2)] {\includegraphics[height=0.17\textheight]{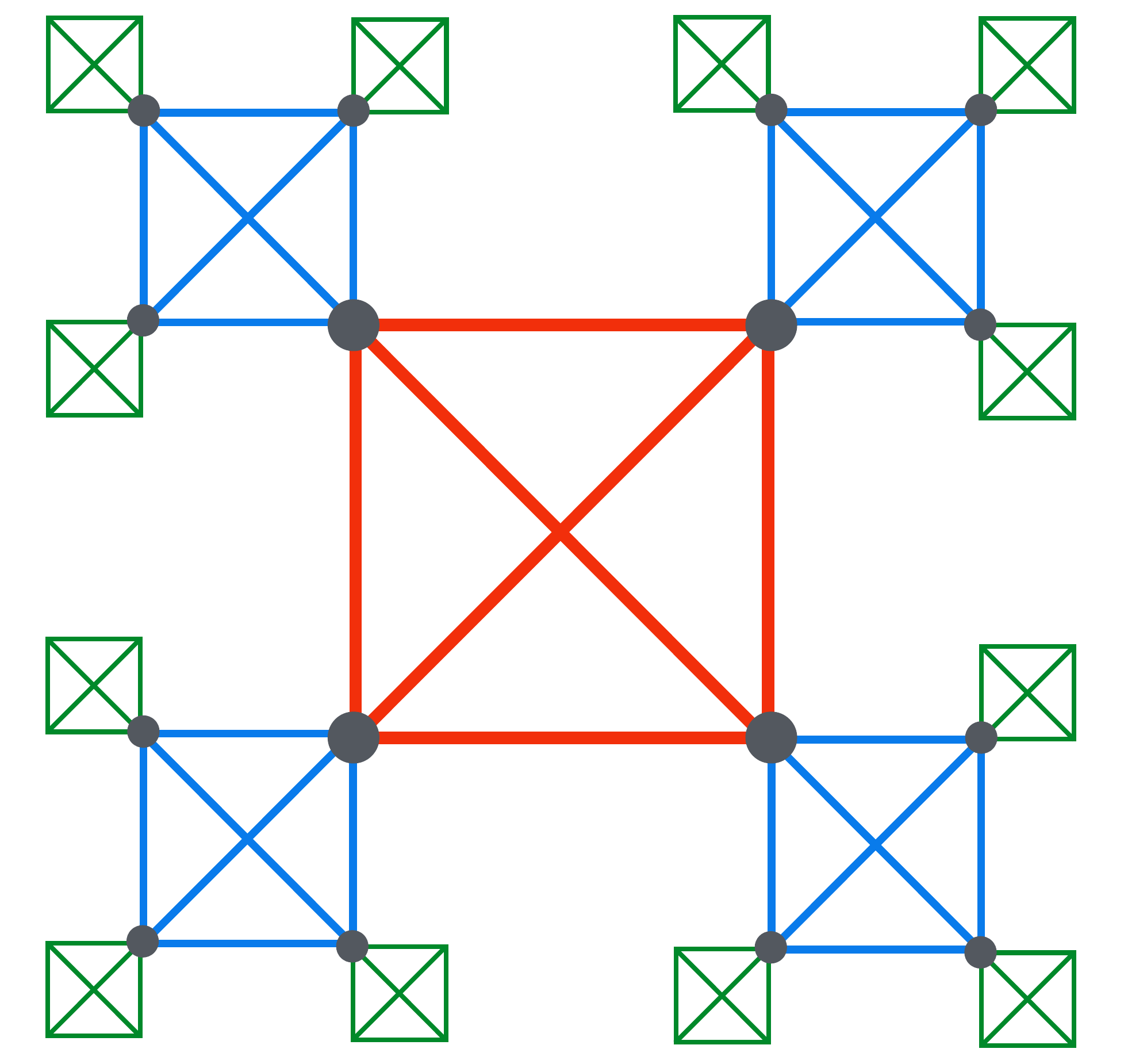}}
	\caption{
	Husimi Trees for various corner sharing lattices.
    	Frustrated cells from real lattices [Fig.~\ref{fig:MC_Lattices}] are arranged on the Husimi 
    	tree (HT), with the central cell in red, the 1$^{st}$ shell in blue and the 2$^{nd}$ shell in green. 
    	(a) HT(3,2): for the kagome and hyperkagome lattice, with corner sharing triangular plaquettes.
	(b) HTS: for the square-kagome lattice.
    	(c) HT(3,3): for the triangular and trillium lattice, where three triangular plaquettes share one corner. 
    	(d) HT(4,2): for the checkerboard, ruby and pyrochlore lattice, which is made of corner-sharing square/tetrahedron plaquettes.
   	 }
\label{fig:HT_Lattices}
\end{figure*}

%%%%%%  outline paper
%
The remainder of this Article is structured as follows. 
In Sec.~\ref{sec:Model}, we introduce the models of classical spin liquids,
defined with Ising spins on a variety of frustrated lattices in two and three dimensions [Fig.~\ref{fig:MC_Lattices}]. 
These models will be solved numerically with classical Monte Carlo simulations and analytically 
on their corresponding Husimi trees [Fig.~\ref{fig:HT_Lattices}].

In Sec.~\ref{sec:Curie crossover}, we present thermodynamic quantities for all spin liquids introduced in Sec~\ref{sec:Model} and discuss analogies and signatures of their Curie-law crossover. In particular, we discuss the reason for the very good match between Monte-Carlo simulations and Husimi-tree calculations, despite the different lattice structure. 

In Sec.~\ref{sec:ansatz}, we discuss the limitations of the conventional Curie-Weiss fit, showing the advantage to use the reduced susceptibility $\chi T$. We introduce and benchmark the Husimi Ansatz [Eq.~(\ref{eq:HT.exp.fit_0})] to numerical simulations of spin-liquid models with Ising and continuous Heisenberg spins.

In Sec.~\ref{sec:experiments}, we apply this Ansatz to experimental data for the pyrochlore NaCaNi$_2$F$_7$ [\onlinecite{Krizan2015}], the square-kagome KCu$_6$AlBiO$_4$(SO$_4$)$_5$Cl [\onlinecite{fujihala20a}] and the spiral spin liquid FeCl$_{3}$ [\onlinecite{Gao2022}].

In Sec.~\ref{sec:conclusions}, we conclude with a brief summary and implications for 
future experiments on spin liquid materials.  

Details on the lattice geometries, Monte Carlo simulations, Husimi tree calculations, connection to Coulomb gauge field theory, and structure factors are given in Appendices \ref{sec:AppSpinDef}, \ref{sec:AppMC}, \ref{sec:AppHT}, \ref{sec:gauge} and \ref{sec:App.Sq} respectively.

%%%%%%%%%%%%%%%%%%%%%%%%%%%%%%%%%%%%%%%%%%
 %Fig.     LATTICES ON THE HUSIMI TREE
%%%%%%%%%%%%%%%%%%%%%%%%%%%%%%%%%%%%%%%%%%
%
%%%%%%%%%%%%%%%%%%%%%%%%%
%

%%%%%%%%%%%%%%%%%%%%%%%%%%%%%%%%%%%%%%%%%%%%%
%
%			   		METHOD / MODEL
%
%%%%%%%%%%%%%%%%%%%%%%%%%%%%%%%%%%%%%%%%%%%%%
\section{Models and methods}
 \label{sec:Model}
%%%%%%%%%%%%%%%%%%%%%%%%%%%%%%%%%%%%%%%%%%
%
%%%%%%%%%%%%%%%%%%%%%%%%%%%%%%%%%%%%%%%%%%
\subsection{The Ising model}
%%%%%%%%%%%%%%%%%%%%%%%%%%%%%%%%%%%%%%%%%%
%

In Sec.~\ref{sec:Model} and \ref{sec:Curie crossover}, we focus on thermodynamic properties of minimal spin-liquid models,
%
%%%%%%%%%%%%%%%%%%%%%%%%%%%
\begin{equation}
	\mathpzc{H}  = J \sum_{\langle ij \rangle}  \vec S_i \cdot \vec S_j  \,  ,
\label{eq:Ham}
\end{equation}
%%%%%%%%%%%%%%%%%%%%%%%%%%%
%
for Ising spins $\vec{S}_i = \sigma_i \vec{e}_i$, with $\sigma_i = \pm 1$, and nearest-neighbour 
coupling $J$, applied to a variety of lattices, as shown in Fig.~\ref{fig:MC_Lattices}. 
We shall consider two types of Ising spins, either collinear along the same global \mbox{$z$-axis}
$\vec e_{z}$, or oriented along their local easy axis $\vec e_{i}$ attached to the sublattice of site $i$.
The latter is motivated from single-ion anisotropy, as found, for example, in kagome materials like Dy$_3$Mg$_2$Sb$_3$O$_{14}$ 
\cite{Paddison2016} and spin ices like Dy$_{2}$Ti$_{2}$O$_{7}$ and Ho$_{2}$Ti$_{2}$O$_{7}$ on the pyrochlore lattice 
\cite{Bramwell2020, SpinIce2021}, and EuPtSi \cite{Adroja1990, Hopkinson2006, Redpath2010} on the trillium lattice. 
We shall refer to each system as ``global-axis'' and ``local-axis'' models, as illustrated in Fig.~\ref{fig:axes}.
All local easy axes relevant for this work are defined in Appendix~\ref{sec:AppSpinDef}. Global-axis and local-axis models are equivalent, up to a simple rescaling of the coupling constant 
$J$ \cite{Bramwell1998,Moessner1998}
%%%%%%%%%%%%%%%%%%%%%%%%%%%
\begin{equation}
	J_{\rm local} = J_{\rm global} \; (\vec e_i \cdot \vec e_j) \,  ,
\label{eq:Jscaling}
\end{equation}
%%%%%%%%%%%%%%%%%%%%%%%%%%%
where $i$ and $j$ are two neighbouring sites. 
For lattices considered here, the scalar product $(\vec e_i \cdot \vec e_j)$ is the same for all neighbouring pairs,
which means that the energy, specific heat and entropy of the two models are the same up to rescaling (\ref{eq:Jscaling}). 
However, magnetic quantities such as the susceptibility differ. 
In this work, the exchange coupling is always antiferromagnetic $J_{\rm global} > 0$ 
(ferromagnetic $J_{\rm local} < 0$) for global-axis (local-axis) models, in order to stabilise a spin-liquid ground state.
From now on, all energies and temperatures are given in units of $J_{\rm global}=1$, understanding that the rescaling of Eq.~(\ref{eq:Jscaling}) is always applied for local-axis models.

In Sec.~\ref{sec:ansatz}, the Hamiltonian of Eq.~(\ref{eq:Ham}) will also be considered for continuous Heisenberg spins.

%%%%%%%%%%%%%%%%%%%%%%%%%%%%%%%%%%%%%%%%%%
 %Fig.     GLOBAL VS LOCAL AXES
%%%%%%%%%%%%%%%%%%%%%%%%%%%%%%%%%%%%%%%%%%
%
\begin{figure}[b]
\centering
\captionsetup[subfigure]{justification=justified, singlelinecheck=false, position=top}
	\subfloat[kagome \ global axis]{\includegraphics[height=0.11\textheight]{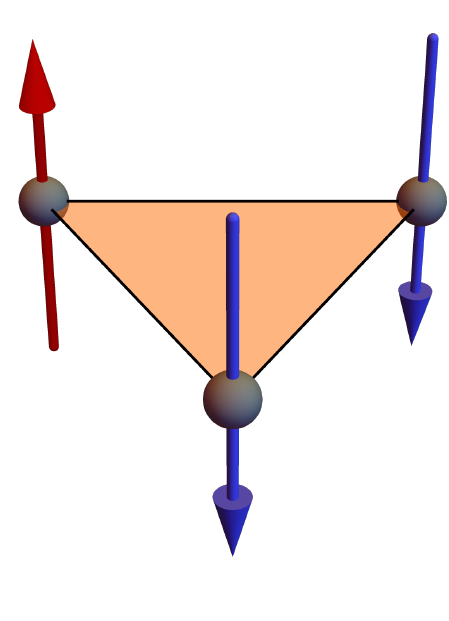}}
	\hspace{-0.0cm}
	\subfloat[kagome \\ local axis]  {\includegraphics[height=0.11\textheight]{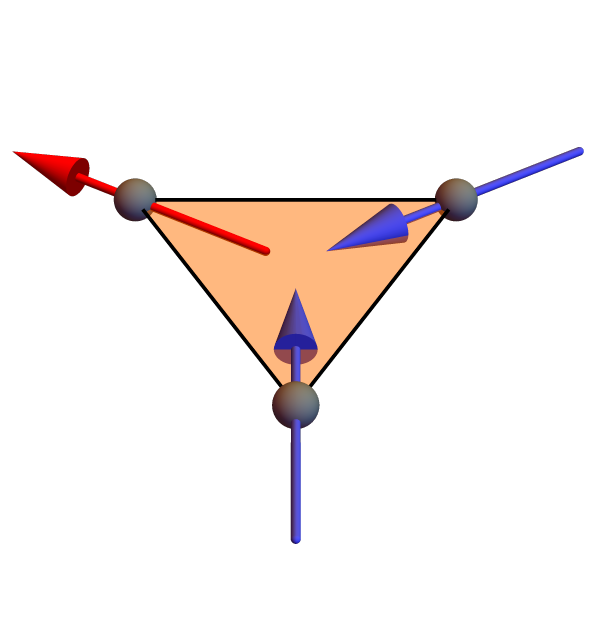}}
	\hspace{-0.1cm}
	\subfloat[pyrochlore  \ global axis]{\includegraphics[height=0.11\textheight]{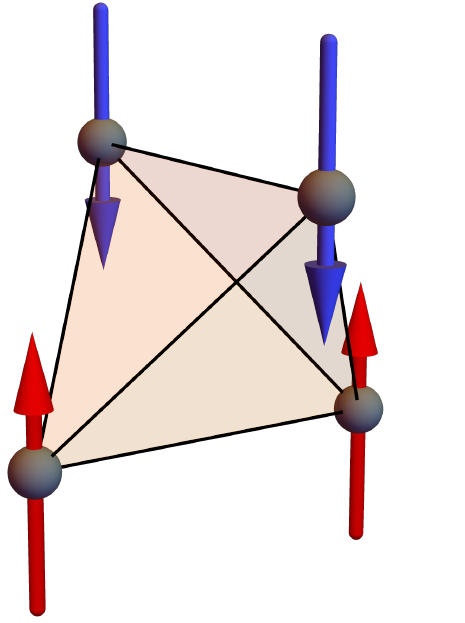}}    
	\hspace{-0.1cm}
	\subfloat[pyrochlore \  local axis]  {\includegraphics[height=0.11\textheight]{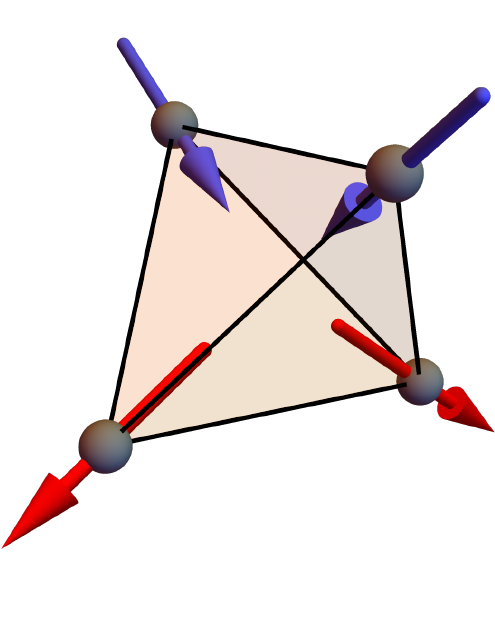}}
	\caption{
	While Ising models usually consider collinear spins (a),(c), the crystal field in materials may 
	impose a local easy axis (b), (d) respecting the symmetry of the magnetic-ion environment,
	as illustrated here for the kagome and pyrochlore lattice. 
	All local easy axes are defined in Appendix \ref{sec:AppSpinDef}.
	}
\label{fig:axes}
\end{figure}
%%%%%%%%%%%%%%%%%%%%%%%%%

%%%%%%%%%%%%%%%%%%%%%%%%%%%%%%%%%%%%%%%%%%
% Tab.    COMPARISON MONTE CARLO & HUSIMI TREE 
%%%%%%%%%%%%%%%%%%%%%%%%%%%%%%%%%%%%%%%%%%
%
%%%%%%%%%%%%%%%%%%%%%%%%%%
\newcolumntype{C}{>{}c<{}} 
\renewcommand{\arraystretch}{3}
\begin{table*}
	\def\arraystretch{1.7}
	\centering
		\begin{tabular}{||C||C|C|C||C|C||}
			\hhline{|t:======:t|}		
			Lattice			&\multicolumn{3}{C||} {S $\big|_{T \to 0}$ }	&\multicolumn{2}{C||}{$C_{0}\equiv\chi T \big|_{T \to 0}$}	\\
			\hhline{||------||}
							& Monte Carlo	& Husimi Tree & other methods	& 	Monte Carlo	& 	Husimi Tree \\
			\hhline{||======||}
			kagome 			& $ 0.502(1) $ 	& $  \frac{1}{3} \ln{\frac{9}{2}} \approx 0.5014 $  \cite{Wills2002}& $0.50183$ \cite{Kano1953}	& \makecell{$0.201(1)$\\ 1.988(1)} & \makecell{$1/5$\\2} \\
			\hhline{||------||}
			hyperkagome 		& $ 0.502(1) $ 	& $ \frac{1}{3} \ln{\frac{9}{2}} \approx 0.5014 $  \cite{Wills2002}& n/a	& 	\makecell{$0.200(1)$\\ 1.500(1)}	& \makecell{$1/5$ \\ 3/2}\\
			\hhline{||------||}	
			square-kagome		& $ 0.504(1) $  \cite{Pohle2016}	& $ \frac{1}{6} \ln{\frac{41}{2}} \approx 0.5034$  \cite{Pohle2016}& n/a	& 	$0.203(1)$ 	& 	$0.2028 $ 		\\
			\hhline{||------||}
			triangular			& $ 0.323(2) $ 	& $ \ln{\frac{3}{2}} \approx 0.4055 $	\cite{Redpath2010}& $0.323066 \cite{Wannier1950,Wannier73a}$	& 	$0.162(8)$ 	& 	$1/7$ 			\\
			\hhline{||------||}
			trillium 			& $ 0.392(1) $ \cite{Redpath2010}& $ \ln{\frac{3}{2}} \approx 0.4055 $	 \cite{Redpath2010}& n/a	& \makecell{$0.135(1)$\\ 0.969(1)}	& \makecell{$1/7$\\ 1}	\\
			\hhline{||------||}
			ruby 				& $ 0.194(1) $ 	& $ \frac{1}{2} \ln{\frac{3}{2}} \approx 0.2027 $ \cite{pauling35a}& n/a	& 	$0.0$ 		& 	$0$ 				\\
			\hhline{||------||}
			checkerboard		& 0.216(1) & $ \frac{1}{2} \ln{\frac{3}{2}} \approx 0.2027   \cite{pauling35a}$ & $ \frac{3}{4} \ln{\frac{4}{3}} \approx 0.2158$ \cite{Lieb67a}& 	0.0 	& 	$0 $ 		\\
			\hhline{||------||}
			pyrochlore 		& $ 0.206(1) $ 	& $ \frac{1}{2} \ln{\frac{3}{2}} \approx 0.2027  \cite{pauling35a}$ & 0.205006(9) \cite{Nagle1966}& 	\makecell{$0.0$\\ 2.002(1)\cite{Isakov04}}& \makecell{$0$\\2}\\
			\hhline{|b:======:b|}
		\end{tabular}
	\caption{ 
	Residual entropy $S \big|_{T \to 0}$ and spin liquid Curie constant $C_{0}$, obtained from Monte Carlo (MC) 
	simulations and Husimi tree (HT) calculations. 
	Table cells including two lines for $C_{0}$ display results for global-axis (first line) and local-axis models (second line). HT calculations of  $C_{0}$ are detailed in Appendix \ref{sec:AppHT.reducedSus}.
	The column ``other methods'' compares the  HT estimate, also known as Pauling entropy, to exact results when available (except for the 3D 
	pyrochlore lattice obtained from series expansion). 
	As a side remark, one should be aware that the Pauling entropy is not always a lower bound of the residual entropy on the corresponding 
	real lattices (see Ref.~[\onlinecite{Lieb1972}] and Appendix \ref{sec:Pauling}). 
	}	
\label{tab:HT_MC_zeroT}
\end{table*}
%%%%%%%%%%%%%%%%%%%%%%%%
%

%%%%%%%%%%%%%%%%%%%%%%%%%%%%%%%%%%%%%%%%%%
% Fig.    THERMODYNAMIC COMPARISON 
%%%%%%%%%%%%%%%%%%%%%%%%%%%%%%%%%%%%%%%%%%
%
%%%%%%%%%%%%%%%%%%%%%%%%%%
\begin{figure*}[t]
\centering
\captionsetup[subfigure]{justification=justified, singlelinecheck=false, position=top}
	\subfloat[\label{fig:x} HT(3,2)] {\includegraphics[height=0.5\textheight]{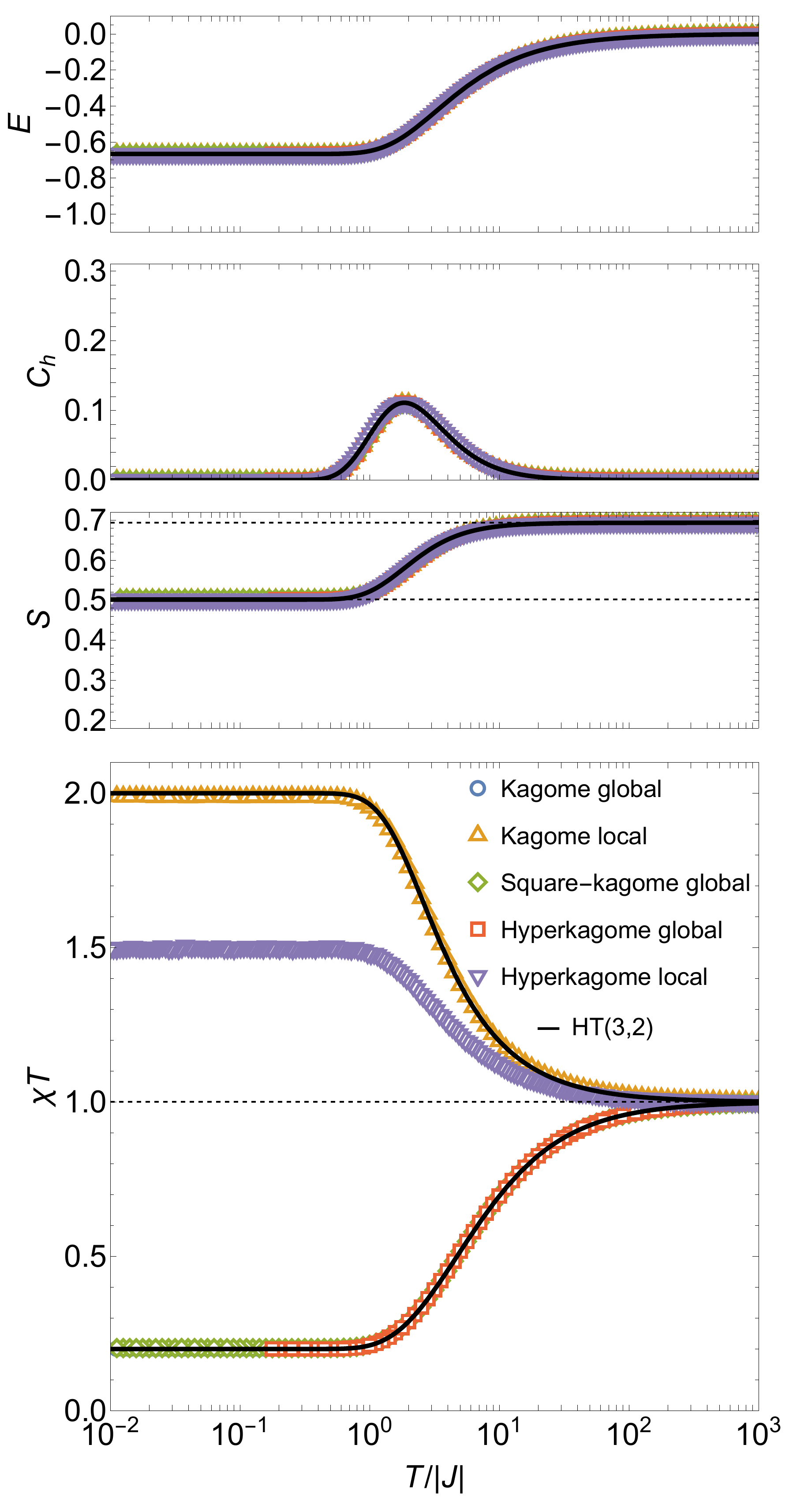}}
	\subfloat[\label{fig:y} HT(3,3)] {\includegraphics[height=0.5\textheight]{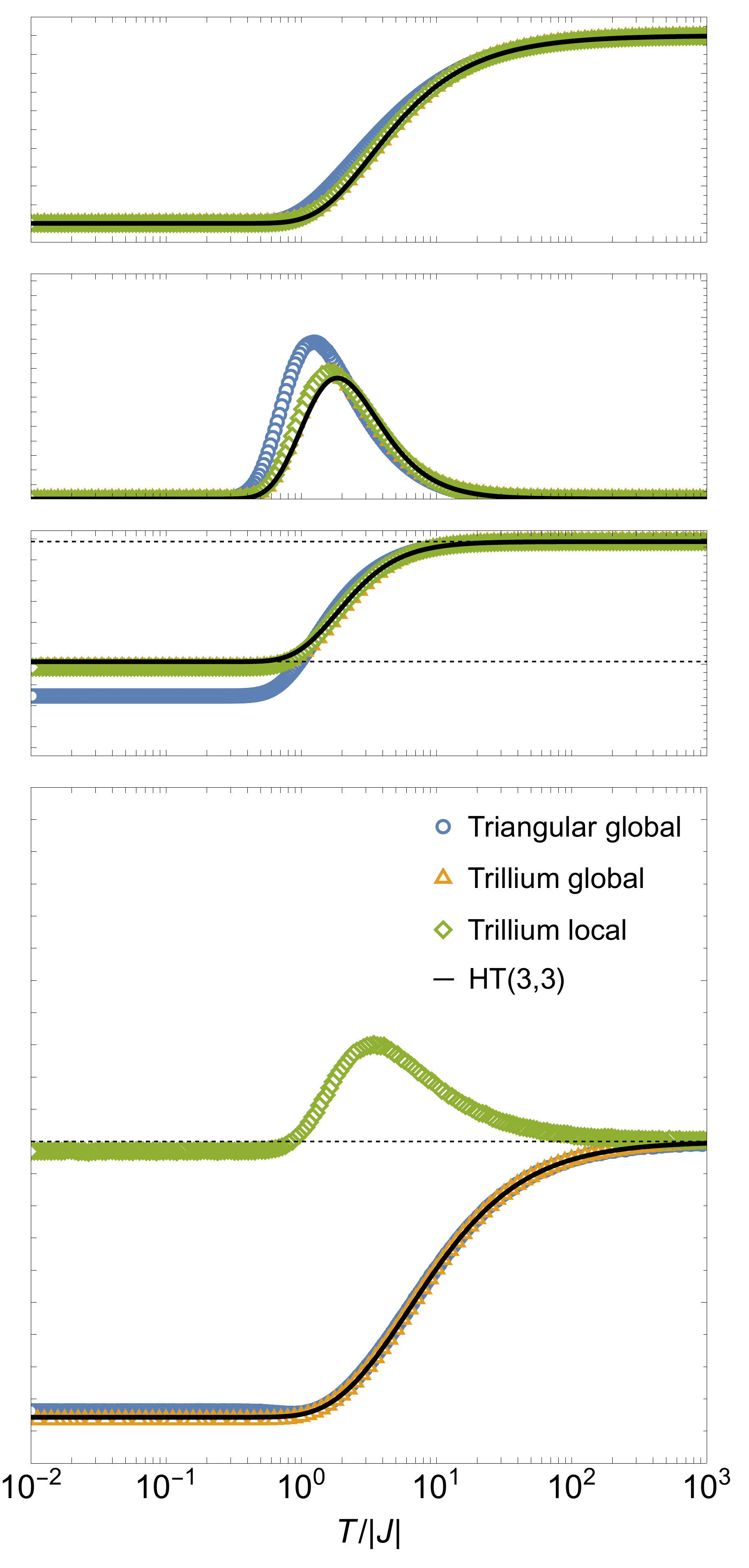}}	
	\subfloat[\label{fig:z} HT(4,2)] {\includegraphics[height=0.5\textheight]{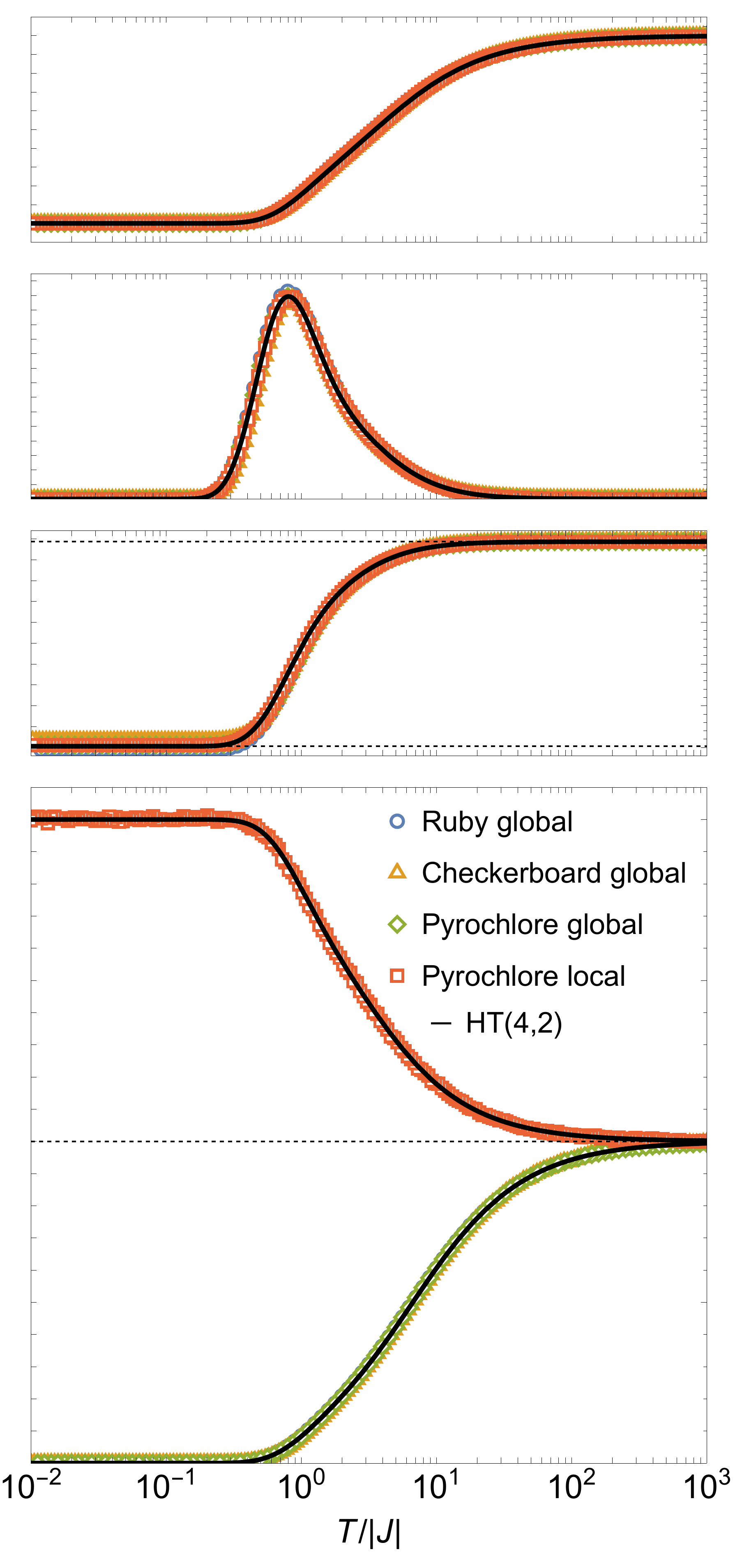}} 
	\caption{
    	Thermodynamic signatures of the Curie-law crossover in spin liquids.
 	Comparison of the normalized energy $E$, specific heat $C_h$, entropy $S$ and reduced 
    	susceptibility $\chi T$ per spin for results obtained from classical Monte Carlo simulations (open symbols) on the physical lattices [Fig.~\ref{fig:MC_Lattices}] and analytical calculations (solid black lines) on their corresponding Husimi trees [Fig.~\ref{fig:HT_Lattices}].
	Observables are shown on a semi-logarithmic plot.
   	(a) Lattices with triangular cells, where each site belongs to two frustrated cells, 
    	HT(3,2):  kagome, square-kagome and hyperkagome.
  	(b) Lattices with triangular cells,  where each site belongs to three frustrated cells,
    	HT(3,3): trillium and triangular. 
   	(c) Lattices with tetrahedral cells,
   	HT(4,2): checkerboard, ruby and pyrochlore.
   	Results are given for global-axis and local-axis models, respectively labeled  ``global'' and ``local'', 
	as explained in Sec.~\ref{sec:Model}. 
    	All systems perform a crossover from a high-temperature paramagnetic regime into a 
    	low-temperature classical spin-liquid regime.
  	This is seen by two different Curie laws at high and low temperatures in the reduced susceptibility 
   	$\chi T$ [Eq.~(\ref{eq:BulkSus0})].
	Technical details on simulations and analytics are given in Appendix \ref{sec:AppMC} 
	and \ref{sec:AppHT}, respectively.
  	}
	\label{fig:MC_HT_Thermodyn}
\end{figure*}
%%%%%%%%%%%%%%%%%%%%%%%%%%

%%%%%%%%%%%%%%%%%%%%%%%%%%%%%%%%%%%%%%%%%%%%%
%
%		  		PHYSICAL LATTICE vs. HUSIMI TREE
%
%%%%%%%%%%%%%%%%%%%%%%%%%%%%%%%%%%%%%%%%%%%%%
\subsection{Spin liquids on the Husimi tree}
\label{sec:Husimi}  
%%%%%%%%%%%%%%%%%%%%%%%%%%%%%%%%%%%%%%%%%%%%%

The frustrated Ising model  [Eq.~(\ref{eq:Ham})] on corner-sharing lattices 
[see Fig.~\ref{fig:MC_Lattices}] can be solved, regardless of its physical dimension, by 
numerical methods such as classical Monte Carlo simulations [see Appendix~\ref{sec:AppMC}].
On the analytical front, however, the question is more delicate. 
Since correlations play a major role, one needs a method beyond standard mean-field 
theory, but nonetheless valid for frustrated models across different dimensions. 
The Husimi-Tree (HT) calculation precisely fits our needs, by incorporating the local frustrated constraints of the lattice, irrespectively of its dimension. HT recursively extends from a central frustrated cell -- e.g. a triangle or a tetrahedron -- into a non-reciprocal lattice, without any internal loop beyond the frustrated cell [Fig.~\ref{fig:HT_Lattices}]. As a consequence, its boundary is of comparable size to the volume of the bulk \cite{Bethe1935, Baxter2007} and the HT remains a mean-field approach. It is thus inaccurate at critical points, except above their upper critical dimensions \cite{Wada98a,Jaubert2008,Jaubert10a}. But since we explicitly study models away from phase transitions, we expect pertinent analytical insights from the HT, spurred on by encouraging results on frustrated systems in the literature \cite{Wada98a,monroe98a,Yoshida2002,Jaubert2008,Jaubert2009b,Jaubert10a,Udagawa2010,Jaubert2013,Jurcisinova14a,Jurcisinova14b,Jurcisinova14c}. Technical aspects of the HT method are explained in Appendix~\ref{sec:AppHT} including the explicit expressions of the thermodynamic quantities and how to include the non-collinear easy-axes of the different sublattices.

We will compare a variety of physical lattices, with different numbers of internal loops and frustrated unit cells [Fig.~\ref{fig:MC_Lattices}], to their pseudo-lattice counterparts on the Husimi tree, which do not have any internally closed loops [Fig.~\ref{fig:HT_Lattices}].
Let us define $\mathcal{L}$ as the smallest internal loop formed by frustrated cells on the physical lattice. 
We relate all physical lattices, as introduced in Fig.~\ref{fig:MC_Lattices}, to their corresponding HT trees, 
according to the number of sites per frustrated cell and their connectivity:
\begin{enumerate}[label=(\roman*)]
	\item HT(3,2) [Fig.~\ref{fig:HT_Lattices}(a)] contains three sites in the frustrated cell, where each site is connected
 	between two cells. 
	It relates to the kagome ($\mathcal{L} = 6$), square-kagome ($\mathcal{L} = 4$) and  hyperkagome ($\mathcal{L} = 10$)
	lattice. 
	Considering the complexity of the frustrated cell in the square-kagome lattice, we also included the Husimi tree HTS 
	[Fig.~\ref{fig:HT_Lattices}(b)] to improve the mean-field approximation. 
	\item HT(3,3) [Fig.~\ref{fig:HT_Lattices}(c)] contains three sites in the frustrated cell, where each site is connected 
	between three cells, and relates to the triangular ($\mathcal{L} = 3$) and trillium ($\mathcal{L} = 5$) lattice. 
	\item HT(4,2) [Fig.~\ref{fig:HT_Lattices}(d)] contains four sites in the frustrated cell, where each site is connected between 
	two cells, and relates to the checkerboard ($\mathcal{L} = 4$), ruby ($ \mathcal{L} = 3$) and pyrochlore ($\mathcal{L} = 6$) 
	lattice. 
\end{enumerate}

The similarity between a given lattice and its Husimi tree, taken individually, makes sense -- except 
maybe for the triangular lattice, which will be discussed separately in Section~\ref{sec:triang}. 
In this set up we shall investigate the Curie-law crossover by comparing thermodynamic quantities between 
the physical lattice (as obtained by classical Monte Carlo simulations) and their corresponding pseudo lattice 
on the Husimi tree in the next section.

%%%%%%%%%%%%%%%%%%%%%%%%%%%%%%%%%%%%%%%%%%%%%
%			   		CURIE LAW CROSSOVER 
%%%%%%%%%%%%%%%%%%%%%%%%%%%%%%%%%%%%%%%%%%%%%
\section{The Curie-law crossover}
\label{sec:Curie crossover}  
%%%%%%%%%%%%%%%%%%%%%%%%%%%%%%%%%%%%%%%%%%
%

The Curie-law crossover describes the evolution of the magnetic susceptibility between 
two different Curie laws \cite{Jaubert2013}, whose origin becomes obvious when considering 
the reduced susceptibility $\chi\,T$:
%
%%%%%%%%%%%%%%%%%%%%%%%%%%%
\begin{eqnarray}
	\chi\,T &=&\frac{1}{N} \sum_{i,j} \left[ \langle \vec S_i  \cdot \vec S_j \rangle\,
			- \langle \vec S_{i} \rangle \langle \vec S_{j} \rangle \right]        \nonumber  \\
			&=& 1 + \dfrac{1}{N} \sum_{i \neq j}\langle  \vec S_i  \cdot \vec S_j \rangle  \, .
	\label{eq:BulkSus}
\end{eqnarray}
%%%%%%%%%%%%%%%%%%%%%%%%%%%
%
In magnetically disordered systems, as studied here, $\langle \vec S_{i} \rangle=0$ for all temperatures, while translational invariance implies additionally that
%
%%%%%%%%%%%%%%%%%%%%%%%%%%%
\begin{eqnarray}
	\chi\,T	= 1 + \sum_{i \neq 0}\langle  \vec S_0  \cdot \vec S_i \rangle
	= \sum_{i}\langle  \vec S_0  \cdot \vec S_i \rangle   \, ,
	\label{eq:BulkSus0}
\end{eqnarray}
%%%%%%%%%%%%%%%%%%%%%%%%%%%
%
where $\vec S_{0}$ is an arbitrary ``central'' spin.  In a paramagnet with uncorrelated spins, Eq.~(\ref{eq:BulkSus0}) gives the Curie constant 
%
%%%%%%%%%%%%%%%%%%%%%%%%%%%
\begin{eqnarray}
	C_{\infty} \equiv \chi T \big|_{T \to \infty} = 1\ .
	\label{eq:Cinf}
\end{eqnarray}
%%%%%%%%%%%%%%%%%%%%%%%%%%%
%
At zero temperature, the Curie constant is renormalised by the correlations of the spin liquid
%
%%%%%%%%%%%%%%%%%%%%%%%%%%%
\begin{eqnarray}
	C_{0} \equiv \chi T \big|_{T \to 0} = \sum_{i}\langle  \vec S_0  \cdot \vec S_i \rangle_{T \to 0}\ .
	\label{eq:C0}
\end{eqnarray}
%%%%%%%%%%%%%%%%%%%%%%%%%%%
%
In fact, $C_{0}$ is nothing less than the integration of spin correlations in real space, with $C_{0}$ smaller (greater) than 1 for dominating antiferromagnetic (ferromagnetic) correlations.

%%%%%%%%%%%%%%%%%%%%%%%%%%%%%%%%%%%%%%%%%%%%%
\subsection{Thermodynamics}
\label{sec:Thermodyn}  
%%%%%%%%%%%%%%%%%%%%%%%%%%%%%%%%%%%%%%%%%%
%

Fig.~\ref{fig:MC_HT_Thermodyn} displays thermodynamic observables: normalized energy $E$, specific heat $C_h$, entropy $S$ and reduced susceptibility $\chi T$, obtained by simulating the Hamiltonian $\mathpzc{H}$ [Eq.~(\ref{eq:Ham})] 
with classical Monte Carlo simulations for the physical lattices [Fig.~\ref{fig:MC_Lattices}] and analytical 
calculations on their corresponding Husimi trees [Fig.~\ref{fig:HT_Lattices}]. 
As explained in the introduction, these systems have often been studied in the literature; see e.g. Refs.~ \cite{stephenson64a,rastelli77a,Isoda08a,Macdonald11a,Pohle2016,jurcisinova18a,richter2022,Henley2010,Rehn17a,Redpath2010,Hopkinson07b,Isakov04,Jaubert2013,Bovo2013,Bovo18a,Wills2002,Kano1953,Wannier1950,Wannier73a,pauling35a,Lieb67a,Nagle1966}. Our interest here is not to study them individually, but to see how their thermodynamic properties compare to each other. In particular, classical spin liquids are known for their residual entropy as $T \to 0^+$, that measures the degeneracy of the spin-liquid ground state. It can be 
categorized into three groups \cite{Wills2002,Kano1953,Wannier1950,Wannier73a,pauling35a,Lieb67a,Nagle1966} (corresponding to the three columns in Fig.~\ref{fig:MC_HT_Thermodyn}), 
(i) kagome, square-kagome and hyperkagome lattices with $S(T \to 0) \approx 0.5$, 
(ii) triangular and trillium lattice with $S(T \to 0) \sim 0.3-0.4$, and 
(iii) ruby, checkerboard and pyrochlore lattices with $S(T \to 0) \approx 0.2$.
The HT estimate of the residual entropy is also known as Pauling entropy, which, as a side-note, is not always a lower bound [see Appendix~\ref{sec:Pauling}].

The behavior of the entropy is accompanied by a change in magnetic correlations from a high-temperature regime with $C_{\infty} = 1$ to a model-dependent value $C_{0}$ at low temperatures [see also Table~\ref{tab:HT_MC_zeroT}]. The low-temperature Curie constant $C_{0}$ is not universal, making its value a characteristic property of the underlying spin liquid.

%%%%%%%%%%%%%%%%%%%%%%%%%%%%%%%%%%%%%%%%%%
% Fig.    SPIN-SPIN CORRELATIONS IN REAL SPACE 
%%%%%%%%%%%%%%%%%%%%%%%%%%%%%%%%%%%%%%%%%%
%
\begin{figure*}[t]
\centering
\captionsetup[subfigure]{justification=justified, singlelinecheck=false, position=top}
	\subfloat[Semi-log plot]{\includegraphics[width=0.48\textwidth]{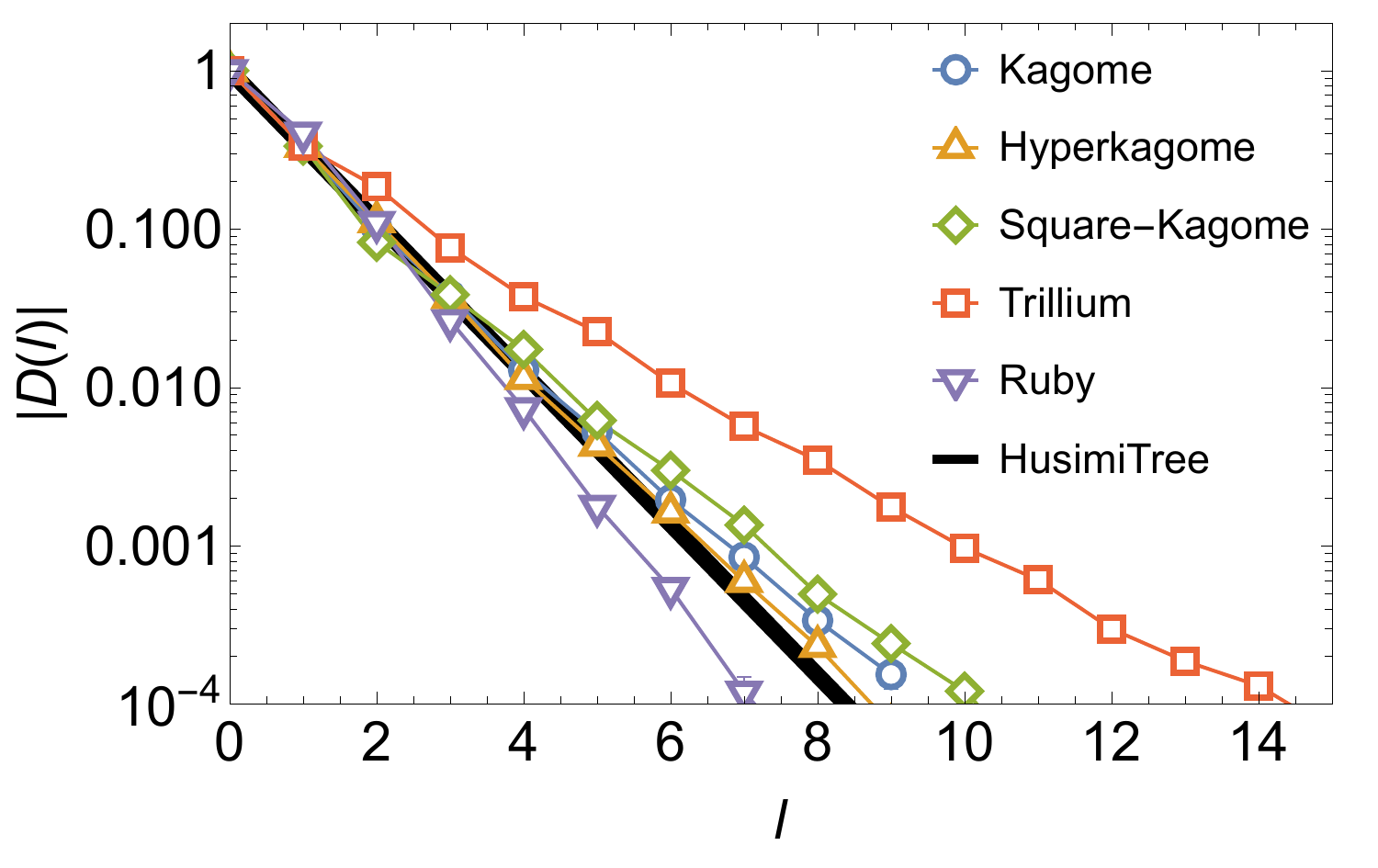}} \quad
	\subfloat[Log-log plot]{\includegraphics[width=0.48\textwidth]{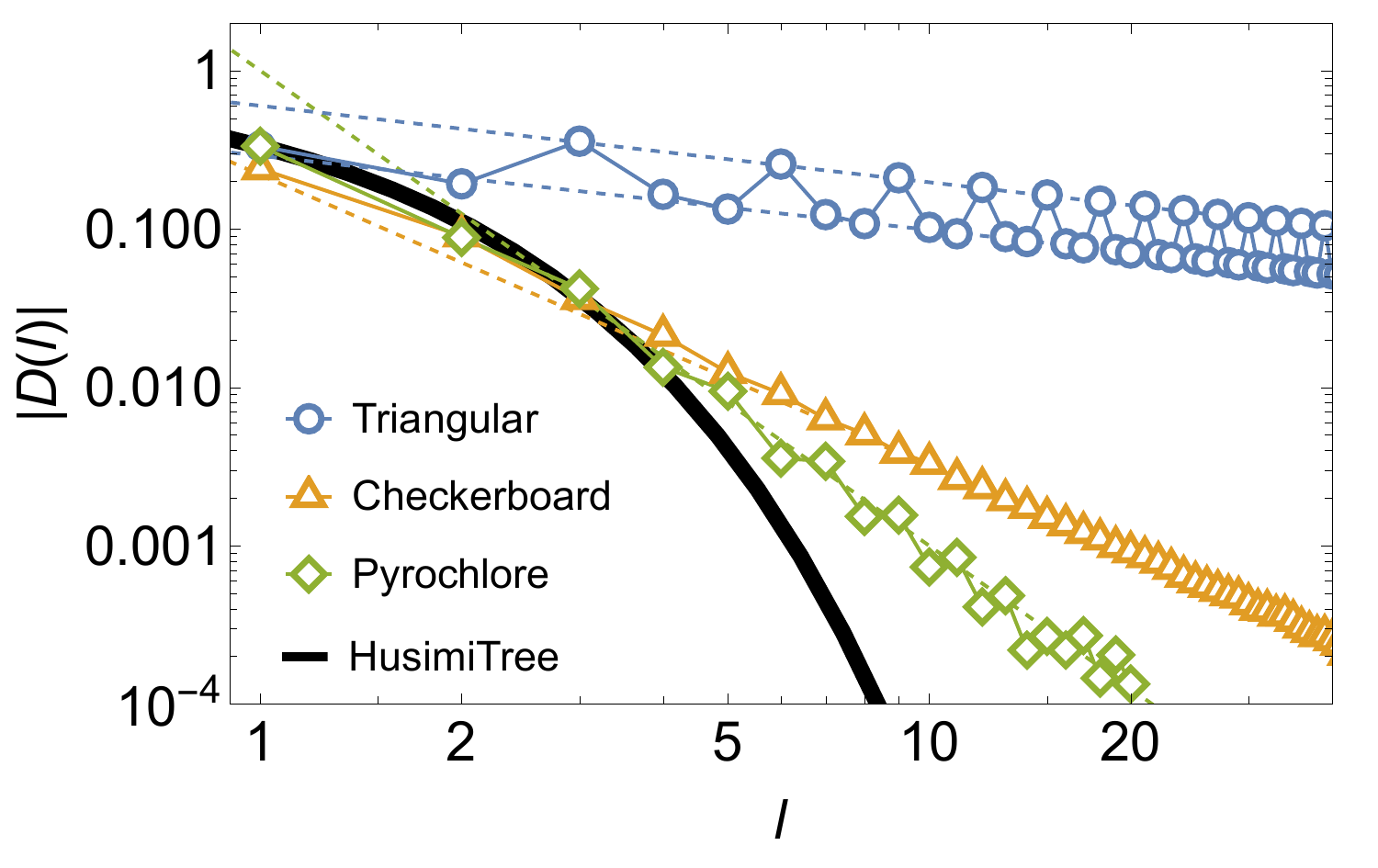}}
	\caption{
	Absolute value of the real-space spin-spin correlation length $|D(\ell)|$ at low temperature, deep in the spin liquid regime of $\mathpzc{H}$ [Eq.~(\ref{eq:Ham})], as obtained from classical Monte Carlo simulations for global-axis Ising spins on their physical lattices [Fig.\ref{fig:MC_Lattices}], and their corresponding Husimi tree (HT).
	(a) Exponential decay $D(\ell) \sim e^{-\ell/\xi}$ on the  
	kagome ($\xi = 1.10(2)$),
	hyperkagome ($\xi = 1.03(2)$),
	square-kagome ($\xi = 1.18(2)$) and 
	ruby lattice ($\xi = 0.76(3)$)
	compare semi-quantitatively well with the exponential decay on the Husimi tree ($\xi_{\rm HT}=0.91$) [Eq.~(\ref{eq:corrlapp})].
	(b) Algebraic decay $D(\ell) \sim 1/\ell^{\alpha}$ on the 
	triangular ($\alpha = 0.476(10) $), 
	checkerboard ($\alpha = 2 $), and 
	pyrochlore ($\alpha = 3 $), 
	lattice.
	On the checkerboard and pyrochlore lattices, correlations are known to decay algebraically 
	(see Section \ref{sec:HTflat}) but follow the HT exponential decay for the first three or four nearest 
	neighbours, i.e. over a distance larger than $\xi_{\rm HT}$. 
	Correlations on the trillium ($\xi = 1.55(2)$) and triangular lattice deviate more strongly from HT expectations, which we believe causes the small, but visible, mismatch of thermodynamic quantities in Fig.~\ref{fig:MC_HT_Thermodyn}(b).
	Note that for a relevant comparison between physical lattices and HT, we used the Manhattan distance 
	$\ell$, defined on paths for each lattice as shown in Fig.~\ref{fig:MC_Lattices}. 
	For lattices in panel (b), the Manhattan distance is also the Euclidian distance. 
	Panels (a) and (b) are respectively on a semi-log and log-log plot.
   	 }
\label{fig:SpinCorr.realSpace}
\end{figure*}
%%%%%%%%%%%%%%%%%%%%%%%%%
%

In some models, the value of $C_{0}$ is easy to understand.
For the ruby, checkerboard and pyrochlore lattice with global axis spins, $C_{0}$ is zero [Fig.~\ref{fig:MC_HT_Thermodyn} (c)]. 
This is because their ground state respects the so-called ice rules \cite{Bramwell1998,Moessner1998} with two up spins 
and two down spins per frustrated cell. 
The magnetization $M = | \sum_i \vec{e}_z \sigma_i |$ is thus not only globally zero on average, $\langle M \rangle =0$, but 
also locally zero for all frustrated units. 
No fluctuations of the magnetization are allowed in the spin liquid, resulting in $\langle M^{2} \rangle =0$ and thus a vanishing 
reduced susceptibility. 
In other words, we get $C_{0} = 0$ as can be expected for any system with a zero-magnetization plateau.
For triangular frustrated units, the opposite reasoning applies because the magnetization cannot be canceled with three 
collinear Ising spins. 
Magnetic fluctuations persist down to zero temperature, and $\langle M^{2}\rangle$ and $\chi T$ remain finite.\\

Remarkably, thermodynamic observables match well within each group of lattices, despite their different physical dimensions. It was already known\cite{Wills2002,Kano1953,Wannier1950,Wannier73a,pauling35a,Lieb67a,Nagle1966} that some models had very similar residual entropies as $T\rightarrow 0^{+}$. Here this similarity is further illustrated with the value of the spin-liquid Curie law $C_{0}$ [see Table \ref{tab:HT_MC_zeroT}]. But more importantly, thermodynamic quantities are essentially the same for \textit{all} temperatures within each group of models. For example, the 2D square-kagome model compares well with the 2D kagome, as recently noticed for quantum spins$-1/2$ [\onlinecite{richter2022}], but also the 3D hyperkagome, while the 2D ruby model matches with 3D pyrochlore for all temperatures.
Furthermore, thermodynamic observables for each group are well reproduced by their corresponding HT, suggesting that correlations barely depend on the physical dimension of the lattice. 
In the following we will try to understand why.

%%%%%%%%%%%%%%%%%%%%%%%%%%%%%%%%%%%%%%%%%%%%%
\subsection{Husimi tree sets the correlation length}
\label{sec:HTcorr}  
%%%%%%%%%%%%%%%%%%%%%%%%%%%%%%%%%%%%%%%%%%%%%

As seen in Eq.~(\ref{eq:C0}), $C_{0}$ corresponds to the integration of real-space correlations in the spin liquid [see also Table \ref{tab:HT_MC_zeroT}]. 
Let us consider HT(3,2) whose $C_{0} = 1/5$. 
This value deviates from Monte-Carlo results on the kagome and hyperkagome lattice within less than 1 \%.
For the square-kagome lattice, the mismatch drops from 2\% to 0.1\% by including a more evolved version 
of the Husimi tree (see HTS in Fig.~\ref{fig:HT_Lattices}(b)), which contains a larger frustrated unit cell
and includes internal loop lengths of four sites. 
Such a trend suggests the presence of a particularly small correlation length $\xi$ in these systems. 

To confirm our suggestion, let us define spin-spin correlations on the HT:
%
%%%%%%%%%%%%%%%%%%%%%%%%%%%
\begin{eqnarray}
	D(\ell)=\langle  \vec S_0  \cdot \vec S_\ell \rangle = \langle  \sigma_0\;  \sigma_\ell \rangle \ ,
\label{eq:Dl}
\end{eqnarray}
%%%%%%%%%%%%%%%%%%%%%%%%%%%
%
assuming that all spins are collinear along a global axis $\vec{S}_i = \sigma_i \vec{e}_z$.
The fact that (i) there is no closed loop in the HT (beyond the size of the frustrated unit), (ii) the Hamiltonian is invariant 
under time-reversal symmetry, and (iii) the HT is by definition locally the same at each vertex, allows us to 
formulate an exact expression for the spin-spin correlations
%
%%%%%%%%%%%%%%%%%%%%%%%%
\begin{eqnarray}
D(\ell) = \langle  \sigma_0\;  \sigma_\ell \rangle
	&=& \langle  \sigma_0\;\sigma_1^{2}\; \sigma_2^{2}\; ... \sigma_{\ell-1}^{2}\; \sigma_\ell \rangle\nonumber\\
	&=& \langle  \sigma_0\;\sigma_1 \rangle \langle\sigma_1\;\sigma_2\rangle ... \langle\sigma_{\ell-1}\; \sigma_\ell \rangle\nonumber\\
	&=& \langle  \sigma_0\;\sigma_1 \rangle^{\ell}	\, .
\label{eq:corrdecoupl}
\end{eqnarray}
%%%%%%%%%%%%%%%%%%%%%%%%
%
The nearest-neighbor spin-spin correlation averaged over the ensemble of ground states 
can be easily calculated. 
And it turns out to be the same result for the three kinds of Husimi trees, 
HT(3,2), HT(3,3) and HT(4,2): 
%
%%%%%%%%%%%%%%%%%%%%%%%%%%%
\begin{equation}
	\langle  \sigma_0\;\sigma_1 \rangle=-1/3 \, .
\label{eq:corr.NN.HT}
\end{equation}
%%%%%%%%%%%%%%%%%%%%%%%%%%%
%
This means that correlations decay exponentially on Husimi trees, following the formula 
%
%%%%%%%%%%%%%%%%%%%%%%%%%%%
\begin{equation}
	D(\ell) =  \bigg(-\frac{1}{3} \bigg)^\ell =  (-1)^\ell  \ e^{- \ell \ln{3}}	\; ,
\label{eq:corrCl}
\end{equation}
%%%%%%%%%%%%%%%%%%%%%%%%%%%
%
for all Husimi trees considered here, giving a correlation length
%
%%%%%%%%%%%%%%%%%%%%%%%%%%%
\begin{equation}
	\xi_{\rm HT} = \Big(\ln{3} \Big)^{-1} \approx 0.91	\;.
\label{eq:corrlapp}
\end{equation}
%%%%%%%%%%%%%%%%%%%%%%%%%%%
%
More generally, for a Husimi tree whose frustrated units are made of $N_{u}$ Ising spins fully connected between each other via antiferromagnetic couplings, the correlation length in the degenerate 
ground state is
%
%%%%%%%%%%%%%%%%%%%%%%%%%%%
\begin{eqnarray}
	\xi_{\rm HT} = &\dfrac{1}{\ln N_{u}},\quad &\textrm{if $N_{u}$ is odd,}\\
	\xi_{\rm HT} = &\dfrac{1}{\ln (N_{u}-1)},\quad &\textrm{if $N_{u}$ is even.}
\end{eqnarray}
%%%%%%%%%%%%%%%%%%%%%%%%%%%
%

Eq.~(\ref{eq:corrlapp}) means that correlations decay typically over the nearest-neighbor distance in Husimi trees. 
This length scale is smaller than any loop in the real lattice, suggesting that correlations in real lattices may decay 
in a similar way at short distances. 
Monte Carlo simulations confirm this assumption on the kagome, hyperkagome \cite{Hopkinson07b}, square-kagome \cite{Pohle2016} and ruby \cite{Rehn17a} 
lattice at low temperatures [Fig.~\ref{fig:SpinCorr.realSpace}(a)], whose correlation lengths are roughly the same as $\xi_{\rm HT}$. 
Since the correlation length is expected to decrease upon heating, this short correlation length is consistent with the success of the 
HT approximation over the whole temperature range for global- and local-axis models alike.

For the sake of clarity, we should point out that the value of $C_0$ is \textit{not} coming from a cutoff of the correlations beyond $\xi_{\rm HT}$. Indeed, it would be tempting to see classical spin liquids as an ensemble of independent clusters of superspins (on each triangle or tetrahedron), and the spin-liquid Curie law as a form of superparamagnetism, as observed with ferromagnetic nanoparticles \cite{Neel47a}. However, we cannot recover the value $C_0=0.2$ for kagome-type systems from such an argument. Appendix \ref{sec:AppHT.reducedSus} shows that the resulting error scales like $-\dfrac{6}{5}\left(-\dfrac{2}{3}\right)^{L+1}$ on a Husimi tree of $L$ layers. Even if correlations ultimately vanish at long distance, the cutoff necessary to approximately recover the value of $C_0$ is much larger than $\xi_{\rm HT}$. To understand the similarity between simulations and analytics, it would be more accurate to see the paths connecting the central spin to the many spins on layer $L$ on the infinite-dimensional Husimi tree of Fig.~\ref{fig:HT_Lattices} as virtual paths of correlations connecting a pair of $L^{\rm th}$ nearest neighbors on the corresponding real lattice of Fig.~\ref{fig:MC_Lattices}. This picture is nearly exact up to the $n^{\rm th}$ nearest neighbor before closing the minimal loop of size $\mathcal{L}$ on the real lattice ($n=\mathcal{L}/2-1$), which is why deviations between Monte Carlo and Husimi tree grow inversely with $\mathcal{L}$ in Fig.~\ref{fig:SpinCorr.realSpace}: first hyperkagome, then kagome and finally square-kagome.

%%%%%%%%%%%%%%%%%%%%%%%%%%%%%%%%%%%%%%%%%%%%%
\subsection{Coulomb field theory and flat bands}
\label{sec:HTflat}  
%%%%%%%%%%%%%%%%%%%%%%%%%%%%%%%%%%%%%%%%%%%%%

On the other hand, correlations on the checkerboard and pyrochlore lattices are algebraic at low temperature, scaling like $1/r^{d}$ [\onlinecite{isakov2004}], with $d$ the physical dimension of the lattice. Their angular dependence is dipolar though, which means that the integration of these correlations over the entire system in Eq.~(\ref{eq:C0}) does not diverge, and $C_{0}$ is well defined. The dipolar nature of these correlations comes from the fact that their ground states are ice models, described by an emergent Coulomb field theory \cite{Henley2010}. With respect to the exponential decay of the HT, these algebraic correlations only differ beyond the third or fourth neighbour; see the comparison to the black curve on Fig.~\ref{fig:SpinCorr.realSpace}(b). In that sense the correlation length $\xi_{\rm HT}$ remains effectively relevant at short distances. That being said, one would have been forgiven to expect larger corrections to $C_{0}$ coming from the long-range algebraic tail. Here again we are left with the question: why are these corrections so small ?\\

For models with a global axis, $C_{0}$ is known to be exactly zero (see discussion in Section \ref{sec:Thermodyn}); the ice rules prevent magnetic fluctuations for all tetrahedra, and thus conveniently prevent any corrections. But this does not explain the match of Fig.~\ref{fig:MC_HT_Thermodyn}(c) for the local-axis pyrochlore model, a.k.a. spin ice, where $C_{0}\approx 2.0$. Magnetic fluctuations are allowed in the spin-ice ground state. Additionally, since the spin-ice model is ferromagnetic, the sum of Eq.~(\ref{eq:BulkSus0}) contains mostly positive terms, as opposed to the alternating series encountered for the integration of correlations in antiferromagnets [Appendix \ref{sec:AppHT.reducedSus}]. For the latter, potential corrections coming from algebraic correlations would partially cancel out; while they would a priori add up in the ferromagnetic model. This suggests that an alternative point of view is necessary.

%%%%%%%%%%%%%%%%%%%%%%%%%%%%%%%%%%%%%%%%%%
% Fig.     Sq -- PINCH-POINT EVOLUTION
%%%%%%%%%%%%%%%%%%%%%%%%%%%%%%%%%%%%%%%%%%
%
\begin{figure}[t]
\centering
	\captionsetup[subfigure]{justification=centering, singlelinecheck=false, position=top} \hspace{-1cm}
	\subfloat[ $T/J$ = 0.01] {\includegraphics[height=0.12\textheight, trim=0cm 0cm 0cm 1.5cm ]{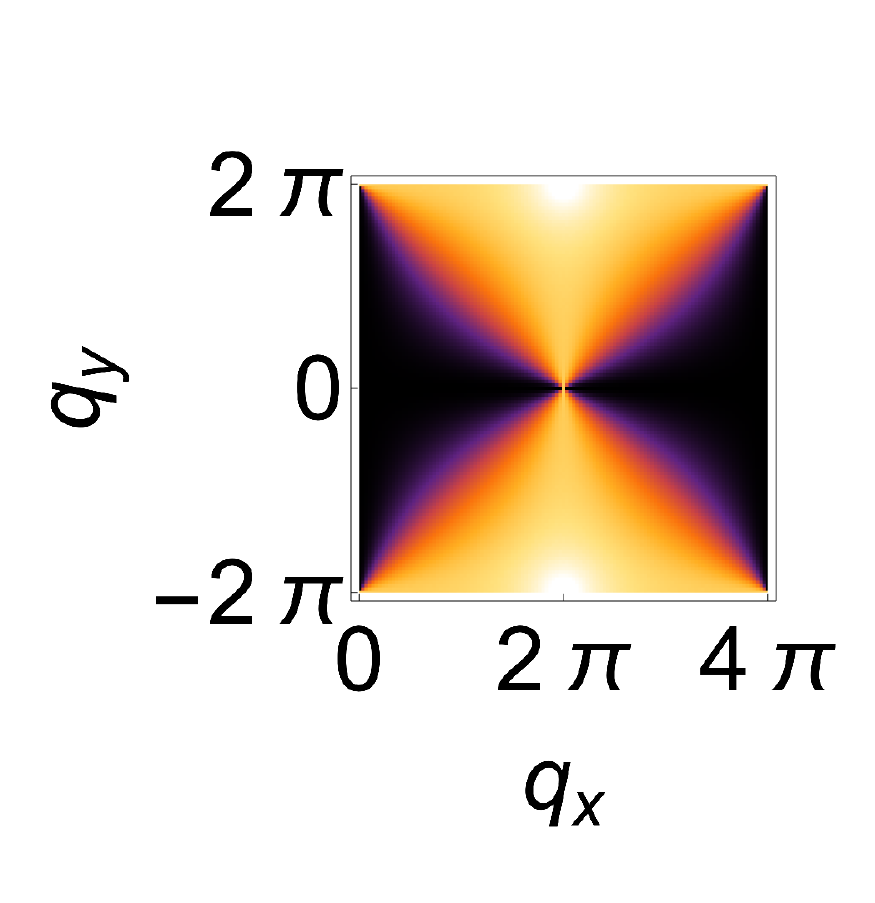}} 
	\subfloat[ $T/J$ = 1] {\includegraphics[height=0.12\textheight, trim=3.cm 0cm 1cm 1.5cm ]{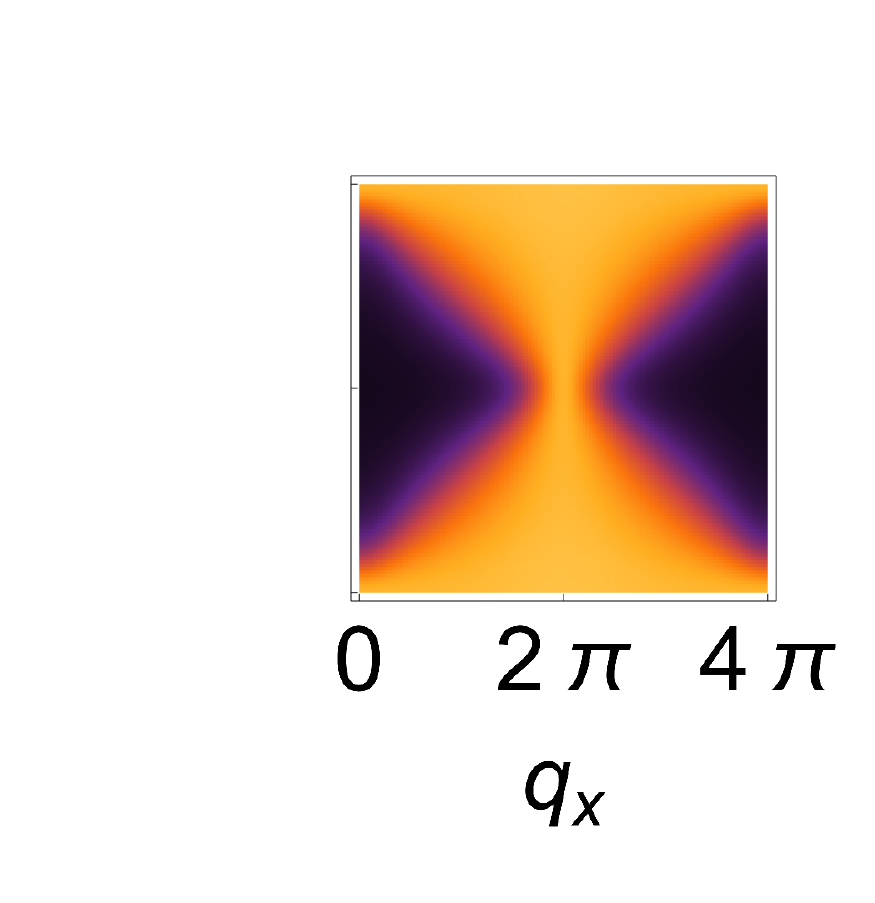}}  	
	\subfloat[ $T/J$ = 5] {\includegraphics[height=0.12\textheight, trim=2.cm 0cm 2cm 1.5cm ]{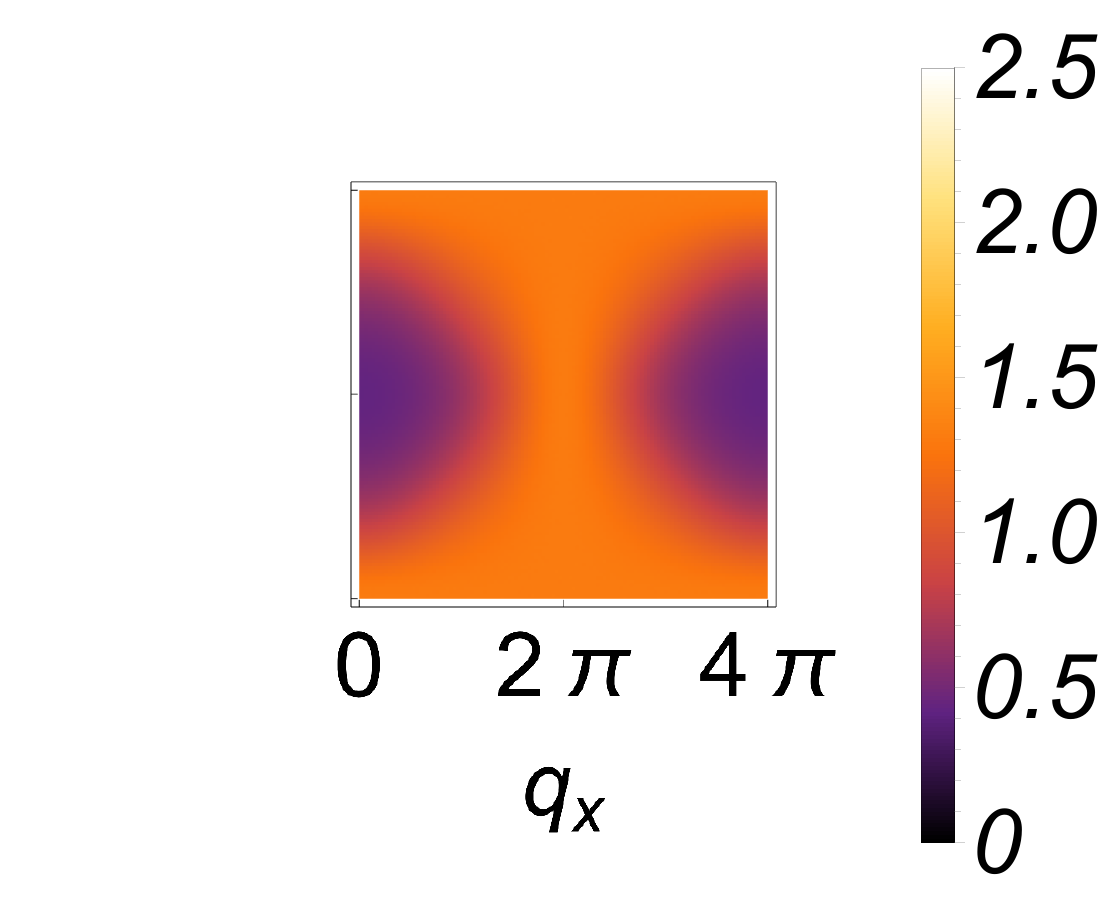}} \\
	\subfloat[FWHM] {\includegraphics[width=0.49\textwidth]{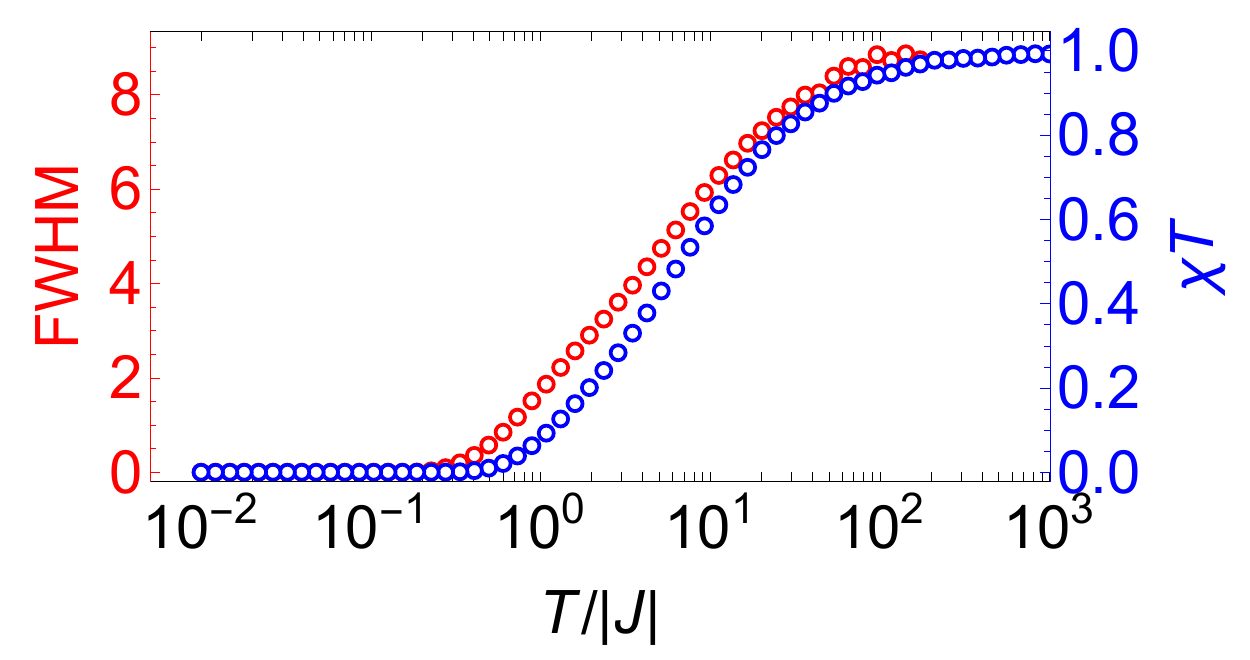}}	
	\llap{\makebox[\wd1][l]{\raisebox{-2.95cm}{\hspace{-7.55cm}{\includegraphics[width=0.155\textwidth]{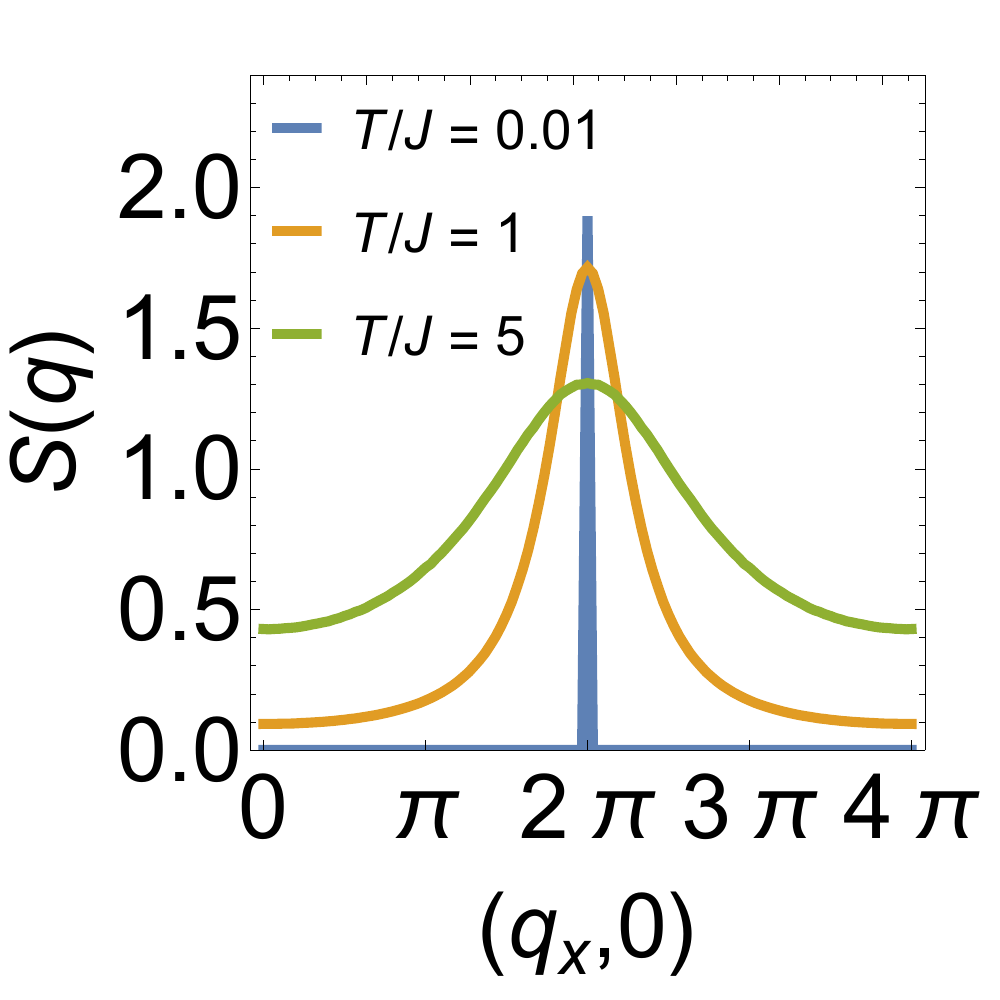}}}}} 	\\
	\caption{
	Signatures of the Curie-law crossover in coulombic spin liquids. 
	(a)--(c) equal-time structure factor $S({\bf q})$ [Eq.~(\ref{eq:Sq})] of $\mathpzc{H}$ [Eq.~(\ref{eq:Ham})] for Ising spins in their global axis 
	on the checkerboard lattice, obtained from classical Monte Carlo simulations. 
	The pinch-point gets broader upon heating. 
	(d) Temperature-dependent full width at half maximum (FWHM) of pinch-points.
	FWHM has been obtained from a Lorentzian fit for line cuts of the pinch point along its singular $q_x$ direction 
	(see inset of (d)). 
	The FWHM illustrates the Curie-law crossover in a similar way as the reduced susceptibility $\chi T$.
}
\label{fig:Sq.pinchpoint}
\end{figure}
%%%%%%%%%%%%%%%%%%%%%%%%%%

Let us temporarily step away from Husimi trees and consider the other facet of spin ice, as a U(1) Coulomb gauge field. As mentioned previously this gauge-field texture comes from the ice rules, that can be rewritten as a divergence-free constraint on the magnetisation field $\vec M$ [\onlinecite{Henley2010}]. But spin ice is not the only model supporting this type of texture. The ground state of the pyrochlore antiferromagnet with classical Heisenberg spins is a U(1)$\times$U(1)$\times$U(1) Coulomb gauge field that has often been described as three copies of spin ice \cite{Canals2001,Garanin02a,isakov2004,Henley2010}. The susceptibility of these divergence-free fields is readily available using the Self-Consistent Gaussian Approximation (SCGA). It means that with the proper normalisation, SCGA offers an alternative approach to compute $C_{0}$ and $C_{\infty}$ [see Appendix \ref{sec:gauge}]. In particular it tells us that the ratio $C_{\infty} / C_{0}$ is due to the topology of the magnetic band structure of the pyrochlore lattice \cite{reimers91a,conlon10a,conlon10b,benton14a}; the ground state is composed of two degenerate flat bands, accounting for \textit{half} ($C_{\infty} / C_{0} = 1/2$) of the total number of bands.

To summarise, since $C_{0}$ comes from the integration of correlations [Eq.~(\ref{eq:C0})], it is remarkable that algebraic correlations in real lattices give almost the same result as exponential correlations in Husimi trees [see pyrochlore and checkerboard results in Fig.~\ref{fig:MC_HT_Thermodyn}]. This is because $C_{0}=0$ is protected by the absence of local fluctuations for global-axis models, while $C_{0}\approx 2.0$ is a direct consequence of the topology of the band structure for local-axis models.\\

Before closing our discussion on the checkerboard and pyrochlore lattices, let us take advantage of these dipolar correlations, whose signatures in the equal-time structure factor present sharp, singular features known as pinch points \cite{youngblood81a, Harris1997, Henley2010}. Upon heating, these singular features broaden as topological-charge excitations disrupt the Coulomb field [Fig.~\ref{fig:Sq.pinchpoint} (a)--(c)] \cite{Fennell09a}. By measuring their breadth, pinch points offer a quantitative way to measure the establishment of the spin liquid. Fig.~\ref{fig:Sq.pinchpoint}(d) shows the full width at half maximum (FWHM) of the pinch point on the checkerboard lattice as a function of temperature. Our point is that the Curie-law crossover, as seen in $\chi T$, is able to grasp the evolution of FWHM, i.e. the build up of the spin liquid. And while only a fraction of spin liquids have characteristic, singular, patterns such as pinch points, the Curie-law crossover is a generic property of all spin liquids. This vindicates the Curie-law crossover as a useful signature of the onset of a spin liquid, and the reduced susceptibility $\chi T$ as a suitable observable to measure it.

%%%%%%%%%%%%%%%%%%%%%%%%%%%%%%%%%%%%%%%%%%%%%
\subsection{The triangular and trillium lattice}
\label{sec:triang}  
%%%%%%%%%%%%%%%%%%%%%%%%%%%%%%%%%%%%%%%%%%
%

\begin{figure}[h]
    \centering
    \includegraphics[width=0.45\textwidth]{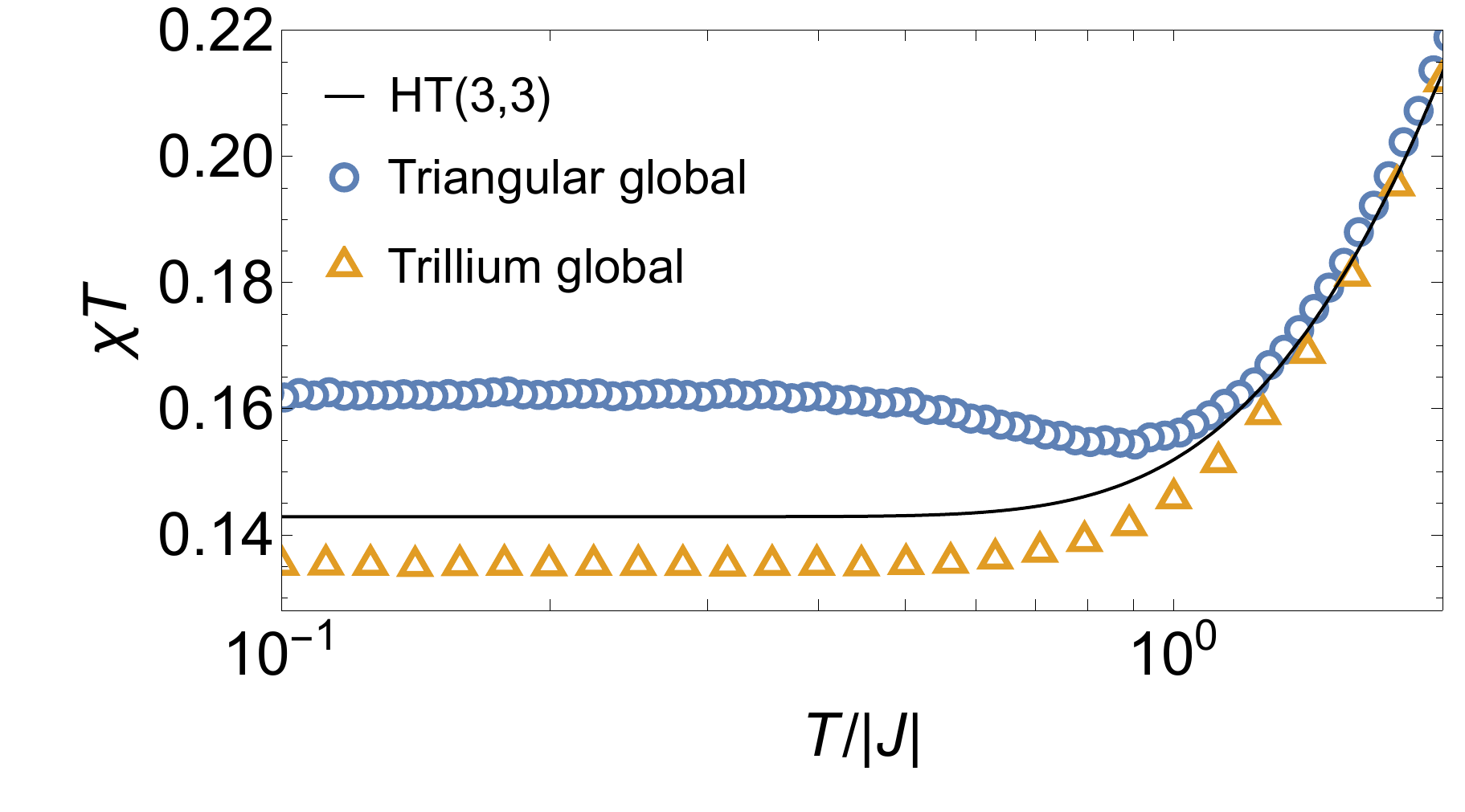}
    \caption{
    Reduced susceptibility $\chi T$ of the Ising antiferromagnet on the triangular and trillium 
    lattice, emphasising the difference between the Husimi tree (HT) and 
    Monte Carlo (MC) simulations. 
    The global-axis triangular antiferromagnet possesses a non-monotonic reduced susceptibility $\chi T$, with a small but distinct minimum at $T =0.9$. 
    }
    \label{fig:Tri_ZoomIn}
\end{figure}

Let us now consider two systems with noticeably different geometries; 
the triangular and trillium lattice. 
While the latter is three dimensional and made of corner-sharing triangles, 
the former is two dimensional and usually 
seen as made of edge-sharing triangles.
From the view point of Husimi trees, HT(3,3) is clearly a reasonable 
approximation for the trillium lattice, with each spin belonging to three triangles.
But, even if less conventional, it can also be used for the triangular 
lattice \cite{Monroe1998,Jurcisinova14c}, since each spin can similarly be 
seen as shared by three triangles (see colored lattice in Fig.~\ref{fig:MC_Lattices}(a)).
The obvious caveat of this choice of Husimi tree (made of 3 spins) is that loops 
that are ignored, are of the same size than the frustrated triangular unit cell itself. 
However, by direct comparison between MC and HT(3,3) results in Fig.~\ref{fig:MC_HT_Thermodyn}(b), 
the reduced susceptibility, $\chi T$, of the two antiferromagnetic models 
overlap with a quantitative difference appearing only below $T \lesssim 1$ 
[Fig.~\ref{fig:Tri_ZoomIn}]. 
%

%%%%%%%%%%%%%%%%%%%%%%%%%%%%%%%%%%%%%%%%%%
% Fig.    ZOOM IN FOR TRIANGULAR LATTICE  
%%%%%%%%%%%%%%%%%%%%%%%%%%%%%%%%%%%%%%%%%%
%
%%%%%%%%%%%%%%%%%%%%%%%%%%
%%%%%%%%%%%%%%%%%%%%%%%%%%

The excellent match above $T \gtrsim 1$ is in part due to the fact that the 
nearest-neighbour correlations in the degenerate ground state 
is $\langle \sigma_{0} \sigma_{1}\rangle=-1/3$ [\onlinecite{stephenson64a}], 
for triangular and trillium systems in accordance with their corresponding Husimi tree 
[see Eq.~(\ref{eq:corr.NN.HT})].
Indeed, the ground state energy is 
$E_{\rm gs}=-N_{\rm bond} J / 3= N_{\rm bond} \langle \sigma_{0} \sigma_{1}\rangle$,
where $N_{\rm bond}$ is the number of nearest-neighbour bonds. 
For $T \lesssim 1$, correlations beyond nearest neighbours apparently start to play a role on
the real lattices. From Fig.~\ref{fig:Tri_ZoomIn}, the deviation from the HT curve indicates a dominant antiferromagnetic (resp. ferromagnetic) contribution for the trillium (resp. triangular) lattice [Eq.~(\ref{eq:BulkSus0})]. 
In the triangular case, the third nearest-neighbour correlations are known to be 
strongly ferromagnetic \cite{stephenson64a, rastelli77a}, with 
$\langle \sigma_{0} \sigma_{3}\rangle > |\langle \sigma_{0} \sigma_{1}\rangle|$,
as $T\rightarrow 0^{+}$. 
It is likely that this increase of ferromagnetic correlations in the ground state causes 
an upturn of the reduced susceptibility [Fig.~\ref{fig:Tri_ZoomIn}].
Accordingly, integrated correlations in the triangular Ising antiferromagnet are more antiferromagnetic at \textit{finite} temperature, for $T\approx 0.9$, than in the spin-liquid ground state.
Such a non-monotonic behavior of the reduced susceptibility $\chi T$ is unusual, but not rare.

It is even more pronounced for the trillium lattice with easy axes. The reduced susceptibility $\chi T$ of easy-axes models necessarily increase upon cooling from high temperature, because nearest neighbor correlations are always ferromagnetic (the scalar product in Eq.~(\ref{eq:Jscaling}) is always negative). For the trillium lattice, however, one can show that $C_0\approx C_\infty=1$ [see Appendix \ref{sec:AppHT.tri2L}]. It means that $\chi T$ has to decrease at low temperature.

The phenomenon of reentrance with bond-dependent interaction anisotropy is yet another example of non-monotonic $\chi T$, and discussed in detail elsewhere [\onlinecite{Pohle2016},\onlinecite{Schmidt17a}].

%%%%%%%%%%%%%%%%%%%%%%%%%%%%%%%%%%%%%%%%%%%%%
%
%			   		CURIE-LAW ANSATZ
%
%%%%%%%%%%%%%%%%%%%%%%%%%%%%%%%%%%%%%%%%%%%%%
\section{Husimi Ansatz for the Curie-law crossover}
\label{sec:ansatz}
%%%%%%%%%%%%%%%%%%%%%%%%%%%%%%%%%%%%%%%%%%

%%%%%%%%%%%%%%%%%%%%%%%%%%%%%%%%%%%%%%%%%%
\subsection{Limitation of the Curie-Weiss fit}
\label{sec:Limit-CWfit}  
%%%%%%%%%%%%%%%%%%%%%%%%%%%%%%%%%%%%%%%%%%

As mentioned in the introduction, the Curie-Weiss temperature $\theta_{\rm cw} = -z\, J$ is a mean-field estimate of the transition temperature $T_{c}$ for a system with connectivity $z$, where the Curie-Weiss law is a consequence of critical scaling invariance with critical exponent $\gamma=1$. Even though the concept of conventional order does not apply to spin liquids, $\theta_{\rm cw}$ does represent a meaningful quantity, as a measure of interaction strength. The practical question is, how accurately can this quantity be measured in experiments ?

Best estimates can only be made at high temperatures, since $\theta_{\rm cw}$ is the first-order correction to the Curie law
\begin{equation}
	\frac{1}{\chi} = \frac{T}{C} \left[ 1 - \frac{\theta_{\text{cw}}}{T} 
				+ \mathcal{O}\left(\frac{1}{T^{2}}\right) \right]	\;.
	\label{eq:Curie-Weiss.HighT}
\end{equation}
And here is the main issue with the Curie-Weiss temperature $\theta_{\rm cw}$. 
In magnets, the high-temperature regime is frequently not accessible, since it 
is two or three orders of magnitude larger than the characteristic exchange coupling 
$J$.
For example in magnets with $3d$ valence electrons, $J$ is often of the order of $\sim100$ K 
and the high-temperature regime is inaccessible because it lies above the melting point 
of the crystal. 
On the other hand for magnets with $4f$ valence electrons, $J$ is much smaller, of the order of $\sim 1$ K. 
But $4f$ ions have a large single-ion degeneracy, lifted by the local crystal field. 
This crystal field introduces a single-ion anisotropy with an associated energy scale, which 
varies a lot from one material to another, but the lowest single-ion excitation is usually 
of the order of $10-100$ K. 
The high-temperature region is thus difficult to access because the nature of magnetic 
moments changes with temperature \cite{Li2021}.
We refer the reader to the useful tutorial written by Mugiraneza \& Hallas [\onlinecite{mugiraneza2022}] for a practical, step-by-step, application of the Curie-Weiss fit.

The susceptibility measures the evolution of the spin-spin correlations [Eq.~(\ref{eq:BulkSus0})]. And the problem is that, as we have seen throughout this paper, this evolution from paramagnetism to spin liquid takes place over several orders of magnitude in temperatures. It is thus naturally best seen on a logarithmic scale. Applying the Curie-Weiss law, which is a linear fit, can be dangerous. What appears to be a reasonable temperature window on a linear scale might actually only measure a small evolution of the spin-spin correlations. The Curie-Weiss fit will always give a result of course, but the outcome will depend on the window of measurement [Fig.~\ref{fig:CurieLaw.Crossover}]. And if the high-temperature regime is not available, then it is not possible to check if the value is correct or not, causing a potentially (largely) inaccurate estimate of $\theta_{\rm cw}$.

%%%%%%%%%%%%%%%%%%%%%%%%%%%%%%%%%%%%%%%%%%
% Fig.    EMPIRICAL FITTING FUNCTION
%%%%%%%%%%%%%%%%%%%%%%%%%%%%%%%%%%%%%%%%%%
%
\begin{figure*}[t]
\centering
\captionsetup[subfigure]{justification=justified, singlelinecheck=false, position=top}
	\subfloat[Pyrochlore Ising global]{\includegraphics[width=0.25\textwidth]{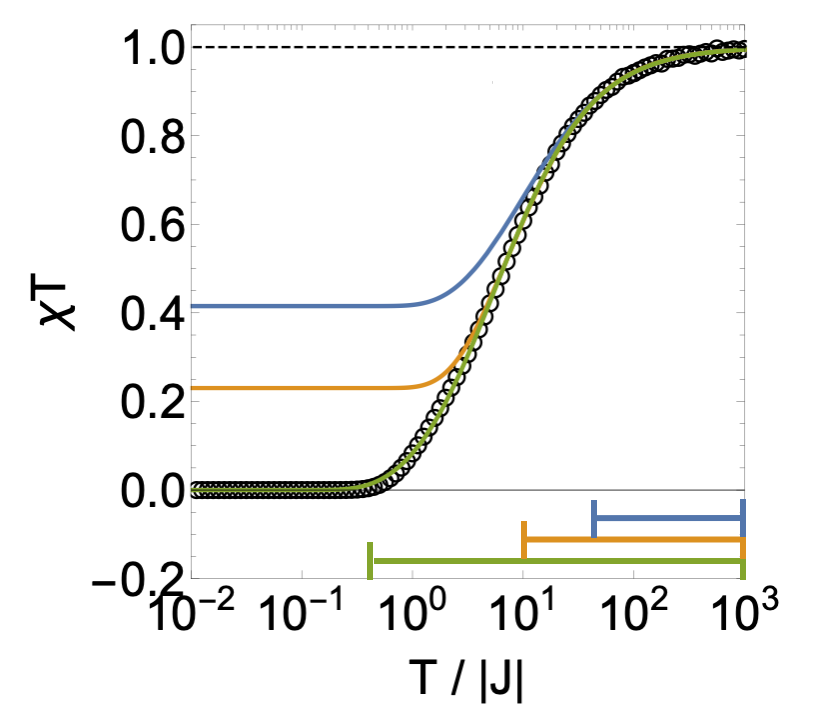}}
	\subfloat[Pyrochlore HAF]{\includegraphics[width=0.25\textwidth]{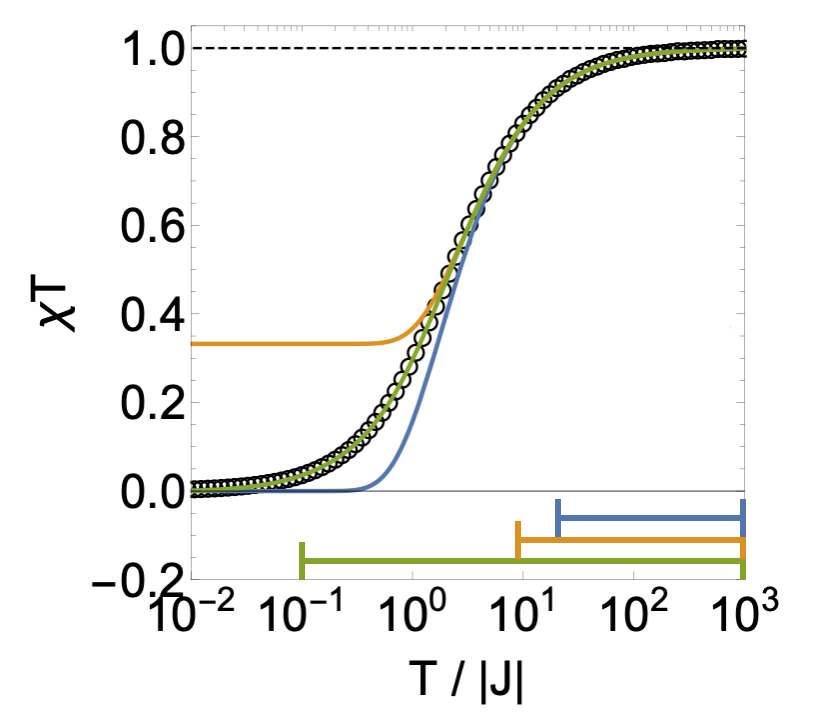}}
	\subfloat[Pyrochlore Ising local]{\includegraphics[width=0.25\textwidth]{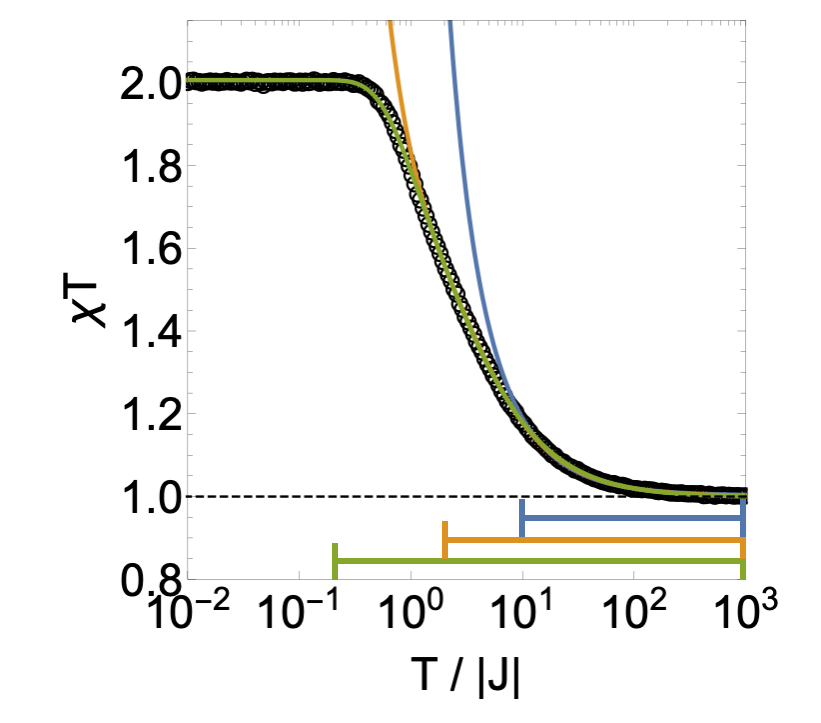}}
	\subfloat[Pyrochlore pHAF]{\includegraphics[width=0.25\textwidth]{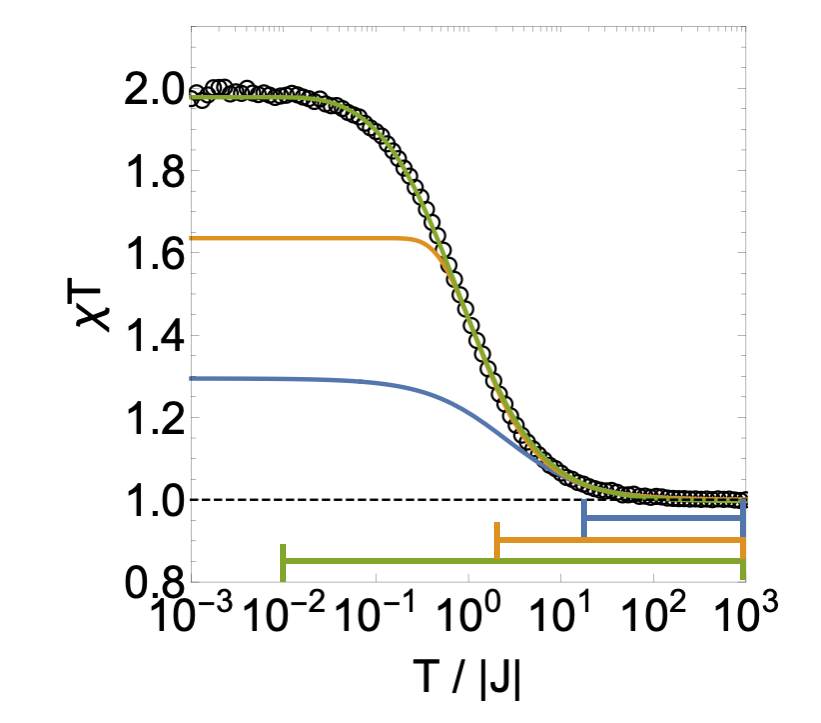}}	\\
	\captionsetup[subfigure]{labelformat=empty}
	\vspace{-0.8cm}
	\subfloat[]{\includegraphics[width=0.25\textwidth]{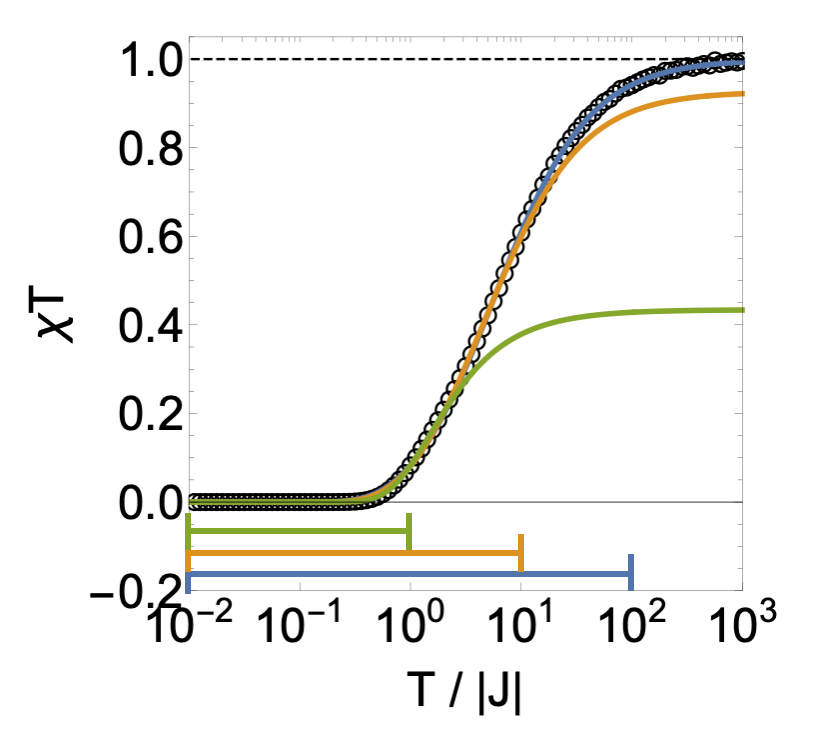}}
	\subfloat[]{\includegraphics[width=0.25\textwidth]{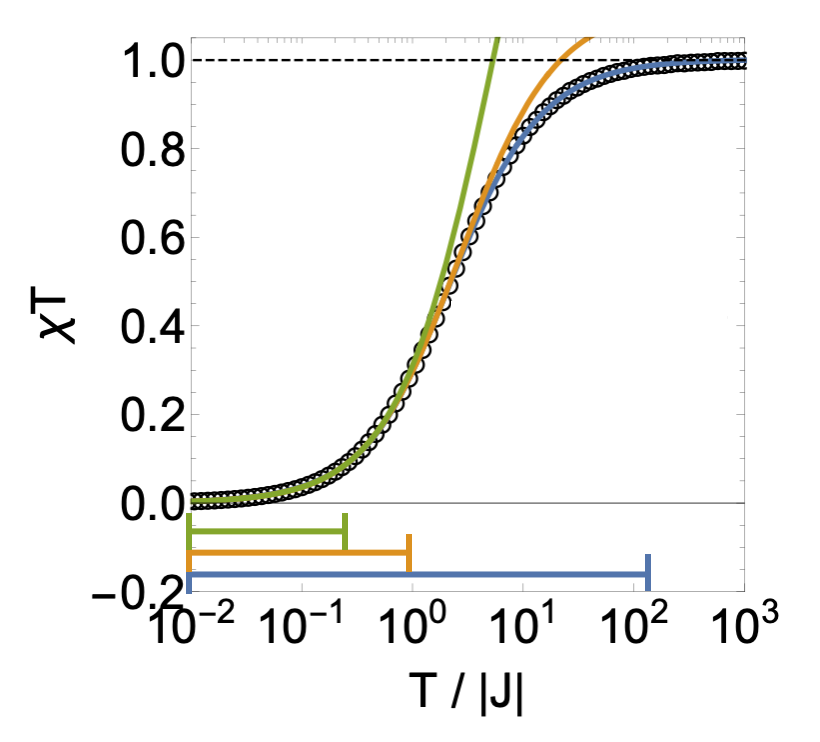}}
	\subfloat[]{\includegraphics[width=0.25\textwidth]{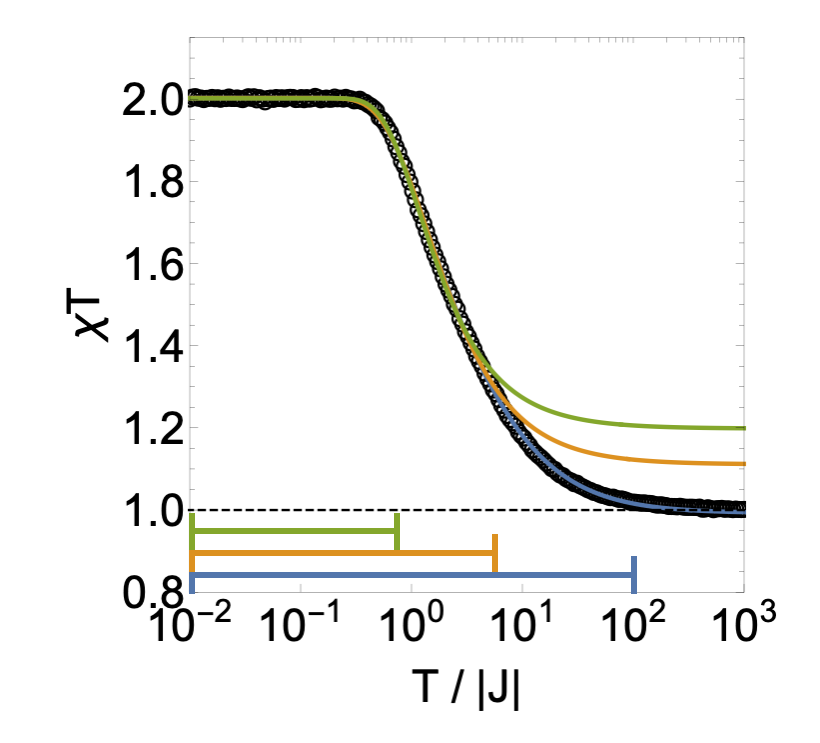}}
	\subfloat[]{\includegraphics[width=0.25\textwidth]{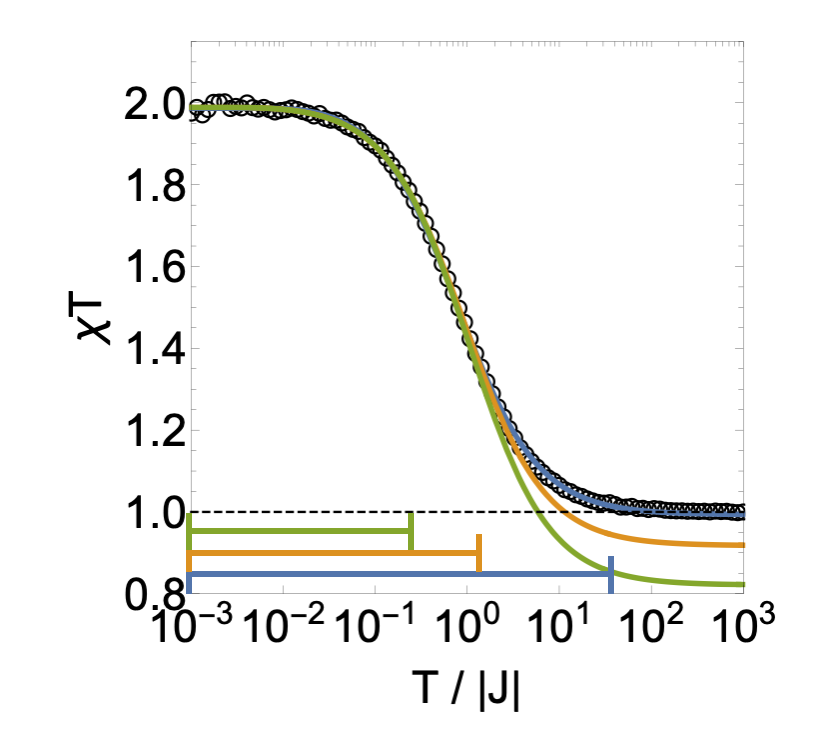}}	\\
	\vspace{-0.8cm}
	\subfloat[]{\includegraphics[width=0.25\textwidth]{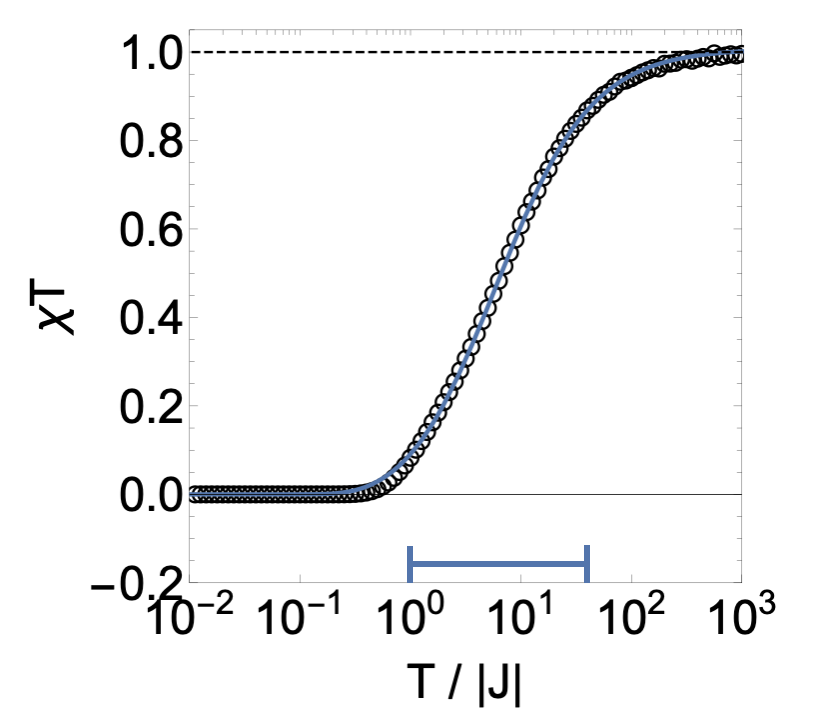}}
	\subfloat[]{\includegraphics[width=0.25\textwidth]{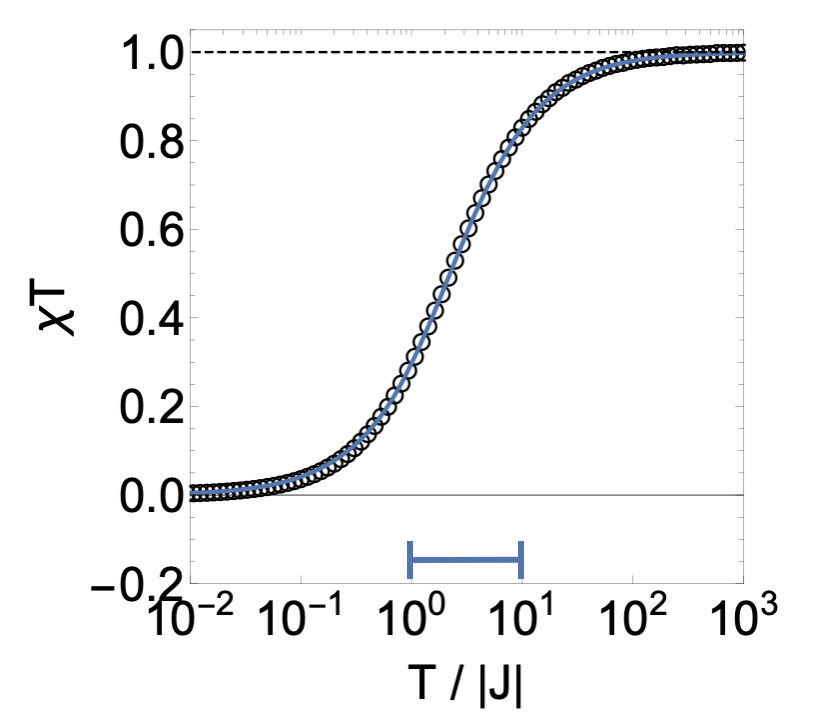}}
	\subfloat[]{\includegraphics[width=0.25\textwidth]{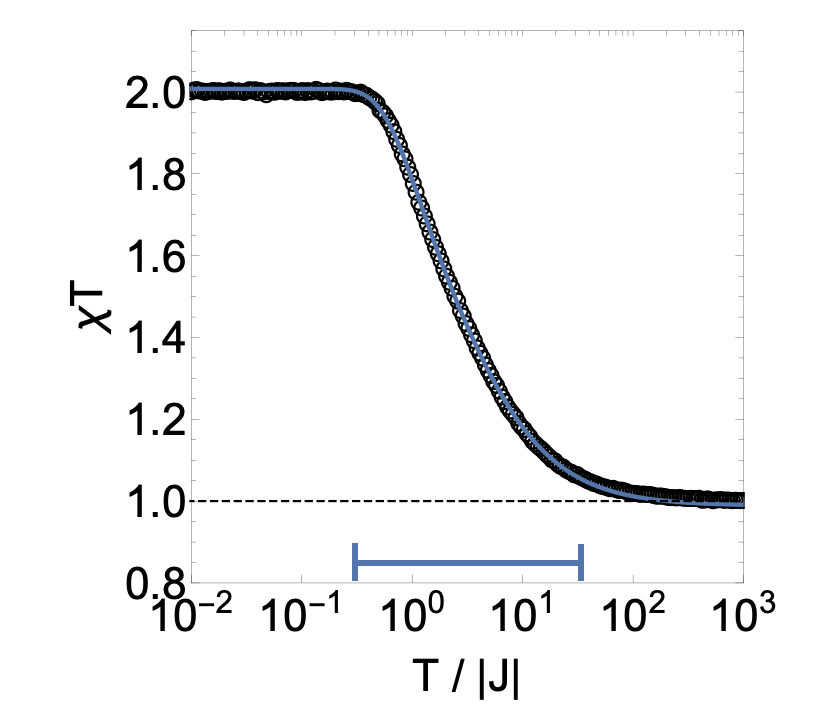}}
	\subfloat[]{\includegraphics[width=0.25\textwidth]{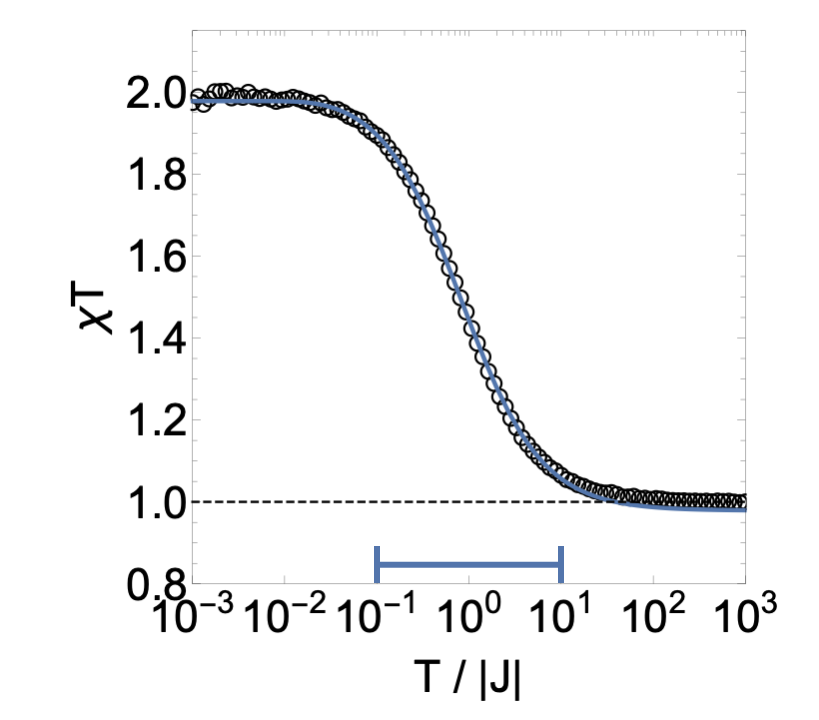}}	\\
	\caption{
	Empirical fit [Eq.(\ref{eq:HT.exp.fit_1})] of the reduced susceptibility $\chi T$ on the pyrochlore 
	lattice [Fig.\ref{fig:MC_Lattices}(h)], as obtained from classical Monte Carlo simulations
	of $\mathpzc{H}$ in Eq.~(\ref{eq:Ham}) for both Ising and continuous Heisenberg spins and $\mathpzc{H}_{\sf XXZ}$ in Eq.~(\ref{eq:HamXXZ}). 
	(a) pyrochlore Ising, global axis and (b) pyrochlore Heisenberg antiferromagnet show dominant 
	antiferromagnetic correlations $\chi T|_{T\to0} = 0$, while (c) pyrochlore Ising, local axis (spin ice) and 
	(d) the pyrochlore pseudo-Heisenberg antiferromagnet show dominating ferromagnetic 
	correlations $\chi T|_{T\to0} = 2$. The fitting windows are given by the coloured bars at the bottom of each figure.
	Examples for different fitting windows are shown for high-temperature (1$^{\sf st}$  row) and low-temperature regions (2$^{\sf nd}$ row).
	The last row shows for each model a minimal fitting window, which is sufficient 
	to reproduce $\chi T$ over the full range of temperatures.
	Simulations were done for system sizes of linear dimensions $L=16$, i.e.\ $N = 65\,536$ spins. 
	} 
\label{fig:EmpiricalFit}
\end{figure*}
%%%%%%%%%%%%%%%%%%%%%%%%%
%

%%%%%%%%%%%%%%%%%%%%%%%%%%%%%%%%%%%%%%%%%%
\subsection{The Husimi Ansatz}
\label{sec:ansatzformula}  
%%%%%%%%%%%%%%%%%%%%%%%%%%%%%%%%%%%%%%%%%%

To measure the Curie-Weiss temperature in spin liquids, a complementary approach, relying on data points within an experimentally accessible temperature region, would be welcome.

While very high-temperatures are often physically not accessible, very low-temperatures are also not ideal.
Irrespectively of the possible difficulty to thermalise the sample, perturbations beyond 
the spin-liquid Hamiltonian usually set a temperature scale $T^{*}$ below which the 
physics of the spin liquid is lost; the system may order or fall out-of-equilibrium.
The most appropriate window in experiments is thus at intermediate temperatures, 
precisely where the crossover between the two Curie laws takes place. 
And while low- and high-temperature expansions are the least accurate in this
regime, Section~\ref{sec:Thermodyn} has shown that HT calculations are quantitatively  
reliable over the entire temperature region for corner-sharing lattices.

Appendix \ref{sec:AppHT2} gives the analytical formula of the susceptibility for different geometries of the Husimi tree. We notice that the reduced susceptibility is always of the form
%
%%%%%%%%%%%%%%%%%%%%%%%%%%
\begin{equation}
\chi T|_{\sf HT} = \frac{\sum_{i} \alpha_i \ e^{\kappa_i / T}}{\sum_{i} \alpha'_i \ e^{\kappa'_i / T}}   \ .
 \label{eq:chiTHT}
\end{equation}
%%%%%%%%%%%%%%%%%%%%%%%%%%
%
This expression is sufficiently generic that it should be able to fit any form of  $\chi T$. But as it is written, Eq.~(\ref{eq:chiTHT}) is unpractical. Fortunately, it turns out that only a few terms are usually necessary. The simplest pertinent form of Eq.~(\ref{eq:chiTHT}) is
%
%%%%%%%%%%%%%%%%%%%%%%%%%%
\begin{equation}
	\chi T|_{\rm HA} = \frac{1 + b_1 \ {\sf exp}[c_1 / T] }{a + b_2 \ {\sf exp}[c_2 / T]}   \ .
 \label{eq:HT.exp.fit_1}
\end{equation} 
%%%%%%%%%%%%%%%%%%%%%%%%%%
%
We shall refer to Eq.~(\ref{eq:HT.exp.fit_1}) as the \textit{Husimi Ansatz}. In this form the Curie constant and Curie-Weiss temperature can be directly extracted from the fitting parameters:
%
%%%%%%%%%%%%%%%%%%%%%%%%%%
\begin{align}
	C_{\infty}^{{\rm HA}} &= \frac{1+ b_1}{a + b_2}\, ,	
	\label{eq:Exp.values.fit_1a}  \\
 	\theta_{\text{cw}}^{{\rm HA}} &= \frac{b_1 c_1}{1 + b_1}  - \frac{b_2 c_2}{a + b_2}	\, . 
	\label{eq:Exp.values.fit_1b}
\end{align}
%%%%%%%%%%%%%%%%%%%%%%%%%%
%
Eq.~(\ref{eq:HT.exp.fit_1}) will be our primary phenomenological Ansatz for the rest of this paper. Intuitively, we understand that the $c_{1}$ and $c_{2}$ parameters correspond to effective energy scales in the Boltzmann factor. However, two energy scales might be too minimal to describe the physics of some models, especially if different types of couplings are involved. This is why we will also consider an extended Ansatz to fit $\chi T$
%
%%%%%%%%%%%%%%%%%%%%%%%%%%
\begin{equation}
	\chi T|_{\rm HA} = \frac{1 + b_1 \ {\sf exp}[c_1 / T] }{a + b_2 \ {\sf exp}[c_2 / T] + b_3 \ {\sf exp}[c_3 / T]}   \ ,
 \label{eq:HT.exp.fit_2}
\end{equation} 
%%%%%%%%%%%%%%%%%%%%%%%%%%
%
where the Curie constant and Curie-Weiss temperature become
%
%%%%%%%%%%%%%%%%%%%%%%%%%%
\begin{align}
	C_{\infty}^{{\rm HA}} &= \frac{1+ b_1}{a + b_2 + b_3}	\, ,	\label{eq:Exp.values.fit_2a}  \\
 	 \theta_{\text{cw}}^{{\rm HA}} &= \frac{b_1 c_1}{1 + b_1}  - \frac{b_2 c_2 + b_3 c_3}{a + b_2 + b_3}	\, . 
	\label{eq:Exp.values.fit_2b}
\end{align}
%%%%%%%%%%%%%%%%%%%%%%%%%%
%
%%%%%%%%%%%%%%%%%%%%%%%%%%%%%%%%%%%%%%%%%%%%%
\subsection{Benchmark of the Husimi Ansatz}  
\label{sec:benchmark.fit}
%%%%%%%%%%%%%%%%%%%%%%%%%%%%%%%%%%%%%%%%%%
%

The purpose of this section is to benchmark the Husimi Ansatz of Eq.~(\ref{eq:HT.exp.fit_1}) in a controlled way on various model Hamiltonians. In Fig.~\ref{fig:EmpiricalFit} we fit the Curie-law crossover with Eq.~(\ref{eq:HT.exp.fit_1}) for pyrochlore models with global-axis and local-axis Ising spins. In order to test the Ansatz on a general framework, beyond the Ising models used to build our Husimi-based intuition, we also consider continuous spins on the Heisenberg antiferromagnet (HAF) 
\cite{moessner98a, Canals2001, Henley2005}, 
and pseudo-Heisenberg antiferromagnet (pHAF) 
\cite{Ross2011, Onoda2011, Lee2012, Taillefumier2017}. 
The pHAF is defined on the XXZ model as follows:
%
%%%%%%%%%%%%%%%%%%%%%%%%%%%
\begin{equation}
	\mathpzc{H}_{\sf XXZ}  =  \sum_{\langle ij \rangle} \left[ J_{\sf zz} S^z_i S^z_j 
						- J_{\pm}  \left( S^{+}_i S^{-}_j  +  S^{-}_i S^{+}_j  \right) \right] \,  ,
\label{eq:HamXXZ}
\end{equation}
%%%%%%%%%%%%%%%%%%%%%%%%%%%
%
with $S_i^z$ along the local [111] easy-axis, as defined in 
Tab.~\ref{tab:PyrochloreCoord}, for parameters $J_{\pm}/J_{zz} = -0.5$ \cite{Taillefumier2017}. This model is thermodynamically equivalent to the HAF, but with different magnetic correlations, and thus a distinct evolution of the Curie-law crossover.

Fig.~\ref{fig:EmpiricalFit}(a) and (b) show vanishing $C_0 = 0$, induced by the 
zero-divergence constraint on the ground state manifold, imposing zero magnetisation in all tetrahedra (see discussion in Section \ref{sec:Thermodyn}).
Fig.~\ref{fig:EmpiricalFit}(c) and (d) show $C_0 = 2$, as a result of dominant 
ferromagnetic correlations. 
Entering the spin-liquid regime at low $T$ for (a) and (c) for models with Ising degrees of freedom 
shows a rather sharp kink below $T/|J| \lesssim 1$, while on the opposite, models with continuous degrees of freedom in (b) and (d) enter the low$-T$ regime rather smoothly.

Results were obtained from classical MC simulations (black circles) and have been fitted with the Husimi Ansatz (solid lines) from Eq.~(\ref{eq:HT.exp.fit_1}) over different temperature windows. 
Examples of three different fitting windows are shown for high-temperature (1$^{\sf st}$ row), 
and low-temperature (2$^{\sf nd}$ row) fits.
The range of fitting windows are indicated by blue, yellow and green bars on the bottom of each plot, 
and allow to judge their reliability in comparison to MC data. 
It becomes clear that fitting windows, which include only one Curie-law regime (either at low or high temperature), do not accurately reproduce the Curie-law crossover. This is especially important for Ising models, because of the relatively sharp kink when entering the spin-liquid regime.

On the other hand, fitting windows including the intermediate temperature window, with only the onset of high- and low-temperature regimes quantitatively reproduce $\chi T$ over the full temperature range. The 3$^{\sf rd}$ row of panels shows the ``minimal'' fitting window. By using Eqs.~(\ref{eq:Exp.values.fit_1a}, \ref{eq:Exp.values.fit_1b}) we can precisely extract the Curie constant  $C_{\infty}$ and Curie-Weiss temperature $\theta_{\text{cw}}$ from those fits. Fitted and exact solutions match perfectly within error bars (see Tab.~\ref{tab:empFit}).

%%%%%%%%%%%%%%%%%%%%%%%%%%%%%%%%%%%%%%%%%%
% Tab.   EXTRACTED VALUES FROM EMPIRICAL FITTING FUNCTION
%%%%%%%%%%%%%%%%%%%%%%%%%%%%%%%%%%%%%%%%%%
%
%%%%%%%%%%%%%%%%%%%%%%%%%%
\newcolumntype{C}{>{}c<{}} 
\renewcommand{\arraystretch}{3}
\begin{table}[t]
	\def\arraystretch{1.7}
	\centering
		\begin{tabular}{||C||C|C||C|C||}
			\hhline{|t:=====:t|}		
			model 				&\multicolumn{2}{C||} { $C_{\infty}$ }	& \multicolumn{2}{C||}{ $\theta_{\text{cw}}$ }	\\
			\hhline{||-----||}
								& fit 			& exact		& fit 				& exact		 \\
			\hhline{||=====||}
			Ising global 			& $ 1.00(1) $ 	& $ 1.0 $ 		& $ -6.0(1) $		& $-6$ 		\\
			\hhline{||-----||}
			HAF 					& $ 1.00(2) $ 	& $ 1.0 $  		& $ -2.03(5) $ 		& $-2$ 		\\
			\hhline{||-----||}	
			Ising local 			& $ 1.00(2) $ 	& $1.0 $ 		& $ 2.0(1)$		& $2 $	 	\\
			\hhline{||-----||}
			pHAF				& $ 1.00(1) $ 	& $ 1.0 $ 		& $ 0.65(2) $		& 2/3 	\\
			\hhline{|b:=====:b|}
		\end{tabular}
	\caption{ 
	Curie constant $C_{\infty}$ [Eq.~(\ref{eq:Exp.values.fit_1a})] and Curie temperature $\theta_{\text{cw}}$ [Eq.~(\ref{eq:Exp.values.fit_1b})],
	obtained for the fit of the reduced susceptibility $\chi T$ with minimally sufficient  fitting window, as shown 
	in the 3$^{\sf rd}$ row of Fig.~\ref{fig:EmpiricalFit}.
	}
\label{tab:empFit}
\end{table}
%%%%%%%%%%%%%%%%%%%%%%%%
%
This benchmark shows that {the Husimi} Ansatz correctly reproduces the Curie-law crossover over the full range of temperatures for several distinct models with Ising and continuous spins. It requires a fitting window spanning typically 1 or 2 orders of magnitude in temperature, in the intermediate regime that is usually accessible to experiments [see the bottom row of Fig.~\ref{fig:EmpiricalFit}]. This is a useful theoretical proof of concept, that now needs to be applied to experiments.

%%%%%%%%%%%%%%%%%%%%%%%%%%%%%%%%%%%%%%%%%%%%%
\section{The Husimi Ansatz in experiments}
\label{sec:experiments} 
%%%%%%%%%%%%%%%%%%%%%%%%%%%%%%%%%%%%%%%%%%

Our goal in this section is to show by examples the advantages and limitations of the Husimi Ansatz in real materials, and to encourage its use jointly with the Curie-Weiss fit.

%%%%%%%%%%%%%%%%%%%%%%%%%%%%%%%%%%%%%%%%%%%%%
\subsection{NaCaNi$_2$F$_7$}  
\label{sec:NaCa} 
%%%%%%%%%%%%%%%%%%%%%%%%%%%%%%%%%%%%%%%%%%
%

First, let us consider a material where the Ansatz gives similar results to the Curie-Weiss fit. 
To do so, let us consider one of the closest materials to the canonical HAF.

NaCaNi$_2$F$_7$ is a spin$-1$ pyrochlore material, well described by a weakly perturbed 
nearest-neighbour Heisenberg Hamiltonian \cite{plumb19a,zhang19a}.
It shows spin freezing at $T^* \approx 3.6~{\rm K}$, %probably due to 
which has been assumed to originate from Na$^{1+}$/Ca$^{2+}$ charge disorder, %but
however, no long-range magnetic order has been observed \cite{Krizan2015}.

\begin{figure}
\centering\includegraphics[width=.99\columnwidth]{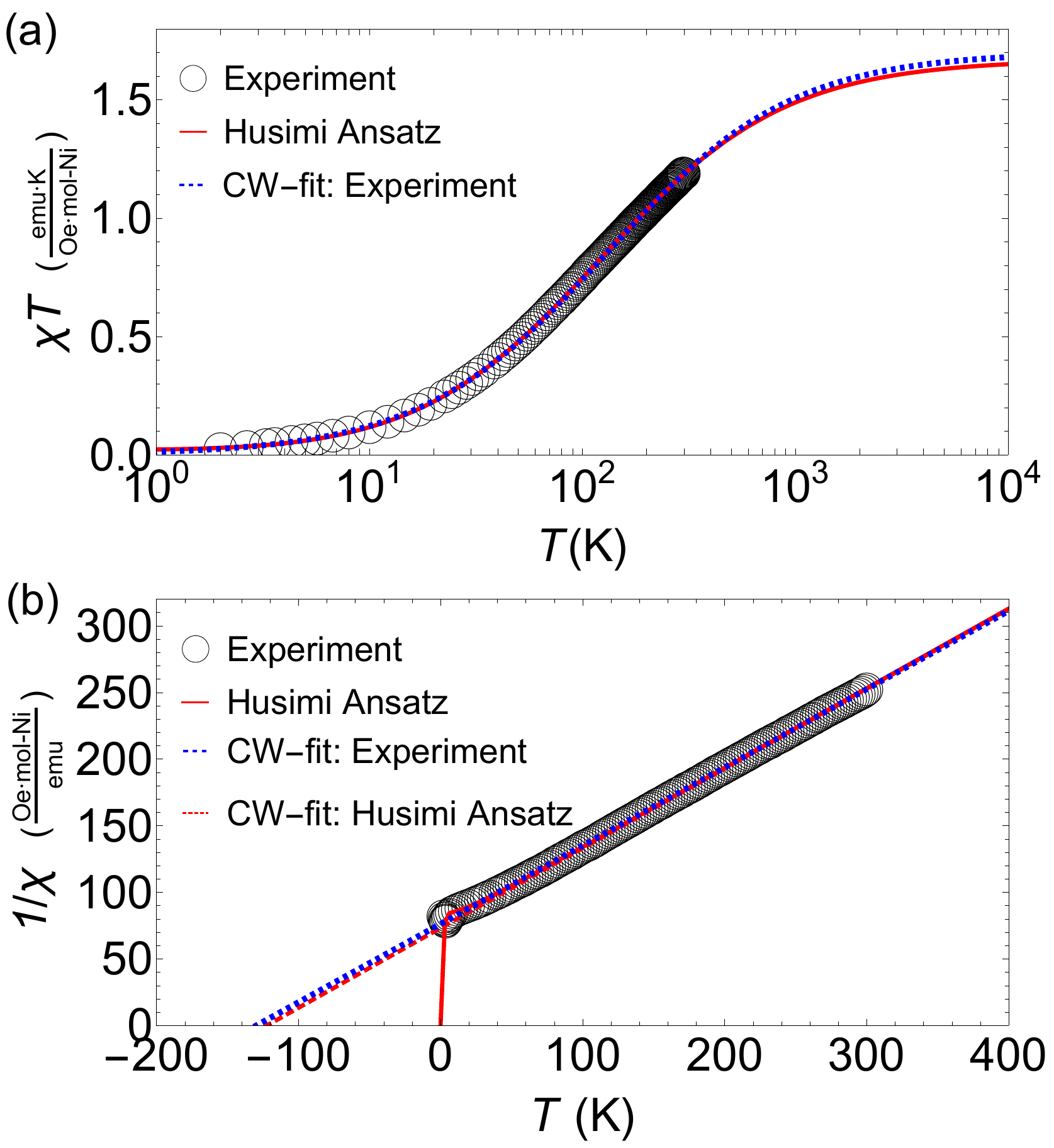}
	\caption{
	Fit of the experimental susceptibility for the pyrochlore material 
	NaCaNi$_2$F$_7$ (black circles) plotted on (a) a semi-logarithmic plot for $\chi T$ 
	and (b) a linear scale for $1/ \chi$. 
	Fitting the Husimi Ansatz [Eq.~\eqref{eq:HT.exp.fit_1}] within a temperature 
	window  $\Delta T = [3.6~{\rm K}, \cdots , 300~{\rm K}]$  (red solid line),
	gives an estimate of $\theta_{\text{cw}}=-122(1) \ {\rm K}$.
	Our result is in good agreement with a standard Curie-Weiss fit (blue dashed line).
	Experimental data were extracted from Ref.~[\onlinecite{Krizan2015}].}
\label{fig:NaCaNi2F2}
\end{figure}

In  Fig.~\ref{fig:NaCaNi2F2}, we plot the magnetic susceptibility of NaCaNi$_2$F$_7$, extracted from Ref.~[\onlinecite{Krizan2015}], on a semi-logarithmic scale for $\chi T$ and on a linear scale for $1/ \chi$.
The data points are well fitted by the Husimi Ansatz of Eq.~(\ref{eq:HT.exp.fit_1}) 
over the whole range of accessible temperatures. 
%

%%%%%%%%%%%%%%%%%%%%%%%%%%%%%%%%%%%%%%%%%%
% Fig.     Chi T - NaCaNi$_2$F$_7$
%%%%%%%%%%%%%%%%%%%%%%%%%%%%%%%%%%%%%%%%%%
%%%%%%%%%%%%%%%%%%%%%%%%%%%%%%%%%%%%%%%%%%
%

We fit the Husimi Ansatz within physically relevant temperatures 
$\Delta T = [3.6~{\rm K}, \cdots , 300~{\rm K}]$, above the 
freezing transition up to the maximally available datapoints, and 
obtain a Curie-Weiss temperature $\theta^{\rm HA}_{\text{cw}}=-122(1)~{\rm K}$, and a
Curie constant $C^{\rm HA}_{\infty} = 1.67(1)$ (emu K)/(Oe mol-Ni), which gives an 
effective magnetic moment of $\mu_{\rm eff}^{\rm HA} = 3.65(1) \mu_{\rm B}/{\rm Ni}$.
All these quantities are in good agreement with a standard Curie-Weiss fit over a temperature 
window $\Delta T = [150~{\rm K}, \cdots , 300~{\rm K}]$, which reveals 
$\theta_{\text{cw}}=-129(1)~{\rm K}$ with $\mu_{\rm eff} = 3.6(1)~\mu_{\rm B}/{\rm Ni}$.
%
%It  
This strongly suggest, as also qualitatively visible from the straight behavior of  $1/\chi$ in Fig.~\ref{fig:NaCaNi2F2}(b), that experimentally measured data points reach the high-temperature regime where a standard Curie Weiss fit becomes a reliable estimate.

%%%%%%%%%%%%%%%%%%%%%%%%%%%%%%%%%%%%%%%%%%%%%
\subsection{KCu$_6$AlBiO$_4$(SO$_4$)$_5$Cl}  
\label{sec:KCu} 
%%%%%%%%%%%%%%%%%%%%%%%%%%%%%%%%%%%%%%%%%%
%

%%%%%%%%%%%%%%%%%%%%%%%%%%%%%%%%%%%%%%%%%%
% Fig.     KCu$_6$AlBiO$_4$(SO$_4$)$_5$Cl
%%%%%%%%%%%%%%%%%%%%%%%%%%%%%%%%%%%%%%%%%%
%
\begin{figure}
\centering\includegraphics[width=0.99\columnwidth]{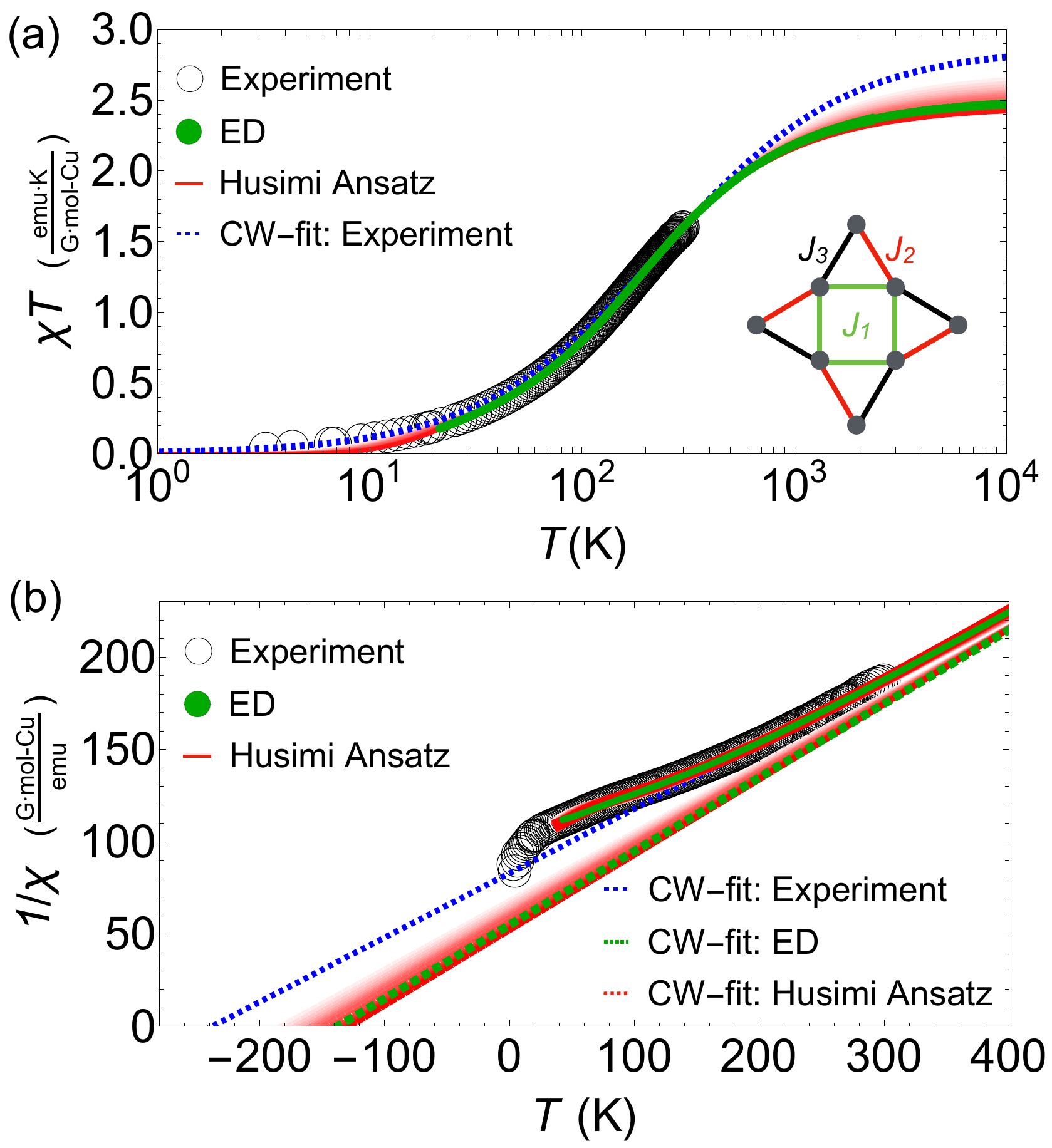}
	\caption{
	Fit of the experimental susceptibility for the $S=1/2$ square-kagome quantum 
	spin liquid candidate KCu$_6$AlBiO$_4$(SO$_4$)$_5$Cl (black circles),
	plotted on (a) a semi-logarithmic plot for $\chi T$ and (b) a linear scale for $1/ \chi$. 
	Experimental data are plotted together with exact diagonalization (ED) data (18 sites, green dots) 
	of an effective $J_1$-$J_2$-$J_3$ Heisenberg model (see inset in (a)), as proposed 
	in [\onlinecite{fujihala20a}].
	We fit experimental data with the Husimi Ansatz [Eq.~\eqref{eq:HT.exp.fit_1}] for  
	$T_{\rm min} \leq T \leq T_{\rm max}$, where we set $T_{\rm max}=300~{\rm K}$ to the 
	highest available temperature from experiment, and vary $T_{\rm min}$ between 
	$10$ and $30~{\rm K}$ (shaded respectively from light red to red).
	Experimental data and ED results were kindly provided by M.~Fujihala [\onlinecite{fujihala20a}].
	}
\label{fig:KCu6AlBiO4(SO4)5Cl}
\end{figure}
%%%%%%%%%%%%%%%%%%%%%%%%%%%%%%%%%%%%%%%%%%
%

%
KCu$_6$AlBiO$_4$(SO$_4$)$_5$Cl is a promising $S=1/2$ quantum spin liquid candidate on 
the distorted square-kagome lattice, as reported by M.~Fujihala {\it et al.} in Ref.\ [\onlinecite{fujihala20a}]. 
Measurement of specific heat and susceptibility did not find any signatures of long-range order down to $1.8~{\rm K}$, while $\mu$SR confirmed the absence of spin order and spin freezing down to $58~{\rm mK}$.

In Fig.~\ref{fig:KCu6AlBiO4(SO4)5Cl}, we plot the magnetic susceptibility of 
KCu$_6$AlBiO$_4$(SO$_4$)$_5$Cl, as kindly provided by M.~Fujihala [\onlinecite{fujihala20a}],
on a semi-logarithmic scale for $\chi T$ and on a linear scale for $1/ \chi$.
In comparison to NaCaNi$_2$F$_7$ in Fig.~\ref{fig:NaCaNi2F2}(b), it becomes
evident that $1 / \chi$ for KCu$_6$AlBiO$_4$(SO$_4$)$_5$Cl shows a rather 
strong deviation from a linear behavior over nearly the whole range of experimentally 
accessible temperatures. 
A Curie-Weiss fit for the high-temperature tail within $\Delta T = [200~{\rm K}, \cdots , 300~{\rm K}]$ 
gives $\theta_{\text{cw}} = -237(2) {\rm K}$ and  $\mu_{\rm eff} = 1.96~\mu_{\rm B}/{\rm Cu}$
with a Land\'e factor $g=2.25$ \cite{fujihala20a}.
The Husimi Ansatz from Eq.~\eqref{eq:HT.exp.fit_1} gives a noticeably different outcome though. 
We find $\theta^{\rm HA}_{\text{cw}} =-154\pm 28~{\rm K}$. The large error bar comes from the choice of the fitting temperature window [$T_{\rm min}, T_{\rm max}$] (see the spread of the red curve in Fig.~\ref{fig:KCu6AlBiO4(SO4)5Cl}), where we fix $T_{\rm max}=300~{\rm K}$ at the highest available temperature, and vary $T_{\rm min}$ between $10$ and $30~{\rm K}$. The non-linearity of $1/\chi$ and spread of the Husimi estimate suggest that $300~{\rm K}$ is too far from the high-temperature limit for a conclusive estimate of $\theta_{\rm cw}$. The noticeable difference between the outcomes of the Curie-Weiss fit and Husimi Ansatz, however, makes us wonder which of the two estimates is more reliable.

From a microscopic analysis in [\onlinecite{fujihala20a}] we understand that 
KCu$_6$AlBiO$_4$(SO$_4$)$_5$Cl is not an ideal square-kagome lattice; the three bonds of a triangle 
in Fig.~\ref{fig:MC_Lattices}(c) are inequivalent.
All triangles are distorted in the same way and form
three distinct ``nearest-neighbour'' couplings, $J_1,J_2,J_3$, on each triangle (see inset in Fig.~\ref{fig:KCu6AlBiO4(SO4)5Cl}.(a)). 
M.~Fujihala {\it et al.} \cite{fujihala20a} built a microscopic Hamiltonian which describes its magnetic susceptibility at high temperature, using exact diagonalization (ED) and finite-temperature Lanczos methods, as shown on Fig.~\ref{fig:KCu6AlBiO4(SO4)5Cl}. ED results fit the experimental data very well down to $T \approx 40~{\rm K}$, below which finite-size effects make further estimates difficult. M. Fujihala {\it et al.} obtained 
%
%%%%%%%%%%%%%%%%%%%%%%%%%%%%%%%%%%%%%%%%%%%%%
\begin{eqnarray}
	J_1 = -135~{\rm K}, \ J_{2} = -162 ~{\rm K}, \ J_{3}= -115~{\rm K} 	\, ,
\label{eq:J1J2J3.Shuriken}
\end{eqnarray}
%%%%%%%%%%%%%%%%%%%%%%%%%%%%%%%%%%%%%%%%%%%%%
%
with a Land\'e factor $g=2.11$. This high-temperature analysis cannot rule out low-energy perturbations, but it establishes the model of M. Fujihala \textit{et al} as a solid parametrisation of KCu$_6$AlBiO$_4$(SO$_4$)$_5$Cl in the temperature regime relevant to $\theta_{\rm cw}$, which is straightforward to estimate from Eq.~\eqref{eq:J1J2J3.Shuriken}
%
%%%%%%%%%%%%%%%%%%%%%%%%%%%%%%%%%%%%%%%%%%%%%
\begin{eqnarray}
	\theta_{\text{cw}}^{ED} &=& \frac{S(S+1)}{3}\,4\left(\frac{J_{1}+J_{2}+J_{3}}{3}\right) \\
	&=& \left(\frac{J_{1}+J_{2}+J_{3}}{3}\right) \approx -137~{\rm K} \, .
\label{eq:CWKCu}
\end{eqnarray}
%%%%%%%%%%%%%%%%%%%%%%%%%%%%%%%%%%%%%%%%%%%%%
%
Eq.~(\ref{eq:CWKCu}) leads to a couple of remarks. Firstly, the ED results are in better agreement with the Husimi Ansatz than the Curie-Weiss fit, which a posteriori validates the former. Secondly, $\theta_{\rm cw}$ here simply corresponds to the average value of the three inequivalent exchange couplings. $J_{1}, J_{2}, J_{3}$ fit within the energy window set by $\theta_{\rm cw}\pm \delta J$, thus defining the anisotropic energy scale $\delta J=25$ K. Using $T_{\rm min}=\delta J$ as a lower bound of our fitting temperature window, we obtain from the Husimi Ansatz $\theta^{\rm HA}_{\text{cw}} = -136~{\rm K}$ with a Land\'e factor  $g = 2.1$, which is essentially the same result as from ED\footnote{It is worth noting that a Land\'e factor of 2.1 is a more realistic value for Cu(II) ions than the value of 2.25 found with the Curie-Weiss fit.}. This suggests that the main difficulty to estimate $\theta_{\rm cw}$ comes from the lattice anisotropy of KCu$_6$AlBiO$_4$(SO$_4$)$_5$Cl. And while the Curie-Weiss law is not adapted to account for multiple energy scales in this intermediate regime, the Husimi Ansatz has been designed to be a flexible fitting function for the crossover that happens in this intermediate regime. We believe it is the reason why the Husimi Ansatz, albeit its large error bar, gives a better result than the Curie-Weiss fit.

%%%%%%%%%%%%%%%%%%%%%%%%%%%%%%%%%%%%%%%%%%%%%
\subsection{FeCl$_3$}
\label{sec:FeCl3} 
%%%%%%%%%%%%%%%%%%%%%%%%%%%%%%%%%%%%%%%%%%

\subsubsection{Experiments}

As seen from the two previous materials with negative Curie-Weiss temperatures, spin liquids usually show dominant antiferromagnetic couplings. However, there also exist frustrated magnets where the interplay between ferro- and antiferro-magnetism can lead to multiple Curie-law crossovers \cite{Pohle2016,Schmidt17a}.
An important example relevant to materials are spiral spin liquids.
They form a class of classical spin liquids where spiral states compete and form 
a sub-extensive ground state manifold with characteristic ring features in 
momentum space \cite{Niggemann2020, Pohle2021, Yao2021, Huang2022, yan22a}. 

The Van der Waals magnet FeCl$_3$ is a prototype of a spiral spin liquid.
At first, investigated as a member of crystallized anhydrous ferric chlorides \cite{Lallemand1935}, 
the history of FeCl$_3$ dates far back into the 1930's, where it already attracted attention 
due to its unusual magnetic properties at low temperature.
Susceptibility measurements reported a Curie-Weiss temperature of 
$\theta_{\text{cw}} \approx -12~{\rm K}$,
however, noticing already at that time a deviation from the conventional 
Curie-Weiss law \cite{Starr1940}.
Furthermore, inelastic neutron-scattering (INS) measurements \cite{Cable1962}, magnetic susceptibility 
\cite{Jones1969}, M\"oessbauer effect \cite{Stampfel1973}, 
magnetic field \cite{Johnson1981}, and NMR measurements \cite{Kang2014}
confirmed a phase transition into an unusual spiral ground state at about $T_N \approx 10~{\rm K}$. 
But it was only recently, with the work of S.~Gao {\it et al.} [\onlinecite{Gao2022}], 
that continuous ring features around the $\Gamma$-point could be observed in 
INS experiments; a clear evidence of spiral spin liquid physics in FeCl$_3$.

\begin{figure}
\centering\includegraphics[width=0.90\columnwidth]{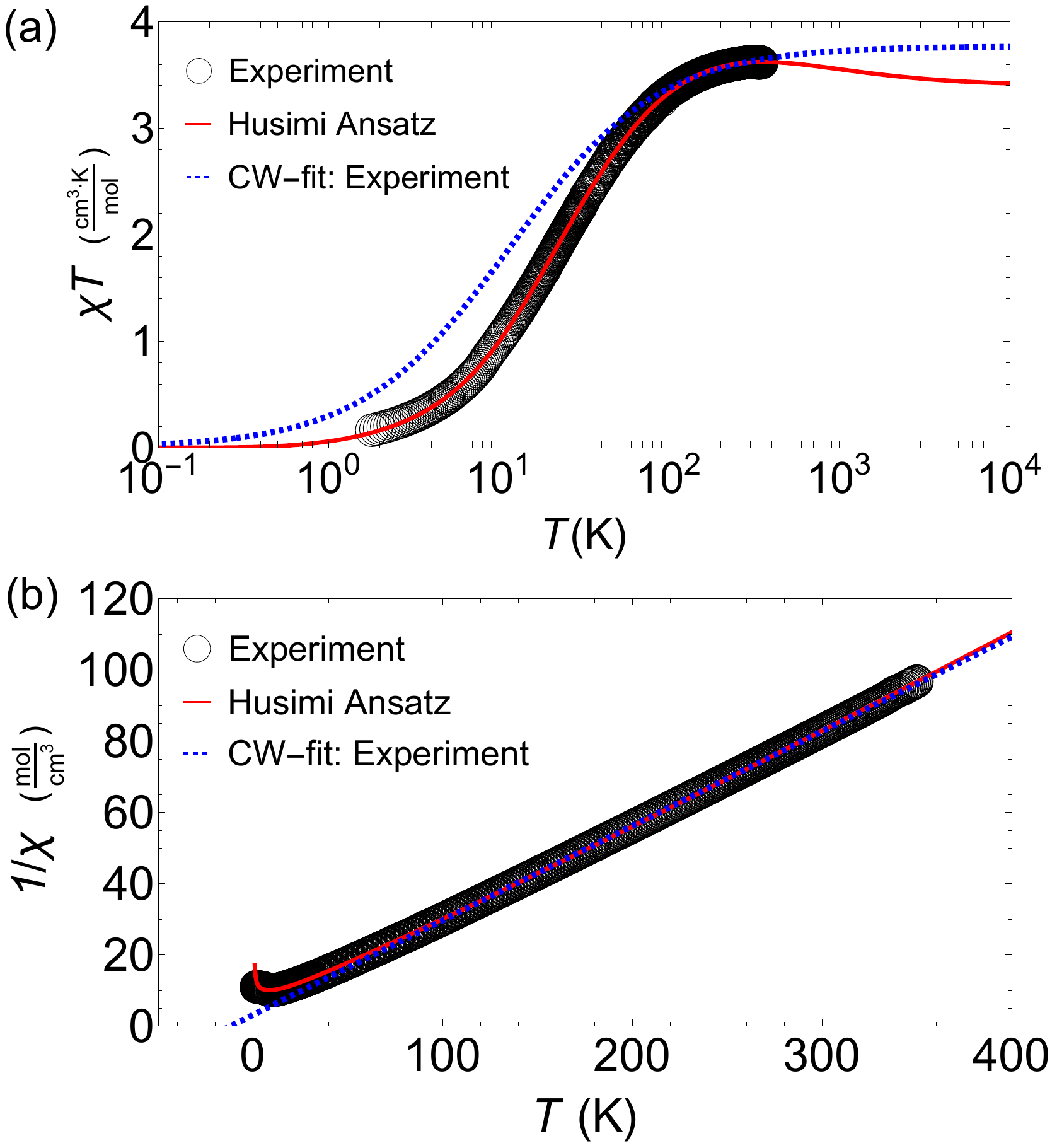}
	\caption{
	Fit of the experimental susceptibility for the $S = 5/2$
	magnet FeCl$_3$ (black circles) plotted on (a) a semi-logarithmic plot for $\chi T$ 
	and (b) a linear scale for $1/\chi$. 
	Fitting the extended Husimi Ansatz [Eq.~\eqref{eq:HT.exp.fit_2}] for all available data points (red solid line) reveals a multistep (non-monotonic) Curie-law crossover with a slight downturn at high temperatures, and hence a positive $\theta^{\rm HA}$. Unfortunately, the absence of data points above 350 K makes it impossible to extract a reliable estimate of $\theta_{\rm cw}$.
	A standard Curie-Weiss fit (blue dashed line) gives a very different result 
	of  $\theta_{\text{cw}} = -11(1)~{\rm K}$, while showing a strong deviation 
	from experimental data points below $50~{\rm K}$.
	Experimental data were kindly provided by {M. McGuire} [\onlinecite{Gao2022}].
	}
\label{fig:FeCl3}
\end{figure}

In Fig.~\ref{fig:FeCl3} we show the magnetic susceptibility of 
FeCl$_3$, as kindly provided by M. McGuire [\onlinecite{Gao2022}],
on a semi-logarithmic scale for $\chi T$ and on a linear scale for $1/ \chi$.
In contrast to the materials above (see Fig.~\ref{fig:NaCaNi2F2} 
and Fig.~\ref{fig:KCu6AlBiO4(SO4)5Cl}), it seems that $\chi T$ 
reaches the plateau of the high-temperature Curie-Weiss regime already at about $300~{\rm K}$.
For the traditional $1/\chi$ vs $T$ plot [Fig.~\ref{fig:FeCl3}.(b)], the Curie-Weiss law shows a good fit over the temperature window $\Delta T = [100~{\rm K}, \cdots , 350~{\rm K}]$, which gives $\theta_{\text{cw}} = -11(1)~{\rm K}$, in agreement with previous measurements  \cite{Starr1940}. However, when plotting the reduced susceptibility $\chi T$ instead [Fig.~\ref{fig:FeCl3}.(a)], the Curie-Weiss law is seen to noticeably deviate from experimental data below 50 K.
In fact, after careful consideration, experimental data show a broad maximum at about $T \approx 300~{\rm K}$, suggesting that the reduced susceptibility $\chi T$ is not monotonic. This motivates us to fit the available experimental data with the extended Husimi Ansatz of Eq.~\eqref{eq:HT.exp.fit_2} which allows for non-monotonic behavior. 
It fits the experimental data quantitatively well over the whole temperature range and indeed presents a slight downturn at high temperatures above $T > 500~{\rm K}$. Unfortunately, FeCl$_3$ is structurally unstable at higher temperatures and there are not enough data points after the downturn of $\chi T$ to extract a reliable estimate of $\theta_{\rm cw}$. And since susceptibility measurements are naturally more noisy at high temperature, one should remain cautious. That being said, the Husimi Ansatz suggests a positive Curie-Weiss temperature in FeCl$_3$ -- as opposed to previous measurements \cite{Lallemand1935, Starr1940, Jones1969} -- and thus a multi-step Curie-law crossover with dominant ferromagnetic interactions, that would justify the anomalous behavior of the susceptibility that has been noticed since 1940 \cite{Starr1940}.

%%%%%%%%%%%%%%%%%%%%%%%%%%%%%%%%%%%%%%%%%%
% Fig.     FeCl_3 in experiment
%%%%%%%%%%%%%%%%%%%%%%%%%%%%%%%%%%%%%%%%%%
%
%%%%%%%%%%%%%%%%%%%%%%%%%%%%%%%%%%%%%%%%%%
%
%
%%%%%%%%%%%%%%%%%%%%%%%%%%%%%%%%%%%%%%%%%%
% Fig.     FeCl_3 modeled with J1-J2-Jc1 Heisenberg model 
%%%%%%%%%%%%%%%%%%%%%%%%%%%%%%%%%%%%%%%%%%
%
\begin{figure}
\centering\includegraphics[width=0.90\columnwidth]{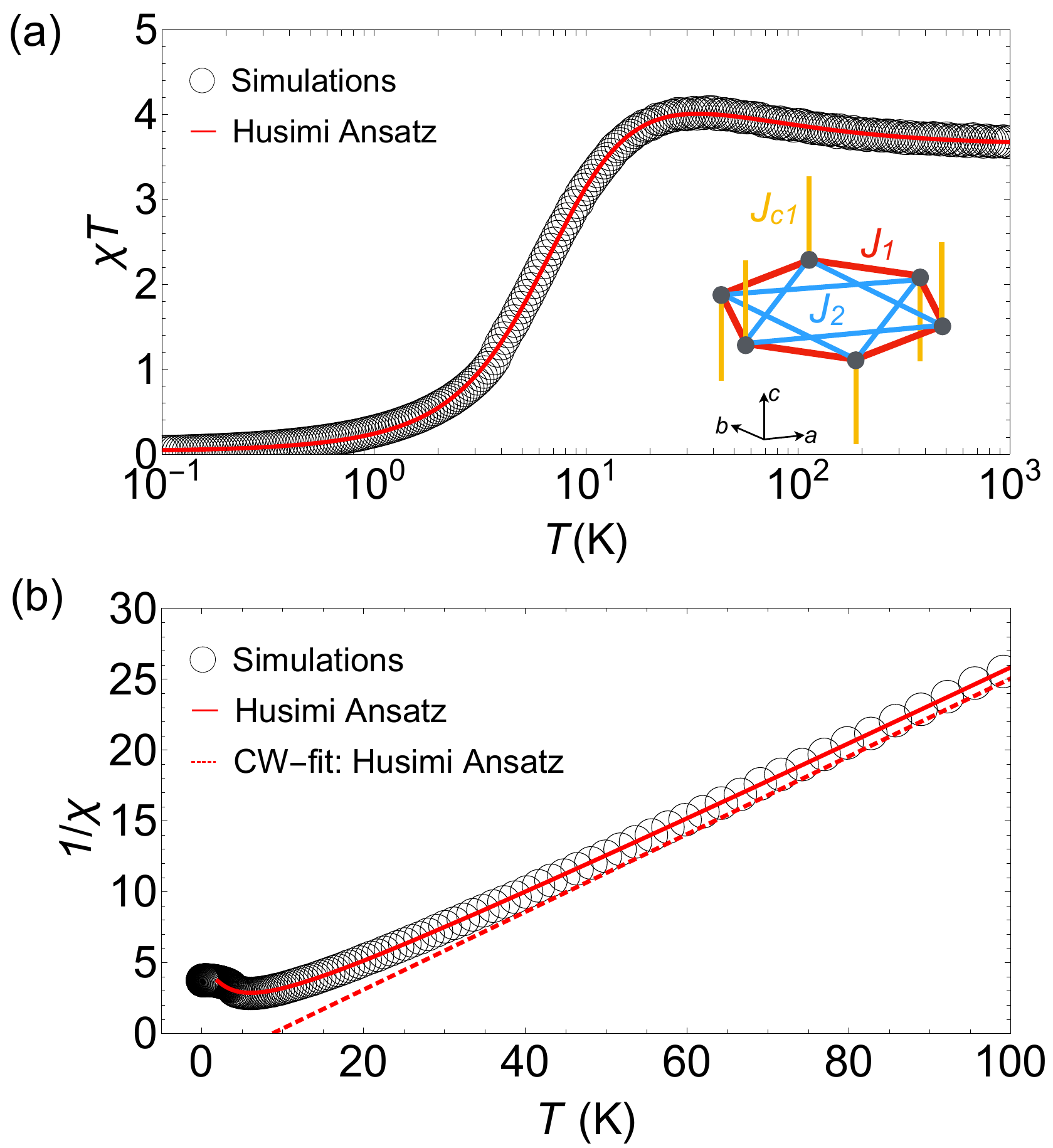}
	\caption{
	Fit of the numerical susceptibility (black circles) plotted on (a) a semi-logarithmic 
	plot for $\chi T$ and (b) a linear scale for $1/\chi$. 	
	Numerical data were obtained from classical Monte Carlo simulations for the 
	$J_1$-$J_2$-$J_{c1}$ Heisenberg model [Eq.~(\ref{eq:Ham.FeCl3})] on the $ABC$ stacked honeycomb 
	lattice (see inset in (a)) for model parameters as proposed for FeCl$_3$ \cite{Gao2022} [Eq.~(\ref{eq:J1J2Jc1.FeCl3})].
	Fitting the extended Husimi Ansatz [Eq.~(\ref{eq:HT.exp.fit_2})] for all available 
	data points (red solid line) clearly reveals a multistep (non-monotonic) Curie-law 
	crossover with a significant downturn at high temperatures, and hence a positive 
	$\theta^{\rm HA}_{\text{cw}} = 8.7(2)~{\rm K}$.
	}
\label{fig:J1J2Jc1}
\end{figure}
%%%%%%%%%%%%%%%%%%%%%%%%%%%%%%%%%%%%%%%%%%
%
\subsubsection{Simulations}
In absence of data points for FeCl$_3$ at high temperature, it is difficult to push the experimental analysis any further. Therefore, to conclude this discussion on the multi-step Curie-law crossover, we shall turn to classical Monte Carlo (MC) simulations.
Magnetic Fe$^{3+}$ ($S=5/2$) ions cover honeycomb layers, which are stacked in 
an $ABC$ arrangement along the $c$ axis.
By comparing LSW theory and SCGA results to INS data, S.~Gao {\it et al.} [\onlinecite{Gao2022}]
proposed a series of models with up to nine coupling parameters. For the sake of simplicity, we will focus on their minimal model, which is able to reproduce the ring features of a spiral spin liquid; the $J_1$-$J_2$-$J_{c1}$ Heisenberg model (see inset in Fig.~\ref{fig:J1J2Jc1}.(a)).
%
%%%%%%%%%%%%%%%%%%%%%%%%%%%
\begin{equation}
	\mathpzc{H}_{\sf J_1J_2J_{c1}}  = J_{1} \sum_{\langle ij \rangle, {\sf nn}}^{\sf intra} {\bf S}_i {\bf S}_j + 
							     J_{2} \sum_{\langle ij \rangle, {\sf nnn}}^{\sf intra}  {\bf S}_i {\bf S}_j +
							     J_{c1} \sum_{\langle ij \rangle, {\sf nn}}^{\sf inter}  {\bf S}_i {\bf S}_j,
\label{eq:Ham.FeCl3}
\end{equation}
%%%%%%%%%%%%%%%%%%%%%%%%%%%
%
where 
%
%%%%%%%%%%%%%%%%%%%%%%%%%%%%%%%%%%%%%%%%%%%%%
\begin{eqnarray}
	J_1 = -0.3~{\rm meV}, \ J_{2} = 0.075 ~{\rm meV}, \ J_{c1}= 0.15~{\rm meV} 	\, .
\label{eq:J1J2Jc1.FeCl3}
\end{eqnarray}
%%%%%%%%%%%%%%%%%%%%%%%%%%%%%%%%%%%%%%%%%%%%%
%
The couplings $J_1$ and $J_2$ respectively account for 
nearest-neighbor and next nearest-neighbor interactions within individual honeycomb layers, while $J_{c1}$ is the nearest-neighbor antiferromagnetic interlayer coupling.

In Fig.~\ref{fig:J1J2Jc1} we show the susceptibility, measured from MC simulations 
of $\mathpzc{H}_{\sf J_1J_2J_{c1}}$ [Eq.~\eqref{eq:Ham.FeCl3}].
Now the multistep Curie-law crossover becomes evident, even from simulation data, and the extended Husimi Ansatz 
from Eq.~\eqref{eq:HT.exp.fit_2} gives $\theta^{\rm HA}_{\text{cw}} = +8.7(2)~{\rm K}$, 
with a Curie constant $C^{\rm HA}_{\infty} = 3.6(1)$. It means that, taken in its extended form, the Husimi Ansatz can also account for a non-monotonic evolution of $\chi T$ due to competing ferro- and antiferromagentic couplings.
%

%%%%%%%%%%%%%%%%%%%%%%%%%%%%%%%%%%%%%%%%%%%%%
\subsection{Summary about experimental comparison}
\label{sec:sumexp} 
%%%%%%%%%%%%%%%%%%%%%%%%%%%%%%%%%%%%%%%%%%

In this section, we analyzed magnetic properties for three spin liquid candidates, namely, the $S=1$ pyrochlore 
fluoride NaCaNi$_2$F$_7$, the $S=1/2$ square-kagome material KCu$_6$AlBiO$_4$(SO$_4$)$_5$Cl,
and the $S=5/2$ spiral spin liquid on the honeycomb lattice FeCl$_3$.
All three materials showed a Curie-law crossover over a wide temperature range, from $\sim 1~{\rm K}$
up to $\sim 10^4~{\rm K}$.
Considering those examples, it becomes clear that a conventional Curie-Weiss fit applied to spin liquids 
can be reliable, but does not always have to.
Depending on the microscopic model parameters and the nature of the underlying spin liquid, the high-temperature 
Curie-Weiss regime might not be practically accessible. 
We showed, that the comparison between the conventional Curie-Weiss fit and the Husimi 
Ansatz, as introduced in Sec.~\ref{sec:ansatzformula}, allows us to quantify whether the high-temperature 
regime of a material is reached or not. 

NaCaNi$_2$F$_7$ is an example where experiments could reach to the high-temperature regime, and results from Husimi Ansatz and Curie-Weiss fit gave nearly the same results. 
On the other hand, KCu$_6$AlBiO$_4$(SO$_4$)$_5$Cl, shows a rather nonlinear behavior of $1/\chi$ in Fig.~\ref{fig:KCu6AlBiO4(SO4)5Cl}(b) for the available temperatures in experiment, which results in a mismatch between standard the Curie-Weiss fit and the Husimi Ansatz. The latter, however, agrees with independent ED results.
Last but not least, FeCl$_3$ is probbaly an example of a multi-step Curie-law crossover. Such non-monotonic behavior of magnetic correlations cannot be described by a conventional Curie-Weiss law, and therefore requires extra caution. By comparison to a minimal Heisenberg model we showed that an extended Husimi Ansatz [Eq.~\eqref{eq:HT.exp.fit_2}] is able to capture such a non-monotonic Curie-law crossover, predicting a Curie-Weiss temperature which is noticeably different compared to the one obtained from a standard Curie-Weiss fit.\\

In addition, some materials support the physics of a spin liquid at low but finite temperature, before ordering (or spin-freezing) at ultra-low temperatures $T^*$. The mechanism is simple. Let us consider a pristine spin-liquid model with a Curie-Weiss temperature $\theta_{\rm cw}$. By definition, this system doesn't order. Now, let us add a relevant, but small, perturbation of energy scale $\Delta$ inducing a transition at $T^*\sim \Delta \ll \theta_{\rm cw}$. One expects the spin-liquid physics to persist for a certain temperature window above $T^*$. But to determine the extent of this window is not easy without a microscopic probe such as neutron scattering, NMR or $\mu$SR. Fortunately, by plotting the reduced susceptibility $\chi T$ on a logarithmic temperature scale, we immediately measure the build-up of magnetic correlations upon cooling [Eq.~(\ref{eq:C0})] and can thus estimate how close we are from the spin-liquid regime. This is the case of NaCaNi$_2$F$_7$ which has a spin-freezing transition at $T^* \approx 3.6~{\rm K}$. On Fig.~\ref{fig:NaCaNi2F2}.(a), we see that $\chi T$ approaches the low-temperature spin-liquid Curie law (with $C_0=0$ here) at $T_p\sim 10-20$K. It means that NaCaNi$_2$F$_7$ supports the spin-liquid physics of the pyrochlore antiferromagnet over the temperature window $T^*\lesssim T \lesssim T_p$. It also means that the Husimi Ansatz can be used successfully even if the system orders at ultra-low temperature.

%%%%%%%%%%%%%%%%%%%%%%%%%%%%%%%%%%%%%%%%%%
%
% 						CONCLUSIONS
%
%%%%%%%%%%%%%%%%%%%%%%%%%%%%%%%%%%%%%%%%%%
\section{Conclusions}    	
\label{sec:conclusions}
%%%%%%%%%%%%%%%%%%%%%%%%%%%%%%%%%%%%%%%%%%

The Curie-Weiss temperature $\theta_{\rm cw}$ is a useful quantity to estimate the strength of frustration in frustrated magnets [Eq.~(\ref{eq:frustindex})].
However, the Curie-Weiss law is an estimate of the magnetic susceptibility close to a mean-field critical point, which -- by definition -- is absent in frustrated magnets. 
In this Article, we have shown that the concept of a Curie-law crossover \cite{Jaubert2013} is a generic feature for spin liquids and a more accurate description of their thermodynamic properties, that can be used to partially classify them.
We systematically study the Curie-law crossover among a variety of frustrated Ising models in two and three dimensions [Fig.~\ref{fig:MC_Lattices}], and motivate its relevance to thermodynamic signatures, as seen in experiments.
Comparing unbiased Monte Carlo simulations with the analytical Husimi-Tree approximation shows that the Curie-law crossover is determined by the type of frustrated unit cell (triangle, tetrahedron, ...) and  the connectivity between them, rather than the physical dimension of the lattice. As a side note, the Husimi-Tree approximation proves to be quantitatively accurate for all temperatures and for many spin-liquid models [Fig.~\ref{fig:MC_HT_Thermodyn}].

As a consequence of the Curie-law crossover, we recommend using the reduced susceptibility $\chi T$, complementary to the usual $1/\chi$ plot, when studying a potential spin liquid. It is often difficult to estimate whether $1/\chi$ has reached the asymptotic linear behavior, while $\chi T$ quickly indicates how far we are from the high-temperature Curie law.

Based on the success of the Husimi-Tree approximation, we propose an empirical Ansatz [Eq.~(\ref{eq:HT.exp.fit_0})] as a useful complement to the Curie-Weiss law. The Husimi Ansatz is easy to use and designed to be a flexible fitting function for the crossover in $\chi T$ that takes place in the temperature regime which is typically accessible to experiments. This means that the Husimi Ansatz can be used on a broader temperature window than the Curie-Weiss fit, which is necessarily limited to the region where $1/\chi$ is linear in $T$. In its extended form [Eq.~\eqref{eq:HT.exp.fit_2}], the Husimi Ansatz can also take into account the competition between ferro- and antiferromagnetic couplings in multi-step Curie-law crossovers.

It should be noted that the approach developed here works for frustrated magnets in general. Frustration doesn't need to be geometric in origin, it may come from further neighbor or anisotropic spin exchange, as present in Kitaev materials. And even if the simulations and calculations were based on classical spins in this paper, the Husimi Ansatz can be applied to quantum materials in the crossover regime, as done in Section \ref{sec:experiments}.

%%%%%%%%%%%%%%%%%%%%%%%%%%%%%%%%%%%%%
%
% 					ACKNOWLEDGEMENTS
%
%%%%%%%%%%%%%%%%%%%%%%%%%%%%%%%%%%%%%
\begin{acknowledgments}
%%%%%%%%%%%%%%%%%%%%%%%%%%%%%%%%%%%%%

The authors thank Harald Jeschke, Elsa Lhotel, Rodolphe Cl\'erac, Claire Lhuillier, Nic Shannon, Benjamin Canals
for fruitful discussions, and
Shang Gao, Jason Krizan and Yukitoshi Motome for critically reading 
the manuscript.
This work was supported by the Theory of Quantum Matter Unit, OIST and 
“Quantum Liquid Crystals” JSPS KAKENHI Grant No. JP19H05825.
L.D.C.J. acknowledges financial support from CNRS (PICS France-Japan MEFLS) and 
from the French "Agence Nationale de la Recherche" under Grant No. ANR-18-CE30-0011-01.
Numerical calculations were carried out using HPC Facilities provided by OIST,
and the Supercomputer Center of the Institute for Solid State Physics, 
the University of Tokyo.

%%%%%%%%%%%%%%%%%%%%%%%%%%%%%%%%%%%%%
\end{acknowledgments}
%%%%%%%%%%%%%%%%%%%%%%%%%%%%%%%%%%%%%

%%%%%%%%%%%%%%%%%%%%%%%%%%%%%%%%%%%%%%%%%%
%
% 						APPENDIX
%
%%%%%%%%%%%%%%%%%%%%%%%%%%%%%%%%%%%%%%%%%%
\appendix	 	
\label{sec:appendix}
%%%%%%%%%%%%%%%%%%%%%%%%%%%%%%%%%%%%%%%%%%

%%%%%%%%%%%%%%%%%%%%%%%%%%%%%%%%%%%%%%%%%%
\section{Definition of local easy axes}			\label{sec:AppSpinDef}
%%%%%%%%%%%%%%%%%%%%%%%%%%%%%%%%%%%%%%%%%%

We provide positions and definitions for the local easy axes $\vec e_{i}$ of Ising 
spins [see Eq.~(\ref{eq:Jscaling})] 
for the kagome (Tab.~\ref{tab:KagomeCoord}), pyrochlore (Tab.~\ref{tab:PyrochloreCoord}), 
hyperkagome (Tab.~\ref{tab:HyperkagomeCoord}) and trillium lattice (Tab.~\ref{tab:TrilliumCoord}). 
Models with global and local axes are equivalent up to a rescaling in the exchange coupling $J$ 
given in each table caption.

%%%%%%%%%%%%%%%%%%%%%%%%%%%%%%%%%%%%%%%%%%
% Tab.    KAGOME LOCAL 
%%%%%%%%%%%%%%%%%%%%%%%%%%%%%%%%%%%%%%%%%%
%
%%%%%%%%%%%%%%%%%%%%%%%%%%
\renewcommand{\arraystretch}{3}
\begin{table}[ht]
\def\arraystretch{1.5}
\centering
\begin{tabular}{ C  C  C }
	\hhline{===}		
	site index $i$ & ${\vec e}_i$					& position\\
	\hhline{---}
	1	& $ \left(0, 1 \right)$					& $ \left( \frac{1}{2},0 \right)$\\
	2	& $ \frac{1}{2} \left(-\sqrt{3}, -1 \right)$		& $\frac{1}{4} \left(3, \sqrt{3} \right)$\\
	3	& $ \frac{1}{2} \left(\sqrt{3}, -1 \right)$	& $\frac{1}{4} \left(1, \sqrt{3} \right)$\\
	\hhline{===}
\end{tabular}
\caption{ 
The three sublattices of the kagome lattice with their local easy axes ${\vec e}_i$ and their positions. 
The corresponding lattice vectors are \mbox{$\vec{a}=(1,0)$}, \mbox{$\vec{b}=\frac{1}{2}(-1, \sqrt{3})$}. 
The rescaling of exchange coupling between local and global axes is 
\mbox{$J^{\rm kagome}_{\rm local} = -2 J^{\rm kagome}_{\rm global}<0$}.
}
\label{tab:KagomeCoord}
\end{table}
%%%%%%%%%%%%%%%%%%%%%%%%
%

%%%%%%%%%%%%%%%%%%%%%%%%%%%%%%%%%%%%%%%%%%
% Tab.    PYROCHLORE LOCAL 
%%%%%%%%%%%%%%%%%%%%%%%%%%%%%%%%%%%%%%%%%%
%
%%%%%%%%%%%%%%%%%%%%%%%%%%
\renewcommand{\arraystretch}{3}
\begin{table}
	\def\arraystretch{1.5}
	\centering
		\begin{tabular}{ C  C  C }
			\hhline{===}		
			site index $i$	& $\vec{e}_i$ 					& position		\\
			\hhline{---}
			1			& $\frac{1}{\sqrt{3}} (+1, +1, +1)$		& $\frac{1}{8} (-3, -3, 1)$	\\
			2			& $\frac{1}{\sqrt{3}} (-1,  +1, -1)$		& $\frac{1}{8} (-1, -3, 3)$ 	\\
			3			& $\frac{1}{\sqrt{3}} (+1, -1, -1)$		& $\frac{1}{8} (-3, -1, 3)$	\\
			4			& $\frac{1}{\sqrt{3}} (-1, -1, +1)$		& $\frac{1}{8} (-1, -1, 1)$	\\
			5			& $\frac{1}{\sqrt{3}} (+1, +1, +1)$		& $\frac{1}{8} (1, -3, -3)$	\\
			6			& $\frac{1}{\sqrt{3}} (-1, +1, -1)$		& $\frac{1}{8} (3, -3, -1)$ 	\\
			7			& $\frac{1}{\sqrt{3}} (+1, -1, -1)$		& $\frac{1}{8} (1, -1, -1)$	\\
			8			& $\frac{1}{\sqrt{3}} (-1, -1, +1)$		& $\frac{1}{8} (3, -1, -3)$	\\
			9			& $\frac{1}{\sqrt{3}} (+1, +1, +1)$		& $\frac{1}{8} (-3, 1, -3)$	\\
			10			& $\frac{1}{\sqrt{3}} (-1, +1, -1)$		& $\frac{1}{8} (-1, 1, -1)$ 	\\
			11			& $\frac{1}{\sqrt{3}} (+1, -1, -1)$		& $\frac{1}{8} (-3, 3, -1)$	\\
			12			& $\frac{1}{\sqrt{3}} (-1, -1, +1)$		& $\frac{1}{8} (-1, 3, -3)$	\\
			13			& $\frac{1}{\sqrt{3}} (+1, +1, +1)$		& $\frac{1}{8} (1, 1, 1)$	\\
			14			& $\frac{1}{\sqrt{3}} (-1, +1, -1)$		& $\frac{1}{8} (3, 1, 3)$ 	\\
			15			& $\frac{1}{\sqrt{3}} (+1, -1, -1)$		& $\frac{1}{8} (1, 3, 3)$	\\
			16			& $\frac{1}{\sqrt{3}} (-1, -1, +1)$		& $\frac{1}{8} (3, 3, 1)$	\\
			\hhline{===}
		\end{tabular}
	\caption{ 
	The 16 sublattices in the cubic unit cell of the pyrochlore lattice with their local easy axes 
	${\vec e}_i$ and their positions. 
	The corresponding lattice vectors respect the cubic symmetry of the lattice \mbox{$\vec{a}=(1,0,0)$}, 
	\mbox{$\vec{b}=(0,1,0)$}, \mbox{$\vec{c}=(0,0,1)$}. 
	The rescaling of exchange coupling between local and global axes is 
	\mbox{$J^{\rm pyrochlore}_{\rm local} = -3 J^{\rm pyrochlore}_{\rm global}<0$}.
	}
	\label{tab:PyrochloreCoord}
\end{table}
%%%%%%%%%%%%%%%%%%%%%%%%
%

%%%%%%%%%%%%%%%%%%%%%%%%%%%%%%%%%%%%%%%%%%
% Tab.    HYPERKAGOME LOCAL 
%%%%%%%%%%%%%%%%%%%%%%%%%%%%%%%%%%%%%%%%%%
%
%%%%%%%%%%%%%%%%%%%%%%%%%%
\renewcommand{\arraystretch}{3}
\begin{table}
\def\arraystretch{1.5}
\centering
\begin{tabular}{ C  C  C }
\hhline{===}		
site index $i$	& ${\vec e}_i$ 			& position		\\
\hhline{---}
1	& $\frac{1}{\sqrt{3}} (+1,+1,+1)$		& $\frac{1}{8} (-3, -3, 1)$	\\
2	& $\frac{1}{\sqrt{3}} ( -1,+1, -1)$	& $\frac{1}{8} (-1, -3, 3)$ 	\\
3	& $\frac{1}{\sqrt{3}} (-1, -1, +1)$		& $\frac{1}{8} (-1, -1, 1)$	\\
4	& $\frac{1}{\sqrt{3}} (+1,+1,+1)$		& $\frac{1}{8} (1, -3, -3)$	\\
5	& $\frac{1}{\sqrt{3}} (+1, -1, -1)$	& $\frac{1}{8} (1, -1, -1)$	\\
6	& $\frac{1}{\sqrt{3}} (-1, -1, +1)$		& $\frac{1}{8} (3, -1, -3)$	\\
7	& $\frac{1}{\sqrt{3}} (+1,+1,+1)$		& $\frac{1}{8} (-3, 1, -3)$	\\
8	& $\frac{1}{\sqrt{3}} ( -1,+1, -1)$	& $\frac{1}{8} (-1, 1, -1)$ 	\\
9	& $\frac{1}{\sqrt{3}} (+1, -1, -1)$	& $\frac{1}{8} (-3, 3, -1)$	\\
10	& $\frac{1}{\sqrt{3}} ( -1,+1, -1)$	& $\frac{1}{8} (3, 1, 3)$ 	\\
11	& $\frac{1}{\sqrt{3}} (+1, -1, -1)$	& $\frac{1}{8} (1, 3, 3)$	\\
12	& $\frac{1}{\sqrt{3}} (-1, -1, +1)$		& $\frac{1}{8} (3, 3, 1)$	\\
\hhline{===}
\end{tabular}
\caption{ 
The twelve sublattices of the hyperkagome lattice with their local easy axes ${\vec e}_i$ and their positions. 
The corresponding lattice vectors respect the cubic symmetry of the lattice \mbox{$\vec{a}=(1,0,0)$},
 \mbox{$\vec{b}=(0,1,0)$}, \mbox{$\vec{c}=(0,0,1)$}. 
The rescaling of exchange coupling between local and global axes is 
\mbox{$J^{\rm hyperK}_{\rm local} = -3 J^{\rm hyperK}_{\rm global}<0$}.
}
\label{tab:HyperkagomeCoord}
\end{table}
%%%%%%%%%%%%%%%%%%%%%%%%
%

%%%%%%%%%%%%%%%%%%%%%%%%%%%%%%%%%%%%%%%%%%
% Tab.    TRILLIUM LOCAL 
%%%%%%%%%%%%%%%%%%%%%%%%%%%%%%%%%%%%%%%%%%
%
%%%%%%%%%%%%%%%%%%%%%%%%%%
\renewcommand{\arraystretch}{3}
\begin{table}
\def\arraystretch{1.5}
\centering
\begin{tabular}{ C  C  C }
\hhline{===}		
site index $i$	& ${\vec e}_i$ 		& position						\\
\hhline{---}
1& $\frac{1}{\sqrt{3}} (+1, +1, +1)$	& $(u, u, u)$				\\
2& $\frac{1}{\sqrt{3}} (+1, -1, -1)$	& $(\frac{1}{2}+u, \frac{1}{2}-u, 1-u)$ \\
3& $\frac{1}{\sqrt{3}} (-1, +1, -1)$	& $(1-u, \frac{1}{2}+u, \frac{1}{2}-u)$	\\
4& $\frac{1}{\sqrt{3}} (-1, -1, +1)$	& $(\frac{1}{2}-u, 1-u, \frac{1}{2}+u)$ \\
\hhline{===}
\end{tabular}
\caption{ 
The four sublattices of the trillium lattice with their local easy axes ${\vec e}_i$ and their positions. 
The corresponding lattice vectors respect the cubic symmetry of the lattice \mbox{$\vec{a}=(1,0,0)$}, 
\mbox{$\vec{b}=(0,1,0)$}, \mbox{$\vec{c}=(0,0,1)$}. 
The rescaling of exchange coupling between local and global axes is 
\mbox{$J^{\rm trillium}_{\rm local} = -3 J^{\rm trillium}_{\rm global}<0$}.
The explicit position of each site within the unit cell is given by the crystal parameter 
$u = 0,138$ in order to compare to previous work \cite{Hopkinson2006, Isakov2008, Redpath2010}.
}
\label{tab:TrilliumCoord}
\end{table}
%%%%%%%%%%%%%%%%%%%%%%%%
%

\newpage

%%%%%%%%%%%%%%%%%%%%%%%%%%%%%%%%%%%%%%%%%%
\section{Monte Carlo simulations}			\label{sec:AppMC}
%%%%%%%%%%%%%%%%%%%%%%%%%%%%%%%%%%%%%%%%%%

Numerical Monte Carlo (MC) simulations of the Hamiltonian $\mathpzc{H}$ 
[Eq.~(\ref{eq:Ham})] for Ising spins (Ising model) were performed by updating 
the site dependent Ising variable $\sigma = \pm 1$ for systems larger than 
$N = 10 000$ spins. 
To account for statistically independent samples at very low temperatures a local 
single-spin flip Metropolis update algorithm has been used in combination with 
parallel tempering \cite{Swendsen1986, Earl2005}, and a worm-update algorithm  
\cite{barkema98a, Prokofev2001} in the case of  the checkerboard, pyrochlore 
and ruby lattice.
A single MC step consists of $N$ local single spin-flip updates on randomly chosen sites, 
and 5 worm updates (checkerboard, pyrochlore and ruby lattice), performed in parallel for replicas at 
100 to 200 different temperatures, with replica-exchange initiated by the parallel tempering 
algorithm every $10^2$ MC step.

MC simulations of the Hamiltonian $\mathpzc{H}$ [Eq.~(\ref{eq:Ham})] for Heisenberg 
spins (Heisenberg model) were performed
by using a local heat-bath algorithm \cite{Olive1986, Miyatake1986}, in combination with 
parallel tempering \cite{Swendsen1986, Earl2005}, and over-relaxation \cite{Creutz1987}. 
Here, a single MC step consists of $N$ local heat-bath updates on randomly chosen sites,
with $N$ over-relaxation steps, flipping the spin direction about their local exchange 
field, and replica-exchange every $10^2$ MC step.

In both cases, simulations for Ising and Heisenberg models, thermodynamic quantities were averaged over $10^6$ statistically independent samples, after $10^6$ MC steps for simulated annealing and $10^6$ MC steps for thermalization.

%%%%%%%%%%%%%%%%%%%%%%%%%%%%%%%%%%%%%%%%%%
\section{Husimi Tree}		\label{sec:AppHT}
%%%%%%%%%%%%%%%%%%%%%%%%%%%%%%%%%%%%%%%%%%
%

%%%%%%%%%%%%%%%%%%%%%%%%%%%%%%%%%%%%%%%%%%
\subsection{Explicit calculations for the kagome Ising antiferromagnet}
\label{sec:AppHT1}
%%%%%%%%%%%%%%%%%%%%%%%%%%%%%%%%%%%%%%%%%%
%

\begin{figure}
\centering
\captionsetup[subfigure]{justification=justified, singlelinecheck=false, position=top}
	\subfloat[\label{fig:HT_Kagome2}]{\includegraphics[height=0.13\textheight]{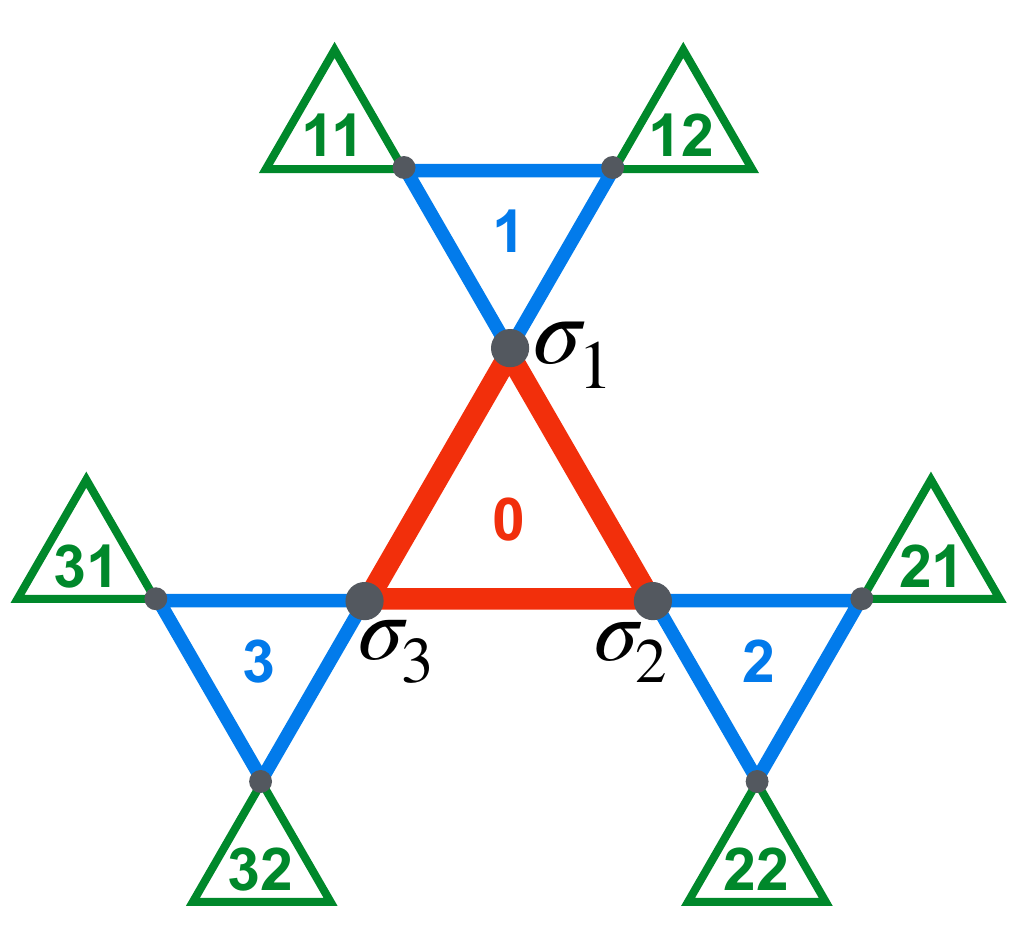}}	
	\subfloat[\label{fig:HT_Configs}]{\includegraphics[height=0.13\textheight]{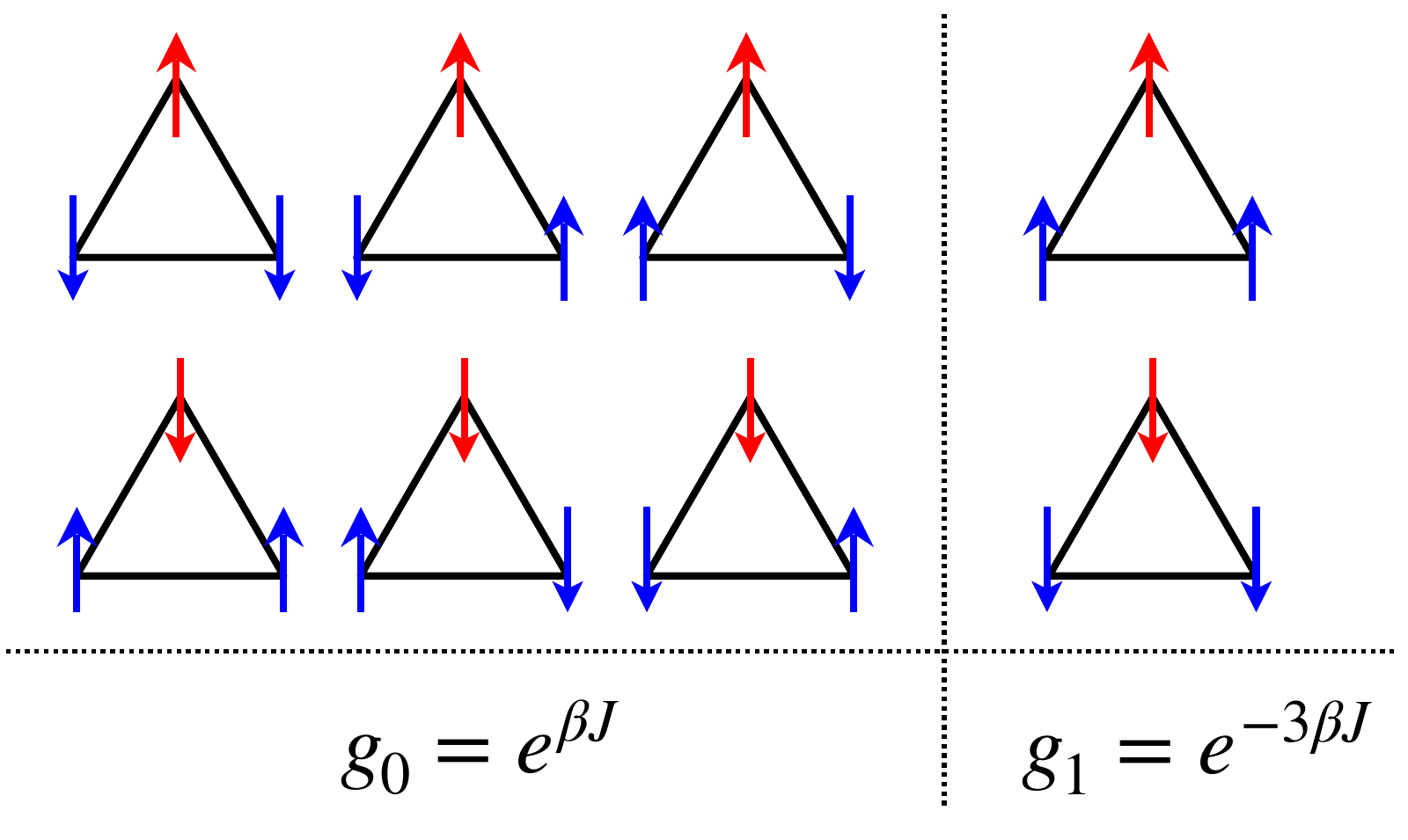}}	
	\caption{
	The Husimi tree HT(3,2) for the kagome lattice.
	(a) Triangular frustrated cells arranged in shells, where shell 0 (red) corresponds to the 
	central unit. 
	The Husimi tree is equivalent to the real kagome lattice up to its 2nd shell (green).
	(b) All possible spin configurations for an isolated triangular cell with corresponding 
	Boltzmann weights $g_{0}$ and $g_{1}$ for global axis Ising spins
	[see Eq.~(\ref{eq:BoltzmannHusimiTree})].
  	}
\label{fig:KagomeHusimi}
\end{figure}

In this section, the Husimi tree calculation shall be explained on the example of HT(3,2) [see Figs.\ref{fig:HT_Lattices}(a) and \ref{fig:KagomeHusimi}(a)]. Branches of nonintersecting triangular cells spread out from the central unit (shell 0, drawn in red). Let us consider the Hamiltonian Eq.~(\ref{eq:Ham}) for Ising spins $\sigma_i$ on sites $i$ with an additional external magnetic field $h$ 
%
%%%%%%%%%%%%%%%%%%%%%%%%%%%
\begin{equation}
	\mathpzc{H}  = J \sum_{\langle ij \rangle}  \sigma_i \sigma_j  - h \sum_i \sigma_i\,  .
\end{equation}
%%%%%%%%%%%%%%%%%%%%%%%%%%%
%
At the end of the calculations, the magnetic field will be taken vanishingly small in order to 
obtain the susceptibility $\chi$. The magnetisation on one of the central site (chosen arbitrarily) is
%
%%%%%%%%%%%%%%%%%%%%%%%%%%
\begin{widetext}
	\begin{align}
		\langle \sigma_1 \rangle 	= \frac{1}{\mathpzc{Z}_0} \sum_{\{\sigma_1, \sigma_2, \sigma_3\}} \sigma_1 
		&\Bigg( \prod_{\langle i j \rangle} g_{ij} \Bigg) \Bigg( \prod_{i=1}^3 \alpha_{i} \Bigg)  \cdot  
		\mathpzc{Z}_1(\sigma_1) \mathpzc{Z}_1(\sigma_2) \mathpzc{Z}_1(\sigma_3)	\; ,	\label{eq:MagHT}	\\
		\textrm{with} \;
		\mathpzc{Z}_0 			=\sum_{\{\sigma_1, \sigma_2, \sigma_3\}} 
		&\Bigg(  \prod_{\langle i j \rangle} g_{ij}  \Bigg) \Bigg( \prod_{i=1}^3 \alpha_{i}  \Bigg)   \cdot  
		\mathpzc{Z}_1(\sigma_1) \mathpzc{Z}_1(\sigma_2) \mathpzc{Z}_1(\sigma_3)	\; ,	\label{eq:PartHT}
	\end{align}
\end{widetext}
%%%%%%%%%%%%%%%%%%%%%%%%%%
%
being the total partition function. 
$\prod_{\langle ij \rangle}$ denotes the product over all nearest-neighbour pairs within the central triangular 
plaquette. 
$\mathpzc{Z}_1(\sigma_{i})$ is the partition function of the branch of the Husimi tree moving outwards and starting 
from the central spin $i$ with orientation $\sigma_{i}$. 
%

%%%%%%%%%%%%%%%%%%%%%%%%%%%%%%%%%%%%%%%%%%
% Fig.: A   HUSIMI TREE ON KAGOME 
%%%%%%%%%%%%%%%%%%%%%%%%%%%%%%%%%%%%%%%%%%
%
%%%%%%%%%%%%%%%%%%%%%%%%%%
%%%%%%%%%%%%%%%%%%%%%%%%%%
%

Let us label $\mathpzc{Z}_{n}(\sigma_{j})$ the partition function starting on site $j$ belonging to the $n^{\rm th}$ 
layer of the tree. 
The  Boltzmann weights are
%
%%%%%%%%%%%%%%%%%%%%%%%%%%
\begin{align}
	g_{ij}		&= e^ {-\beta J \sigma_i \sigma_j} 		\; ,\label{eq:BoltzmannHusimiTree} \\
	\alpha_{i}		&= e^ { \beta h \sigma_i } 				\; ,
\end{align}
%%%%%%%%%%%%%%%%%%%%%%%%%%
%
taking the values $g_0 = e^ {\beta J }$, and $g_1 = e^ {- 3 \beta J }$ here [Fig.~\ref{fig:KagomeHusimi}(b)].
Eq.~(\ref{eq:MagHT}) then gives explicitly 
%
%%%%%%%%%%%%%%%%%%%%%%%%%%
\begin{equation}
	\langle \sigma_1 \rangle 	=	
		\frac{g_0 (Y_1 - Y_1^2) + g_1 (1 - Y_1^3)}{3g_0(Y_1 + Y_1^2) + g_1(1 + Y_1^3)}	\; ,
\label{eq:ExampleKagHT}
\end{equation}
%%%%%%%%%%%%%%%%%%%%%%%%%%
%
where we introduced the ratio between partition functions of a spin on shell $n$,
pointing $\uparrow$  ($\sigma = 1$) and a spin pointing down $\downarrow$ ($\sigma = -1)$ \cite{Baxter2007}
as
%
%%%%%%%%%%%%%%%%%%%%%%%%%%
\begin{equation}
	\frac{\mathpzc{Z}_n(\downarrow)}{ \mathpzc{Z}_n(\uparrow)}  = Y_n \ e^{2 \beta h}	\; ,
\label{eq:PartitionFunctionRatioHT}
\end{equation}
%%%%%%%%%%%%%%%%%%%%%%%%%%
%
and where successive layers of the Husimi tree are related recursively
%
%%%%%%%%%%%%%%%%%%%%%%%%%%
\begin{equation}
	Y_n = \frac{g_0 Y_{n+1}^2 + g_1(1+2Y_{n+1})}{g_0+g_1Y_{n+1} (2 + Y_{n+1})}
\label{eq:PartitionFunctionrecursion}
\end{equation}
%%%%%%%%%%%%%%%%%%%%%%%%%%
%
To solve the Husimi tree, we calculate the limit $Y_{n}\xrightarrow[n \rightarrow +\infty]{} Y$ and replace 
it in Eq.~(\ref{eq:ExampleKagHT}), $Y_{1}=Y$ \footnote{Please note that we use this numbering of 
layers -- increasing outwards -- essentially for pedagogical reasons. Mathematically, the problem is well posed if we 
consider a change of variable: layer 0 would be on the outer boundary, infinitely far away from the central layer which 
would be labeled $n$. Following this notation, this is why the limit $Y$ can be used in Eq.~(\ref{eq:ExampleKagHT}).}. 
In absence of an external magnetic field $Y = 1$, since the disordered system does not prefer any spin direction. 
This gives $\langle \sigma_1 \rangle=0$ as trivially expected. But other observables such as the energy $E$, specific 
heat $C_h$ and entropy $S$ can be derived analytically from the partition function $\mathpzc{Z}_0 $. 
These calculations are relatively straightforward and explicit results for the different Husimi trees are given in Appendix~\ref{sec:AppHT2}.

In this section, we will further show the calculation of the susceptibility. 
An external magnetic field $h$ causes a perturbation $\epsilon$ away from the trivial value
%
%%%%%%%%%%%%%%%%%%%%%%%%%%
\begin{equation}
	Y = 1 - \epsilon		\; ,
\label{eq:HusimiPerturbation}
\end{equation}
%%%%%%%%%%%%%%%%%%%%%%%%%%
%  
which can be used together with Eqs.~(\ref{eq:PartitionFunctionRatioHT}),(\ref{eq:PartitionFunctionrecursion}) to 
obtain $\epsilon$ in first order of $h$
%
%%%%%%%%%%%%%%%%%%%%%%%%%%
\begin{equation}
	\epsilon 	= 2 \beta  h \  \frac{3 g_0 + g_1}{5 g_0 - g_1}	\; .
\label{eq:EpsilonHusimiKagome}
\end{equation}
%%%%%%%%%%%%%%%%%%%%%%%%%%
%
The first-order expansion in $h$ is sufficient to compute the magnetic susceptibility, since higher-order terms vanish as $h \to 0$.
Introducing Eqs.~(\ref{eq:PartitionFunctionRatioHT})--(\ref{eq:EpsilonHusimiKagome}) into Eq.~(\ref{eq:ExampleKagHT}) gives the 
temperature-dependent magnetisation
%
%%%%%%%%%%%%%%%%%%%%%%%%%%
\begin{equation}
	\langle \sigma_1 \rangle 	= \beta  h	\ \frac{ g_0 + 3 g_1 }{5 g_0 - g_1} 	\; .
\end{equation}
%%%%%%%%%%%%%%%%%%%%%%%%%%
%
and the reduced susceptibility
%
%%%%%%%%%%%%%%%%%%%%%%%%%%
\begin{equation}
	\chi T \equiv \frac{1}{\beta}\dfrac{\partial \langle \sigma_1 \rangle}{\partial h}\Big|_{h\rightarrow 0^{+}}= \frac{ g_0 + 3 g_1 }{5 g_0 - g_1} 	\; .
\label{eq:HT.suscept}
\end{equation}
%%%%%%%%%%%%%%%%%%%%%%%%%%
%

%%%%%%%%%%%%%%%%%%%%%%%%%%%%%%%%%%%%%%%%%%
\subsection{Analytic Equations}		\label{sec:AppHT2}
%%%%%%%%%%%%%%%%%%%%%%%%%%%%%%%%%%%%%%%%%%
%
Next to the magnetization and reduced susceptibility [Eq.~(\ref{eq:HT.suscept})], 
thermodynamic observables like energy $E$, specific heat $C_h$ and entropy $S$ are directly 
obtained from the partition function of the Husimi tree [Eq.~(\ref{eq:PartHT})] \cite{Jaubert09c}.
%
%%%%%%%%%%%%%%%%%%%%%%%%%%
\begin{align}
E &= -\frac{1}{\mathpzc{Z}_0} \left( \frac{\partial \mathpzc{Z}_0}{\partial \beta} \right) \Bigg|_{ \{h, \epsilon\} \to 0}	\, ,
\end{align}
\begin{align}
C_h &= - \beta^2 \left( \frac{\partial E}{\partial \beta} \right)  \Bigg|_{\{ h, \epsilon\} \to 0} \, ,
\end{align}
\begin{align}
S &= \beta E + \log{\frac{\mathpzc{Z}_0}{\mathpzc{Z}_1(\uparrow)} A_0}	\, ,   
\end{align}
%%%%%%%%%%%%%%%%%%%%%%%%%%
%
where $A_0$ is fitted such that $S|_{T \to \infty} = \log(2)$.

Here we show analytic expressions for thermodynamic 
observables, as obtained by HT calculations.
All thermodynamic observables are plotted in Fig.~\ref{fig:MC_HT_Thermodyn},  
for $J = 1$, inducing antiferromagnetic correlations between spins. 

Husimi tree HT(3,2) corresponding to the kagome and hyperkagome lattices:
%
%%%%%%%%%%%%%%%%%%%%%%%%%%
\begin{align}
	E 				&= 2 J \ \frac{- g_0 + g_1}{3g_0 + g1} 		\, ,	\\
	C_h 				&=  32 J^2 \beta^2 \ \frac{g_0 g_1}{(3g_0 + g_1)^2}	  	\, ,	\\
	S				&= 2 J \beta \ \frac{- g_0 + g_1}{3g_0 + g1}  
				 	+ \frac{2}{3} \log{\left[ \frac{1}{\sqrt{2}} (3g_0 + g_1)  \right] }		\, ,  \\
	\chi_{\sf glob}T 		&= \frac{ g_0 + 3g_1 }{5g_0 - g_1}	\, ,	
\label{eq:HT.32.a}
\end{align}
%%%%%%%%%%%%%%%%%%%%%%%%%%
%
where $g_0 = e^ {\beta J }$, and $g_1 = e^ {- 3 \beta J}$.

Husimi tree HTS corresponding to the square-kagome lattice:
%
%%%%%%%%%%%%%%%%%%%%%%%%%%
\begin{widetext}
\begin{align}
	E 				&= \frac{2}{3} J \frac{-41 g_0 + 30g_2 + 8g_3 + 3g_4}{41g_0 + 52g_1 + 30g_2 + 4g_3 + g4}   \, , 	\\
	C_h 				&= \frac{16}{3} J^2 \beta^2 \frac{41g_0(26g_1+60g_2+18g_3+11g_4) + 30g_2(2g_3-g_4+26g_1) 
					+ 26g_1(16g_3 + 9 g_4) - g_4(g_4 + 22g_3)}{(41g_0 + 52g_1 + 30g_2 + 4g_3 + g_4)^2}	\\
	S				&= \frac{2}{3} J \beta \frac{-41 g_0 + 30g_2 + 8g_3 + 3g_4}{41g_0 + 52g_1 + 30g_2 + 4g_3 + g4} 
					+ \frac{1}{6} \log{ \left[ \frac{1}{2} (41g_0 + 52g_1 + 30g_2 + 4g_3 + g_4) \right] }		\\
	\chi T 			&= \frac{2}{3} \frac{2g_0 + 7g_1 + 15 g_2 + 5g_3 + 3g_4}{17g_0 + 30 g_1 +16g_2 + 2g_3 - g_4}
					  + \frac{4}{3} \frac{(65g_0 + 381 g_1 + 605 g_2 + 601 g_3) + g_4(275g_1+103g_2 + 15g_3 + 3g_4)}
					  	{(41g_0 + 52g_1 + 30g_2 + 4g_3 + g_4) (17g_1 + 13g_2 + 3g_3 - g_4)}
\end{align}
\end{widetext}
%%%%%%%%%%%%%%%%%%%%%%%%%%
%
where $g_0 = e^ {4 \beta J }$, $g_1 = 1$, $g_2 = e^{-4 \beta J}$, $g_3 = e^{-8 \beta J}$, and $g_4 = e^{-12 \beta J}$.

Husimi tree HT(3,3) corresponding to the triangular and trillium lattice:
%
%%%%%%%%%%%%%%%%%%%%%%%%%%
\begin{align}
	E 		&=  3J \frac{- g_0 + g_1 }{3 g_0 + g_1}		\, , 	\\
	C_h 		&=  48 J^2 \beta^2 \frac{g_0 g_1}{(3g_0 + g_1)^2}	\, ,	\\
	S		&= 3 J \beta \ \frac{- g_0 + g_1}{3g_0 + g1}  + \log{ \left[ \frac{1}{2} ( 3g_0 + g_1 ) \right]}		\, ,	\\
	\chi T 	&= \frac{g_0 + 3g_1}{7g_0 - 3g_1}	\, ,
	\label{eq:trilliumchiT}
\end{align}
%%%%%%%%%%%%%%%%%%%%%%%%%%
%
where $g_0 = e^ { \beta J }$, and $g_1 = e^ {- 3 \beta J}$.

Husimi tree HT(4,2) corresponding to the checkerboard, ruby and pyrochlore lattice:
%
%%%%%%%%%%%%%%%%%%%%%%%%%%
\begin{align}
	E 	 &= - 3 J \ \frac{g_0 -g_2}{3g_0 + 4g_1 + g_2}	\, , 	\\
	C_h 	 &=  24 J^2 \beta^2 \ \frac{g_0g_1 + 4g_0 g_2 + 3g_1 g_2}{(3g_0 + 4g_1 + g_2)^2}	\, ,	\\
	S	&=  - 3 J \ \beta \frac{g_0 -g_2}{3g_0 +4g_1 +g_2} + \frac{1}{2} \log{ \left[ \frac{1}{2} ( 3 g_0 + 4g_1 + g_2 ) \right]}	\, ,	\\
	\chi T &= 2 \frac{ g_1 + g_2 }{3 g_0 + 2g_1 -g_2}	\, ,
\end{align}
%%%%%%%%%%%%%%%%%%%%%%%%%%
%
where $g_0 = e^ {2 \beta J }$, $g_1 = 1$ and $g_2 = e^{-6 \beta J}$.

%%%%%%%%%%%%%%%%%%%%%%%%%%%%%%%%%%%%%%%%%%
\subsection{High-temperature expansion of the susceptibility}
\label{sec:AppHTsusc}
%%%%%%%%%%%%%%%%%%%%%%%%%%%%%%%%%%%%%%%%%%

As discussed in Section \ref{sec:Limit-CWfit}, $ \theta_{\text{cw}} $ contributes to the first order 
correction of the Curie law:
%
%%%%%%%%%%%%%%%%%%%%%%%%%%%%%
\begin{equation}
	\frac{1}{\chi} = \frac{T}{C} \left[ 1 - \frac{\theta_{\text{cw}}}{T} \Big( 1 + \Delta(T) \Big) \right] \; .
\end{equation}
%%%%%%%%%%%%%%%%%%%%%%%%%%%%%
%
The same high-temperature expansion can be done for the results from Husimi tree 
calculations, where second and higher-order terms will account for the deviation from the 
Curie-Weiss law.
Curie-constant $C$, Curie temperature $\theta_{\text{cw}}$ and higher-order corrections
$\Delta(T)	$, extracted from the  inverse susceptibility $1/\chi$ for global axes spins 
are summarised as follows: \\

HT(3,2):  
%%%%%%%%%%%%%%%%%%%%%%%%%%%
\begin{equation}
	\begin{split}
		C		&= 1 	\;, \\
		\theta_{\text{cw}} 	&= -4 J	\;, \\	
		 \Delta(T)			&= \frac{J}{T} - \frac{J^2}{3 T^2} - \frac{5 J^3}{3 T^3}  + \cdots \; .  \\
	\end{split}
 \label{eq:HT.32}
\end{equation}
%%%%%%%%%%%%%%%%%%%%%%%%%%%
%

\textrm{HTS}: 
%%%%%%%%%%%%%%%%%%%%%%%%%%%
\begin{equation}
	\begin{split}
		C		&= 1 	\;, \\
		\theta_{\text{cw}} 	&= -4 J	\;, \\	
		 \Delta(T)			&= \frac{J}{T} - \frac{J^2}{3 T^2} - \frac{4 J^3}{3 T^3}  + \cdots \; .  \\
	\end{split}
 \label{eq:HTS}
\end{equation}
%%%%%%%%%%%%%%%%%%%%%%%%%%%
%

\textrm{HT(3,3)}: 
%%%%%%%%%%%%%%%%%%%%%%%%%%%
\begin{equation}
	\begin{split}
		C		&= 1 	\;,	\\
		\theta_{\text{cw}} 	&= -6 J	\;,	\\	
		\Delta(T)			&=  \frac{J}{T} - \frac{J^2}{3 T^2} - \frac{5 J^3}{3 T^3}  + \cdots \; .  \\
	\end{split}
 \label{eq:HT.33}
\end{equation}
%%%%%%%%%%%%%%%%%%%%%%%%%%%

\textrm{HT(4,2)}: 
%%%%%%%%%%%%%%%%%%%%%%%%%%%
\begin{equation}
	\begin{split}
		C 		&= 1 	\;, \\
		\theta_{\text{cw}} 	&= -6J	\;, \\	
		\Delta(T)			&= \frac{J}{T} - \frac{4J^2}{3 T^2} - \frac{5 J^3}{3 T^3}  + \cdots \; .  \\
	\end{split}
 \label{eq:HT.42}
\end{equation}
%%%%%%%%%%%%%%%%%%%%%%%%%%%
%
%\end{widetext}

Since $J=1$, all models show negative values for $\theta_{\text{cw}}$, indicating dominating antiferromagnetic interactions. 
Furthermore, their absolute values correspond to the number of nearest neighbor sites, and measures the 
local exchange field (Weiss field or molecular field) acting on every individual spin.
The deviation $\Delta(T)$ of $\theta_{\text{cw}} $ scales independently of the type of the Husimi tree with $1/T$ in leading order. 
However, the deviation in second-order terms of $1/T^2$ differs between Husimi trees, made of triangular plaquettes
and square plaquettes.
And from this comparison it becomes evident that HTS shows only a small difference of 2\%  compared to HT(3,2) 
[see Tab.~\ref{tab:HT_MC_zeroT}], since their deviation happens from third-order $1/T^3$.

%%%%%%%%%%%%%%%%%%%%%%%%%%%%%%%%%%%%%%%%%%
\subsection{An alternative way to compute $C_{0}$}
\label{sec:AppHT.reducedSus}
%%%%%%%%%%%%%%%%%%%%%%%%%%%%%%%%%%%%%%%%%%

In Appendix~\ref{sec:AppHT1}, the susceptibility was calculated as the linear response to an external magnetic field $h$, when $h\to0$. At zero temperature it is also possible to calculate it as the sum of spin-spin correlations, following Eq.~(\ref{eq:BulkSus0}). When applied to the ground-state ensemble, this method allows to extract the value of the spin-liquid Curie constant $C_{0}$ as has been done for spin-ice related models \cite{gobush1972,yanagawa1979,Jaubert09c,Macdonald11a,Jaubert2013}. For ease of calculations, let us consider that the Husimi tree is made of $L$ layers of spins, centred around a central site instead of a central frustrated unit. It is then common practice to consider this central spin $\vec S_{0}$ as the spin representative of the bulk of the real lattice. This is because $\vec S_{0}$ is the furthest away from the boundary of the tree, and thus less sensitive to surface effects. For a HT of $L$ layers, the spin-liquid Curie constant of Eq.~(\ref{eq:C0}) becomes
%
%%%%%%%%%%%%%%%%%%%%%%%%%%%
\begin{eqnarray}
	C_{0}(L) = 1 + \sum_{\ell =1}^{L}g_{\ell}\;\langle  \vec S_0  \cdot \vec S_\ell \rangle   \, ,
	\label{eq:BulkSusHT}
\end{eqnarray}
%%%%%%%%%%%%%%%%%%%%%%%%%%%
%
where $\langle  \vec S_0  \cdot \vec S_\ell \rangle$ is the correlation between the central spin and one of the spins on layer $\ell\in[1:L]$, in the ground state. $g_{\ell}$ is the number of spins in this layer.

%%%%%%%%%%%%%%%%%%%%%%%%%%%%%%%%%%%%%%%%%%
\subsubsection{Kagome-type Husimi tree with global axis}
\label{sec:AppHT.kagG}
%%%%%%%%%%%%%%%%%%%%%%%%%%%%%%%%%%%%%%%%%%

For HT(3,2), the number of sites per layer is $g_{\ell}=2\times 2^{\ell}$. Using Eq.~(\ref{eq:BulkSusHT}) and 
Eq.~(\ref{eq:corrCl}) with global Ising axis, one gets
%%%%%%%%%%%%%%%%%%%%%%%%%%%
\begin{align}
	C_{0}(L) & = 1 + \sum_{\ell=1}^{L} 2 \cdot 2^\ell \bigg(-\frac{1}{3} \bigg)^\ell	\nonumber \\
	& = 1 +2\left(\dfrac{1-(-2/3)^{L+1}}{1+2/3}-1\right)\nonumber\\
	& = 0.2 -\dfrac{6}{5}\left(-\dfrac{2}{3}\right)^{L+1} \xrightarrow[L\to \infty]{} 0.2\;.
\label{eq:kagG}
\end{align}
%%%%%%%%%%%%%%%%%%%%%%%%%%%
The value of $0.2$ is recovered in the thermodynamic limit of the alternating 
(antiferromagnetic) series of spin-spin correlations.

%%%%%%%%%%%%%%%%%%%%%%%%%%%%%%%%%%%%%%%%%%
\subsubsection{Trillium-type Husimi tree with global axis}
\label{sec:AppHT.trilG}
%%%%%%%%%%%%%%%%%%%%%%%%%%%%%%%%%%%%%%%%%%

For HT(3,3), the number of sites per layer is $g_{\ell}=(3/2)\times 4^{\ell}$. As a consequence, the series of Eq.~(\ref{eq:BulkSusHT}) becomes alternating divergent, because of the boundary
%%%%%%%%%%%%%%%%%%%%%%%%%%%
\begin{align}
	C_{0}(L) & = 1 + \sum_{\ell=1}^{L} \frac{3}{2} \cdot 4^\ell \bigg(-\frac{1}{3} \bigg)^\ell	\nonumber \\
	& = 1 +\frac{3}{2}\left(\dfrac{1-(-4/3)^{L+1}}{1+4/3}-1\right)\nonumber\\
	& = \frac{1}{7} -\dfrac{9}{14}\left(-\dfrac{4}{3}\right)^{L+1}\;.
\label{eq:trilG}
\end{align}
%%%%%%%%%%%%%%%%%%%%%%%%%%%
If the size of the boundary grows faster than the correlations decay, then the series diverges. That being said, even if the calculation is mathematically ill posed, it is interesting to notice that the constant term, $1/7$,  is the same as the one obtained from the complete Husimi-tree calculation [see Eq.~(\ref{eq:trilliumchiT}) in the limit $\beta\rightarrow+\infty$ and Table \ref{tab:HT_MC_zeroT}].
%

%%%%%%%%%%%%%%%%%%%%%%%%%%%%%%%%%%%%%%%%%%
\subsubsection{Pyrochlore-type Husimi tree with global axis}
\label{sec:AppHT.pyroG}
%%%%%%%%%%%%%%%%%%%%%%%%%%%%%%%%%%%%%%%%%%

For HT(4,2), the number of sites per layer is $g_{\ell}=2\times 3^{\ell}$. As a consequence, the sum of Eq.~(\ref{eq:BulkSusHT}) becomes alternating,
%%%%%%%%%%%%%%%%%%%%%%%%%%%
\begin{align}
	C_{0}(L) & = 1 + \sum_{\ell=1}^{L} 2 \cdot 3^\ell \bigg(-\frac{1}{3} \bigg)^\ell	\nonumber \\
	&  \xrightarrow[L\to \infty]{} 1 + 2\left(\dfrac{1}{1+1}-1\right)=0
\label{eq:pyroG}
\end{align}
%%%%%%%%%%%%%%%%%%%%%%%%%%%
As was the case for HT(3,3), even if the calculation is mathematically ill posed, it is interesting to notice that the outcome is the exact result [Table \ref{tab:HT_MC_zeroT}].
%

%%%%%%%%%%%%%%%%%%%%%%%%%%%%%%%%%%%%%%%%%%
\subsubsection{Kagome Husimi tree with local easy axes}
\label{sec:AppHT.kagL}
%%%%%%%%%%%%%%%%%%%%%%%%%%%%%%%%%%%%%%%%%%

Considering local axes makes the calculation a bit more complex, because spins are not collinear anymore. For the kagome lattice, the local easy axes are given in Table \ref{tab:KagomeCoord}, giving $\vec e_{i}\cdot \vec e_{j}=-1/2$ for spins on different sublattices. Eq.~(\ref{eq:BulkSusHT}) then becomes
%
%%%%%%%%%%%%%%%%%%%%%%%%%%%
\begin{align}
	C_{0}(L) = 1 + \sum_{\ell =1}^{L}u_{\ell}\;\left(-\frac{1}{3} \right)^\ell 
					+ \sum_{\ell =1}^{L}v_{\ell}\;\left(-\frac{1}{2}\right)\;\left(-\frac{1}{3} \right)^\ell    \, .
	\label{eq:BulkSusHT.kagL}
\end{align}
%%%%%%%%%%%%%%%%%%%%%%%%%%%
%
From now on, $u_{\ell}$ (resp. $v_{\ell}$) are the number of spins on layer $\ell$ belonging to the same (resp. a different) sublattice as the central spin of reference, $\vec S_{0}$. By definition, we have $u_{\ell}+v_{\ell}=g_{\ell}=2\times 2^{\ell}$ for HT(3,2). It is not difficult to see that these sequences are related by recursion
%
%%%%%%%%%%%%%%%%%%%%%%%%%%%
\begin{eqnarray}
\begin{split}
u_{\ell +1}&=v_{\ell} \\ v_{\ell +1} &= v_{\ell} + 2 \; u_{\ell}
\end{split}
\end{eqnarray}
%%%%%%%%%%%%%%%%%%%%%%%%%%%
%
which gives
%
%%%%%%%%%%%%%%%%%%%%%%%%%%%
\begin{align}
\begin{split}
u_{\ell}&= \frac{2}{3}\;2^{\ell} + \frac{4}{3}\;(-1)^{\ell}\\
v_{\ell}&= \frac{4}{3}\;2^{\ell} - \frac{4}{3}\;(-1)^{\ell}
\end{split}
\label{eq:BulkSusHT.kagL.ulvl}
\end{align}
%%%%%%%%%%%%%%%%%%%%%%%%%%%
%
Injecting Eq.~(\ref{eq:BulkSusHT.kagL.ulvl}) into Eq.~(\ref{eq:BulkSusHT.kagL}), and taking the limit $L\rightarrow +\infty$, finally gives $C_{0}=2$ for the kagome lattice with local easy axes.

%%%%%%%%%%%%%%%%%%%%%%%%%%%%%%%%%%%%%%%%%%
\subsubsection{Spin-ice Husimi tree with local easy axes}
\label{sec:AppHT.siL}
%%%%%%%%%%%%%%%%%%%%%%%%%%%%%%%%%%%%%%%%%%

For 3D spin ice on the pyrochlore lattice [Table \ref{tab:PyrochloreCoord}], the calculation is very similar. The scalar product between spins on different sublattices is now $\vec e_{i}\cdot \vec e_{j}=-1/3$, and the number of spins belonging to the same, $u_{\ell}$, and different, $v_{\ell}$, sublattices are
%
%%%%%%%%%%%%%%%%%%%%%%%%%%%
\begin{align}
\begin{split}
u_{\ell}&= \frac{1}{2}\;3^{\ell} + \frac{3}{2}\;(-1)^{\ell}\\
v_{\ell}&= \frac{3}{2}\;3^{\ell} - \frac{3}{2}\;(-1)^{\ell}
\end{split}
\label{eq:BulkSusHT.siL.ulvl}
\end{align}
%%%%%%%%%%%%%%%%%%%%%%%%%%%
%
which gives $C_{0}=2$ for the pyrochlore lattice with local easy axes. Please note this is the same value, up to a normalisation, as the one calculated for the dielectric constant of cubic ice \cite{gobush1972,yanagawa1979}.

%%%%%%%%%%%%%%%%%%%%%%%%%%%%%%%%%%%%%%%%%%
\subsubsection{Hyperkagome Husimi tree with local easy axes}
\label{sec:AppHT.hkagL}
%%%%%%%%%%%%%%%%%%%%%%%%%%%%%%%%%%%%%%%%%%

There are four different types of spin orientations in the hyperkagome lattice [see Table \ref{tab:HyperkagomeCoord}], labelled 1, 2, 3, 4. Let us assume that the central spin of reference has orientation 1, at no cost in generality. When posing the problem, one quickly sees that the number of spins with orientation 1 in layer $\ell$ is not obvious to calculate, because there are four types of triangles, with orientations $\{1,2,3\}, \{1,2,4\}, \{1,3,4\}, \{2,3,4\}$. Among the $u_\ell$ sites with orientation 1 on layer $\ell$, we need to make a distinction between:
\begin{itemize}
\item the $a_\ell$ spins that have a site with orientation 1 as second neighbor in the internal layers ($n<\ell$),
\item the $b_\ell$ spins that \textit{do not} have a site with orientation 1 as second neighbor in the internal layers.
\end{itemize}
We have $u_\ell =a_\ell + b_\ell$ and $u_{\ell}+v_{\ell}=g_{\ell}=2\times 2^{\ell}$ sites on layer $\ell$ for HT(3,2). If we impose the local geometry of the hyperkagome lattice on HT(3,2), one gets the following recursion relations
%
%%%%%%%%%%%%%%%%%%%%%%%%%%%
\begin{eqnarray}
\begin{split}
a_{\ell +2}&=b_{\ell +1} + a_{\ell} + b_{\ell} \\
b_{\ell +3}&=2\left(a_{\ell +1} + a_{\ell} + b_{\ell}\right) \\
\end{split}
\end{eqnarray}
%%%%%%%%%%%%%%%%%%%%%%%%%%%
%
Imposing the appropriate initial conditions, one gets
%
%%%%%%%%%%%%%%%%%%%%%%%%%%%
\begin{widetext}
\begin{align}
a_\ell= 2^{\ell-2}+(-1)^{\ell}+2^{\ell/2-2} \left(\frac{3}{\sqrt{7}} \sin \left(\ell \left(\pi -\tan ^{-1}\left(\sqrt{7}\right)\right)\right)- \cos \left(\ell \left(\pi -\tan ^{-1}\left(\sqrt{7}\right)\right)\right)\right)\\
b_\ell=
2^{\ell-2}- 2^{\ell/2-2} \left(\frac{1}{\sqrt{7}} \sin \left(\ell \left(\pi -\tan ^{-1}\left(\sqrt{7}\right)\right)\right)-3 \cos \left(\ell \left(\pi -\tan ^{-1}\left(\sqrt{7}\right)\right)\right)\right)
\end{align}
%%%%%%%%%%%%%%%%%%%%%%%%%%%
%
whose sum can be simplified into
%
%%%%%%%%%%%%%%%%%%%%%%%%%%%
\begin{align}
u_\ell= 2^{\ell-1}+(-1)^\ell+\frac{(-1)^{\ell+1} 2^{\frac{\ell}{2}+\frac{1}{2}} \sin \left((\ell-1) \tan ^{-1}\left(\sqrt{7}\right)\right)}{\sqrt{7}}.
\end{align}
\end{widetext}
%%%%%%%%%%%%%%%%%%%%%%%%%%%
%
Since the easy axes of the hyperkagome lattice give $\vec e_{i}\cdot \vec e_{j}=-1/3$ for spins with different orientations, we get
%
%%%%%%%%%%%%%%%%%%%%%%%%%%%
\begin{align}
C_{0} &= 1
+ \sum_{\ell =1}^{+\infty}u_{\ell}\;\left(-\frac{1}{3} \right)^\ell 
+ \sum_{\ell =1}^{+\infty}v_{\ell}\;\left(-\frac{1}{3}\right)\;\left(-\frac{1}{3} \right)^\ell
\nonumber\\
&= 1 + \frac{2}{3} \sum_{\ell =1}^{+\infty}\left(-\frac{1}{3} \right)^\ell \left(2 \, u_{\ell} - 2^\ell\right)
\nonumber\\
&= \frac{5}{3} - \frac{4\sqrt{2}}{3\sqrt{7}} \sum_{\ell =1}^{+\infty}\left(\frac{\sqrt{2}}{3} \right)^\ell\sin \left((\ell-1) \tan ^{-1}\sqrt{7}\right)
\nonumber\\
&= \frac{5}{3} - \frac{4\sqrt{2}}{3\sqrt{7}} \, \times \,\frac{1}{8}\sqrt{\frac{7}{2}}=\frac{3}{2}
\label{eq:BulkSusHT.hkagL}
\end{align}
%%%%%%%%%%%%%%%%%%%%%%%%%%%
%
for the hyperkagome lattice with local easy axes.

%%%%%%%%%%%%%%%%%%%%%%%%%%%%%%%%%%%%%%%%%%
\subsubsection{Trillium Husimi tree with local easy axes}
\label{sec:AppHT.tri2L}
%%%%%%%%%%%%%%%%%%%%%%%%%%%%%%%%%%%%%%%%%%

There are four sublattices in the minimal unit cell of the trillium lattice, labelled 1, 2, 3, 4. Let us assume that the central spin of reference is on sublattice 1, at no cost in generality. As for the hyperkagome case in Appendix \ref{sec:AppHT.hkagL}, there are four types of triangles, with sublattices $\{1,2,3\}, \{1,2,4\}, \{1,3,4\}, \{2,3,4\}$. Among the $v_\ell$ sites that do not belong to sublattice 1 on layer $\ell$, we need to make a distinction between:
\begin{itemize}
\item the $c_\ell$ spins that were in a triangle with a sublattice-1 site in layer $\ell-1$,
\item the $d_\ell$ spins that were \textit{not} in a triangle with a sublattice-1 site in layer $\ell-1$.
\end{itemize}
We have $v_\ell =c_\ell + d_\ell$ and $u_{\ell}+v_{\ell}=g_{\ell}=\frac{3}{2}\times 4^{\ell}$ sites on layer $\ell$ for HT(3,3). If we impose the local geometry of the trillium lattice on HT(3,3), one gets the following recursion relations
%
%%%%%%%%%%%%%%%%%%%%%%%%%%%
\begin{eqnarray}
\left\{
\begin{split}
u_{\ell +1}&=c_{\ell} + 2 d_{\ell}\\
c_{\ell +1}&=4 u_{\ell} + c_{\ell} + 2 d_{\ell} \\
d_{\ell +1}&=2 c_{\ell}
\end{split}
\right.
\end{eqnarray}
%%%%%%%%%%%%%%%%%%%%%%%%%%%
%
which gives a self-consistent recursion relation for the number of sites in sublattice 1
%
%%%%%%%%%%%%%%%%%%%%%%%%%%%
\begin{eqnarray}
u_{\ell+3}=u_{\ell+2}+8\ u_{\ell+1} + 16\ u_\ell
\label{eq:trilsub1}
\end{eqnarray}
%%%%%%%%%%%%%%%%%%%%%%%%%%%
%
whose solution is
%
%%%%%%%%%%%%%%%%%%%%%%%%%%%
\begin{widetext}
\begin{eqnarray}
u_{\ell} = 3\left(2^{2 \ell-3}+2^{\ell-3}\left(\frac{1}{\sqrt{7}} \sin\left(\ell \left(\pi -\tan ^{-1}\left(\frac{\sqrt{7}}{3}\right)\right)\right)+3  \cos \left(\ell \left(\pi -\tan ^{-1}\left(\frac{\sqrt{7}}{3}\right)\right)\right)\right)\right)
\end{eqnarray}
Since the easy axes of the trillium lattice give $\vec e_{i}\cdot \vec e_{j}=-1/3$ for spins on different sublattices, we get
\begin{align}
C_{0}(L) &= 1
+ \sum_{\ell =1}^{L} \left(-\frac{1}{3} \right)^\ell \left( u_{\ell}-\frac{v_{\ell}}{3}\right)
= 1 + \sum_{\ell =1}^{+\infty}\left(-\frac{1}{3} \right)^\ell \left(\frac{4}{3}  \, u_{\ell} - \frac{4^{\ell}}{2} \right)
\nonumber\\
&= 1 + \frac{1}{2}\sum_{\ell =1}^{L}\left(-\frac{2}{3} \right)^\ell \left(\frac{1}{\sqrt{7}} \sin\left(\ell \left(\pi -\tan ^{-1}\left(\frac{\sqrt{7}}{3}\right)\right)\right)+3  \cos \left(\ell \left(\pi -\tan ^{-1}\left(\frac{\sqrt{7}}{3}\right)\right)\right)\right)
\label{eq:triltruc}
%\label{eq:BulkSusHT.trilL}
\end{align}
\end{widetext}
%%%%%%%%%%%%%%%%%%%%%%%%%%%
%
The sum of Eq.~(\ref{eq:triltruc}) converges to zero for $L\rightarrow +\infty$, which is why the Husimi tree for the trillium lattice with easy axes gives $C_0=1$ [see Table \ref{tab:HT_MC_zeroT}]. However, the first term of the sum is positive (it is $2/3$ for $L=1$), which means that the build up of correlations at short distance is primarily ferromagnetic. This is consistent with the increase of the reduced susceptibility $\chi T$ when cooling from high temperature in Fig.~\ref{fig:MC_HT_Thermodyn}(b).

%%%%%%%%%%%%%%%%%%%%%%%%%%%%%%%%%%%%%%%%%%
\subsection{Comment on the Pauling entropy}
\label{sec:Pauling}  
%%%%%%%%%%%%%%%%%%%%%%%%%%%%%%%%%%%%%%%%%%

For ice problems, the Pauling entropy provides a lower bound on the exact value of the entropy \cite{Lieb1972}. Ice problems are defined as systems of connected vertices, where each link between two vertices has a direction (the spin), and each vertex possesses as many inward as outward links -- the so-called ice rules. The ground state of the checkerboard and pyrochlore lattices are ice problems, and their Pauling entropy are indeed lower than their exact residual entropy [Table \ref{tab:HT_MC_zeroT}]. The ground state of the ruby lattice is, however, not an ice problem \cite{Rehn17a}, even if it is also made of corner-sharing tetrahedra with two spins up and two spins down. This is because the centre of the tetrahedra -- the above-mentioned vertices -- form a kagome lattice, which is not bipartite but tripartite. There are three kinds of tetrahedra on the ruby lattice, labeled for convenience red, green and blue. If an up spin is mapped to an outward (inward) link in a red (green) tetrahedron, what happens in the blue tetrahedra? It is easy to show that all-in/all-out states then appear in the blue tetrahedra, and the ground-state ensemble is thus not an ice problem. The ground state of the Ising ruby antiferromagnet is actually a $\mathbb{Z}_{2}$ spin liquid, as opposed to the U(1) gauge structure on pyrochlore \cite{Rehn17a}. Nevertheless, despite these fundamental differences, the thermodynamic quantities of these three models (ruby, checkerboard and pyrochlore) are semi-quantitatively the same for all temperatures, including their residual entropy.

%%%%%%%%%%%%%%%%%%%%%%%%%%%%%%%%%%%%%%%%%%
\section{$C_{0}$ appearing in Coulomb gauge field theory}
\label{sec:gauge}
%%%%%%%%%%%%%%%%%%%%%%%%%%%%%%%%%%%%%%%%%%
%
The spin-ice ground state is famously known as a U(1) Coulomb gauge field \cite{Henley2010}. This gauge-field texture comes from the ice rules (``2 in - 2 out''), that can be rewritten as a divergence-free constraint on the magnetisation field $\vec M(\vec r)$ at position $\vec r$. At lowest order, the probability distribution of $\vec M(\vec r)$ is \cite{Henley2005}
%
%%%%%%%%%%%%%%%%%%%%%%%%%%%
\begin{equation}
	\mathcal{P} \propto \exp\left( - \frac{\kappa_{0}}{2 v_{\rm cell}} \int \textrm{d}\vec r\; |\vec M|^{2}(\vec r) \right)\, ,
\label{eq:Skappa}
\end{equation}
%%%%%%%%%%%%%%%%%%%%%%%%%%%
%
where $v_{\rm cell}$ is the volume of the primitive unit cell. From Eq.~(\ref{eq:Skappa}), the entropic stiffness $\kappa_{0}$ is also the inverse of the variance of the magnetisation in the spin-ice ground state (up to a prefactor), i.e.
%
%%%%%%%%%%%%%%%%%%%%%%%%%%%
\begin{equation}
	\kappa_{0} \propto \frac{1}{C_{0}} \, ,
\label{eq:kappaC0}
\end{equation}
%%%%%%%%%%%%%%%%%%%%%%%%%%%
%
It means that $C_{0}$ is a measure of the (inverse of) the strength of entropic interactions between topological-charge excitations in spin ice \cite{castelnovo11a}. To conclude, the stiffness is also the Lagrange multiplier appearing in the Self-Consistent Gaussian Approximation (SCGA) that ensures the spin-length constraint on average \cite{conlon10b}. For many models with continuous spins, this Lagrange multiplier can be computed analytically in the limit of zero and infinite temperatures, and thus offers an alternative way to compute the ratio $C_{0}/C_{\infty}$ and to connect it to the number of flat bands in the system (see discussion in Section \ref{sec:HTflat}).

%
%%%%%%%%%%%%%%%%%%%%%%%%%%%%%%%%%%%%%%%%%%
\section{S(q) -- equal-time structure factor}
\label{sec:App.Sq}
%%%%%%%%%%%%%%%%%%%%%%%%%%%%%%%%%%%%%%%%%%

The equal-time (energy-integrated) structure factor is defined as
%
%%%%%%%%%%%%%%%%%%%%%
\begin{equation}
	\begin{aligned}
		S(\vec{q}) &=  \frac{1}{N} \sum_{i, j}^N  \mathrm{e}^{ \mathrm{i} \vec{q} (\vec{r}_{i} - \vec{r}_j ) }\left 
				\langle \vec S_{i} \cdot \vec S_{j} \right\rangle \\
				&=  \frac{1}{N} \left< \left| \sum_{i}^N  \mathrm{e}^{ \mathrm{i} \vec{q}\cdot\vec{r}_{i}} 
				\vec S_{i} \right|^2 \right>.		
	\end{aligned}
	\label{eq:Sq}
\end{equation}
%%%%%%%%%%%%%%%%%%%%%
%

%%%%%%%%%%%%%%%%%%%%%%%%%%%%%%%%%%%%%%%%%%%%%
%
%					BIBLIOGRAPHY
%
%%%%%%%%%%%%%%%%%%%%%%%%%%%%%%%%%%%%%%%%%%%%%
\bibliography{Bibliography}
%%%%%%%%%%%%%%%%%%%%%%%%%%%%%%%%%%%%%%%%%%%%%

\end{document}